\documentclass[twocolumn,aps,prb,groupedaddress,eqsecnum,floatfix]{revtex4-1}
\usepackage{silence}
\usepackage{graphicx}
\usepackage{color}
\usepackage{amsmath}
\usepackage{amsfonts}
\usepackage{enumitem}
\usepackage{tikz}
\usepackage[compat=1.1.0]{tikz-feynman}
\usetikzlibrary{shapes.geometric}
\usetikzlibrary{positioning}
\usepackage{environ}
\usetikzlibrary{arrows}
\usepackage{verbatim}
\usepackage{simpler-wick}

\usepackage[colorlinks=true]{hyperref}
\hypersetup{
    bookmarks=true,         
    unicode=false,          
    pdftoolbar=true,        
    pdfmenubar=true,        
    pdffitwindow=false,     
    pdfstartview={FitH},    
    pdftitle={Transport and chaos in lattice Sachdev-Ye-Kitaev models},    
    pdfauthor={Haoyu Guo et al.},     
    pdfsubject={},   
    pdfcreator={},   
    pdfproducer={}, 
    pdfkeywords={} {} {}, 
    pdfnewwindow=true,      
    colorlinks=true,       
    linkcolor=blue, 
    citecolor=blue,        
    filecolor=blue,      
    urlcolor=blue           
}

\newcommand{\Tr}{{\rm \, Tr\,}}
\renewcommand{\Re}{{\rm \, Re\,}}
\renewcommand{\Im}{{\rm \, Im\,}}
\renewcommand{\bar}{\overline}
\renewcommand{\tilde}{\widetilde}

\newcommand{\rOTOC}{\operatorname{OTOC}}

\newcommand{\rd}{{\rm d}}

\newcommand{\midarrow}{\tikz \draw[-triangle 45] (0,0) -- +(0.1,0);}

\begin{document}
\title{Transport and chaos in lattice Sachdev-Ye-Kitaev models}

\author{Haoyu Guo}
\affiliation{Department of Physics, Harvard University, Cambridge, MA 02138, USA}
\author{Yingfei Gu}
\affiliation{Department of Physics, Harvard University, Cambridge, MA 02138, USA}
\author{Subir Sachdev}
\affiliation{Department of Physics, Harvard University, Cambridge, MA
02138, USA}

\date{\today}
\begin{abstract}
We compute the transport and chaos properties of lattices of quantum Sachdev-Ye-Kitaev islands coupled by single fermion hopping, and with the islands coupled to a large number of local, low energy phonons. We find two distinct regimes of linear-in-temperature ($T$) resistivity, and describe the crossover between them. When the electron-phonon coupling is weak, we obtain the `incoherent metal' regime, where there is near-maximal chaos with front propagation at a butterfly velocity $v_B$, and the associated diffusivity $D_{\rm chaos} = v_B^2/(2 \pi T)$ closely tracks the energy diffusivity. On the other hand, when the electron-phonon coupling is strong, and the linear resistivity is largely due to near-elastic scattering of electrons off nearly free phonons, we find that the chaos is far from maximal and spreads diffusively. We also describe the crossovers to low $T$ regimes where the electronic quasiparticles are well defined.
\end{abstract}
\maketitle
\tableofcontents

\section{Introduction}
\label{sec:intro}

\begin{figure}
  \centering
  \includegraphics[width=2.5in]{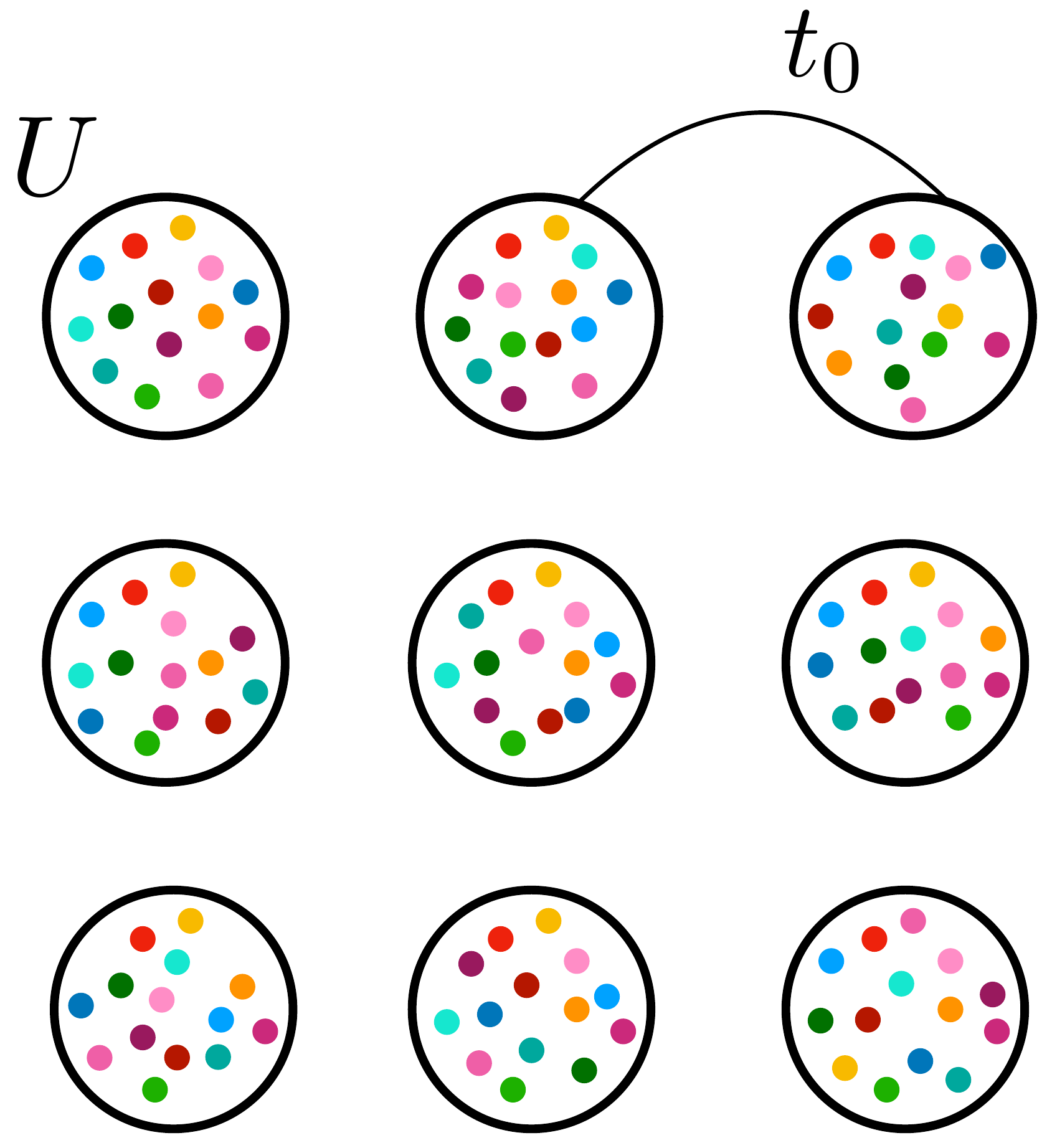}
  \caption{Schematic of the lattice of SYK islands, each with $N$ orbitals with two-body interaction $U$. The islands are coupled with one-body hopping $t_0$.}
  \label{fig:SYKlattice}
\end{figure}

Most strongly correlated metals exhibit ``strange" or ``bad" metal behavior with a linear-in-temperature ($T$) resistivity, with values which can exceed the Mott-Ioffe-Regel limit \cite{Taillefer10}. Recent studies \cite{Gu,Davison17,Song,Zhang2017,Chowdhury,Patel2017,Aurelio2018,Patel19} (and in some earlier related work \cite{PG98}) have shown that such behavior appears naturally in lattice models of coupled `islands', with each island described by a $N$ orbital Sachdev-Ye-Kitaev (SYK) model \cite{SY93,kitaev2015talk,Sachdev2015} of random all-to-all two-body (four-fermion) interactions. When the coupling between the islands is a two-body interaction \cite{Gu,Davison17}, we obtain a non-Fermi liquid metal with a $T$-independent resistivity.
However, with a one-body hopping between islands as in Fig.~\ref{fig:SYKlattice} (the hopping can be random or non-random),\cite{Song,Zhang2017,Chowdhury,Patel2017} we obtain a linear-in-$T$ resistivity for $E_c \ll T \ll U$, where $U$ is the root-mean-square interaction strength within an island, $t_0 \ll U$ is the root-mean-square one-body hopping, and $E_c = t_0^2 /U$.

We note in passing that a pair of SYK islands of Majorana fermions with identical two-body interactions, coupled by one-body hopping, have been used to describe eternal traversable wormholes in a dual gravity theory \cite{MaldacenaQi18,Verbaar19,PZhang19}. We also note that we take the large $N$ limit with $t_0/U$ fixed, with couplings in the Hamiltonian scaled with $N$ as in \eqref{eq:Hamiltonian}, and so implicitly assume {\it e.g.\/} that $t_0 \gg U/N$.

A remarkable consequence of the SYK description is that it opens up insightful connections between strange metal transport and many-body quantum chaos \cite{Shenker13,Maldacena15,kitaev2015talk,Maldacena,KitaevSuh}. The chaos is characterized by a butterfly velocity, $v_B$, and a Lyapunov rate, $\lambda_L$, and it has been argued \cite{Maldacena15} that under certain conditions there is an upper bound on the Lyapunov rate $\lambda_L \leq 2 \pi T$ as $T \rightarrow 0$. We can combine these chaos characteristics to obtain a `chaos diffusion constant' $D_{\rm chaos} = v_B^2 /\lambda_L$. 
Using insights from holographic models, Blake \cite{Blake16,Blake16a} argued that there was a close connection between $D_{\rm chaos}$ and the diffusivities of strange metal transport. Subsequent work noted that while additional parameters appeared in the value of the charge diffusivity \cite{LucasSteinberg}, there was indeed a close connection \cite{Patel2016,Gu,Blake2017} between the values of $D_{\rm chaos}$ and the energy diffusivity, $D_E$. The close connection between chaos and energy diffusion is also a central feature of recent quantum hydrodynamic descriptions \cite{Crossley,BlakeLiu1,BlakeLiu2} of strongly interacting fluids.

In the first part of the present paper, we study the coupled SYK models with one-body hopping introduced by Song {\it et al.} \cite{Song}. We will extend their transport results to computations of out-of-time-order correlators (OTOCs). In extracting the chaos parameters from the OTOCs, we will employ recent insights on the structure of OTOCs by Gu and Kitaev \cite{Gu2}. They argued that large $N$ systems of the type we examine have OTOCs in frequency ($\omega$) and momentum ($q$) space of the form
\begin{equation}
    \rOTOC (q, \omega) \sim\frac{1}{N \cos(\lambda_L(q)/(4T))}\frac{1}{[\omega-i\lambda_L(q)]}\,, \label{OTOC1}
\end{equation}
where the Lyapunov rate $\lambda_L (q)$ is now $q$-dependent. Differing ways of extracting the butterfly velocity, $v_B$, from the $q$ dependence of $\lambda_L (q)$ have been discussed in the literature. Gu and Kitaev argued that in a regime close to maximal chaos, the appropriate method relies on the pole of Eq.~(\ref{OTOC1}) which appears when the Lyapunov rate 
\begin{equation}
    \lambda_L (q_1) = 2 \pi T\,, \label{OTOC2}
\end{equation} 
the maximal value. This happens (as we will show by explicit computation in our model) when the momentum is purely imaginary, $q_1 = i |q_1|$. From the value of $q_1$, we can now define a butterfly velocity and a chaos `diffusion constant' by
\begin{equation}
    v_B = \frac{2 \pi T}{|q_1|} \quad, \quad D_{\rm chaos} = \frac{v_B^2}{2 \pi T} = \frac{2 \pi T}{|q_1|^2} \,. \label{OTOC3}
\end{equation}
We will compute the value of $D_{\rm chaos}$ for the model of Song {\it et al.} \cite{Song}; we find that it closely tracks the energy diffusivity, $D_E$, in the incoherent strange metal regime, as was noticed in earlier models \cite{Patel2016,Blake2017}.

\begin{figure*}
  \centering
 \includegraphics[width=0.85\textwidth]{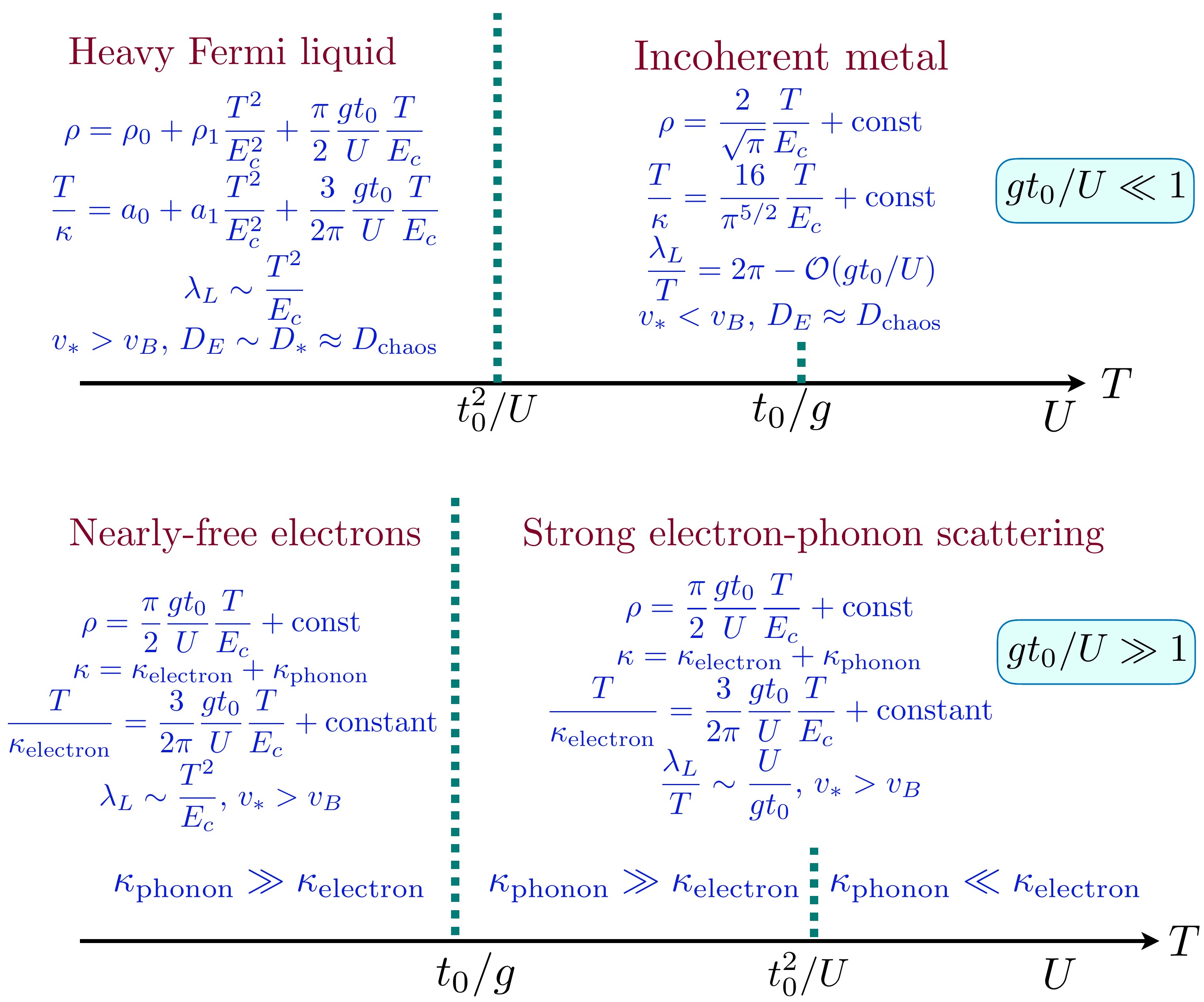}
  \caption{Crossovers as a function $T$ for $gt_0/U \ll 1$ and $gt_0/U \gg 1$, where $g$ is a dimensionless measure of the electron-phonon coupling. The two chaos velocities $v_\ast$ and $v_B$ are defined as in Ref.~\onlinecite{Gu2}. The resistivity is $\rho$ (in units of $h/e^2$), the thermal conductivity is $\kappa$ (in units of $k_B^2 T/\hbar$), and the thermal diffusivity is $D_E$. The chaos exponent $\lambda_L$, and the diffusivities $D_{\rm chaos}$ and $D_\ast$ are defined in Section~\ref{sec:scrambling}. There is near-maximal chaos and front propagation with velocity $v_B$ 
  only in the ``incoherent metal'' regime which has $v_\ast < v_B$. The other regimes have $v_\ast > v_B$ and diffusive propagation of far-from-maximal chaos. Here $\kappa_{\rm phonon}$ refers to the phonon drag correction, discussed in Section~\ref{sec:phonondrag}. The values above do not include the saturation effects discussed in Section~\ref{sec:saturation}.}
  \label{fig:crossovers}
\end{figure*}
The second part of our paper will study the role of phonons in strange metal transport. Our motivation is drawn from recent observations of the thermal diffusivity of a strongly-coupled `electron-phonon soup' in cuprate superconductors \cite{JC16,JC18}. Here we will employ the model of strong electron-phonon coupling introduced by Werman {\it et al.} \cite{Werman,Werman2}, and combine it with the model of strong electron-electron interactions by Song {\it et al.} \cite{Song}. However, our approach does have some limitations, and so a direct contact with observations \cite{JC16,JC18} is not possible at this stage. In our framework, the phonons largely act as a heat bath of free oscillators, which influences the electron dynamics. However, the feedback from the electrons to the phonon dynamics is small, and so it is not appropriate to consider the combined system as a single chaotic soup characterized by a single butterfly velocity. 

We will show that, provided the electron-phonon coupling is not too strong, the phonons do not alter the basic characteristics of the strange metal theory without phonons discussed in Section~\ref{sec:scrambling_tJ}, and summarized in Fig.~\ref{fig:crossovers}. The main influence of the phonons is in altering the slope of the linear-in-$T$ resistivity, and various related numerical prefactors. These corrections are characterized by a single dimensionless parameter $g t_0/U$, where $g$ measures the strength of the electron-phonon coupling. This additional parameter introduces a degree of non-universality in our results, which we expect will be overcome in a theory which uses a more self-consistent approach. 

For large electron-phonon coupling $g t_0/U \gg 1$, the linear-in-$T$ resistivity persists, but the chaos properties are far from maximal and exhibit diffusive chaos propagation. A summary of the crossovers in the transport and chaos properties in a model with both electron-electron and electron-phonon interactions appears in Fig.~\ref{fig:crossovers}. 

The plan of our paper is the following. 
Section~\ref{sec:scrambling} will recall the description of structure of scrambling from Ref.~\onlinecite{Gu2}.
We will start Section~\ref{sec:scrambling_tJ} by reviewing previous results on SYK-based strongly correlated metal (the $t$-$U$ model). Then we present the calculation of the scrambling rate and butterfly velocity in this $t$-$U$ model. We discuss a generalization of the $t$-$U$ model to incorporate phonons in Sections~\ref{sec:phonons}-\ref{sec:phononscrambling}.

\section{Description of Scrambling}
\label{sec:scrambling}

In this section, we review the description of scrambling in a many-body system, following Ref.~\onlinecite{Gu2}. We are going to define three quantities which we will calculate 
in Section~\ref{sec:scrambling_tJ} for the $t$-$U$ model. They are the scrambling rate $\lambda_L$, short-distance scrambling diffusion coefficient $D_*$ and the long-distance scrambling diffusion coefficient $D_{\rm chaos}$.

\subsection{Electron Out of Time Order Correlator}

    We will use the following out-of-time-order correlator (OTOC) to characterize the scrambling:

\begin{equation}\label{eq:OTOC_nonretarded}
    \begin{aligned}
    &\rOTOC(x;t_1,t_3;t_2,t_4)=\\
    &\frac{1}{N^2}\sum_{ab}\Tr\left( y c_{ax}^\dagger(t_1)yc_{b0}(t_3)yc_{ax}(t_2)yc_{b0}^\dagger(t_4)\right)_\text{conn.}, \\
&\qquad \text{where} \qquad
    y^4=\frac{\exp(-\beta H)}{Z}.
    \end{aligned}
\end{equation}
Here $t_1\approx t_2\gg \beta $, $t_3\approx t_4\approx0 $, and the operators are evenly spaced along the imaginary time circle for our convenience.     
    In the time range $\beta \lesssim t \ll \lambda^{-1}_L \ln N$, the OTOC is expected to grow exponentially,
\begin{equation}\label{eq:OTOCgrowth}
    \rOTOC(x;t_1,t_3;t_2,t_4)\propto \frac{1}{N} \exp \left( \lambda_Lt\right),
\end{equation}
where $t$ is the center of mass time separation:
\begin{equation}
    t=\frac{t_1+t_{2}-t_{3}-t_4}{2}\,,
\end{equation}
and $\lambda_L$ is the Lyapunov exponent or scrambling rate. 

We comment on the regularization $y$ in the above definition. 
For a thermalizing system, we expect that the details of the regularization will be washed out after a thermal scale $1/\beta$, and therefore will not affect the Lyapunov exponent \cite{Maldacena15}. 
Technically speaking, the Bethe-Salpeter equation approach we use searches for unstable eigenmodes on double Keldysh countour from a generic initial condition. The shift of regularization will affect the definition of the Wightman propagator $G_W$ at short times (see Appendix~\ref{sec:wightman}), which enters into the Bethe-Salpeter equation. 
Consequently, for the eigenmodes of the equation, only the short time behavior  is affected, not the Lyapunov exponent that characterizes the growth of chaos in a large time window.  For the SYK model, our expectations were verified in Ref.~\onlinecite{romero2019regularization}, and we expect similar results here.

In general, scrambling can propagate in space and thus the OTOC depends on $x$. If we Fourier transform position $x$ to momentum $q$, we will obtain a $q$-dependent scrambling rate $\lambda_L(q)$. For now we are interested in the temporal growth of the OTOC, and consider the translationally invariant scrambling rate $\lambda_L \equiv \lambda_L(q=0)$. Due to the presence of SYK type interactions, we expect that at high temperatures $T\gg E_c$, $\lambda_L$ saturates the chaos bound, {\it i.e.} $\lambda_L/T\approx 2\pi$, and at low temperatures $T\ll E_c$, we expect that $\lambda_L$ is given by the Fermi liquid inelastic scattering rate, $\lambda_L/T\sim T/E_c$ (see Fig.~\ref{fig:crossovers}). 



\subsection{Spatial Propagation of Scrambling and Butterfly Velocity} \label{sec:chaos_propagation}

	We shift the focus to the spatial propagation of scrambling. To discuss the propagation, it is convenient to discuss the Fourier transform of $\rOTOC$ in both space and time 
\begin{equation}\label{eq:f_1_omega}
\begin{split}
    \rOTOC(q,\Omega,\Omega',\omega)=&\int\rd^d x \rd^3 t \rOTOC(x;t_1,t_3;t_2;t_4)\\
\times&e^{-i q\cdot x+i\Omega t_{34}+i\Omega't_{21}+i\omega t},
\end{split}
\end{equation}    
where $t_{ij}=t_i-t_j$, $t=(t_1+t_2-t_3-t_4)/2$, and by time-translation symmetry we have only integrated over three time variables.
As mentioned before, $\lambda_L$ depends on momentum, and this encodes the information about scrambling propagation. The exponential growth in Eq.~\eqref{eq:OTOCgrowth} is translated to a pole singularity $\rOTOC(q,\omega)\sim c/(\omega-i\lambda_L(q))$.
	
As discussed in Section~\ref{sec:intro}, we will analyze the $q$-dependent OTOC using the ladder identity of Ref.~\onlinecite{Gu2}:
the prefactor of the OTOC contains a pole in $q$, as in Eq.~(\ref{OTOC1}):
    \begin{equation}
    		\rOTOC(q,\omega)\sim\frac{1}{N\cos(\lambda_L(q)\beta/4)}\frac{1}{\omega-i\lambda_L(q)}.
    \end{equation}
The pole sits on imaginary $q$-axis at $q_1=i|q_1|$, where $\lambda_L(q_1)=2\pi T$, as noted in Eq.~(\ref{OTOC2}).

To obtain the propagation of scrambling, we Fourier transform the OTOC back to real space:
\begin{equation}\label{eq:OTOCx}
    \rOTOC(x,t)\sim\int\frac{\rd^dq}{(2\pi)^d}\frac{e^{\lambda_L(q)t+iq\cdot x}}{N\cos(\lambda_L(q)\beta/4)},
\end{equation}
where we have evaluated the frequency integral by picking up the $\omega$-pole and omitted regular factors which does not change the qualitative behavior of the OTOC.
\begin{figure}
[t]
\center
\includegraphics[width=0.95 \columnwidth]{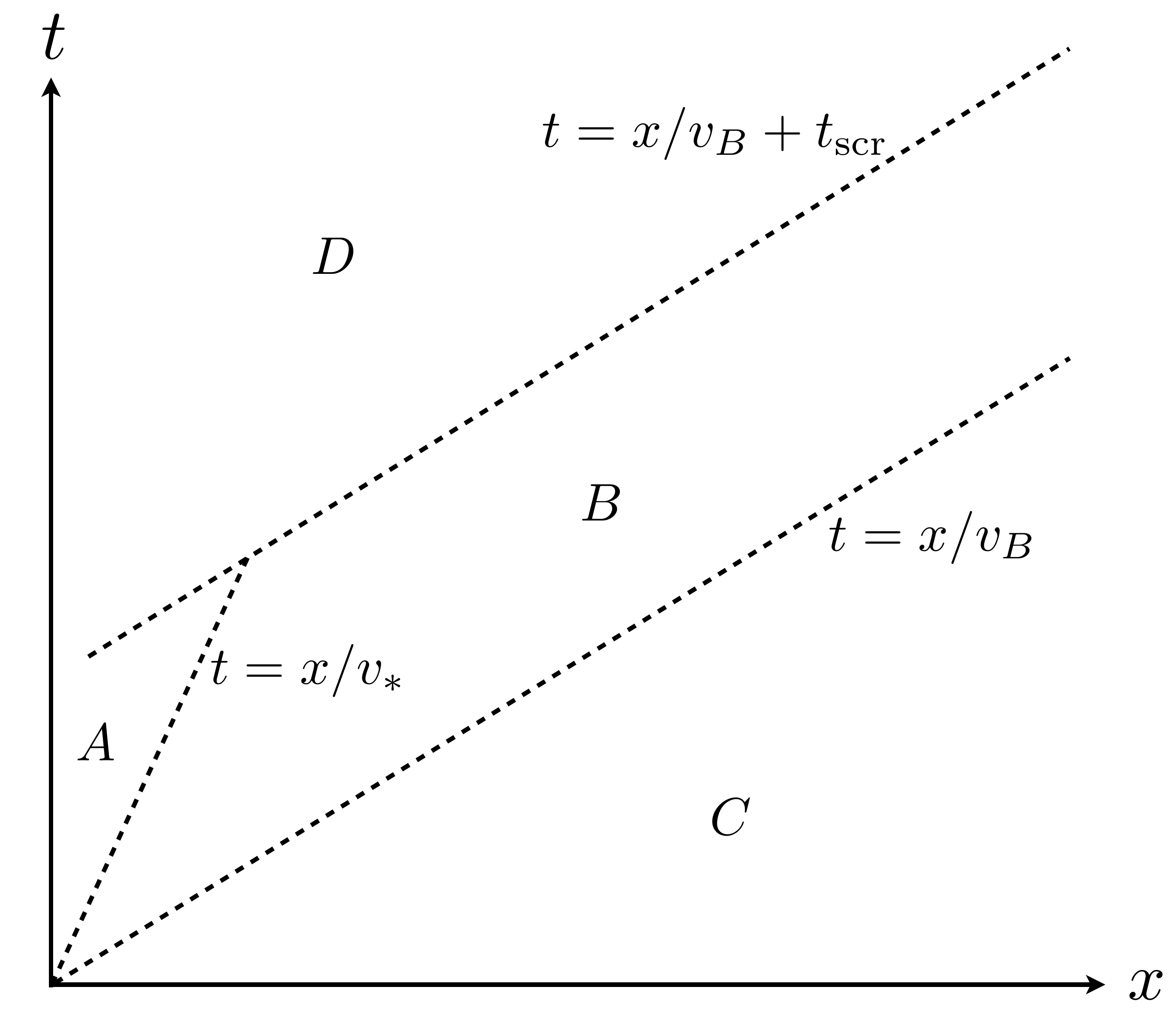}
\caption{
Chaos crossovers in spacetime for $v_\ast < v_B$, adapted from Ref.~\onlinecite{Gu2}. In region A, the OTOC is diffusive, see \eqref{eq:OTOC_diffusive}. In region B, the OTOC shows a wave-front propagation as well as maximal chaos, see \eqref{eq:OTOCfront}. In region C, the OTOC does not grow. In regiond D, the OTOC has saturated.}
  \label{fig:chaos_domains1} 
\end{figure}
\begin{figure}[ht]
\center
\includegraphics[width=0.85 \columnwidth]{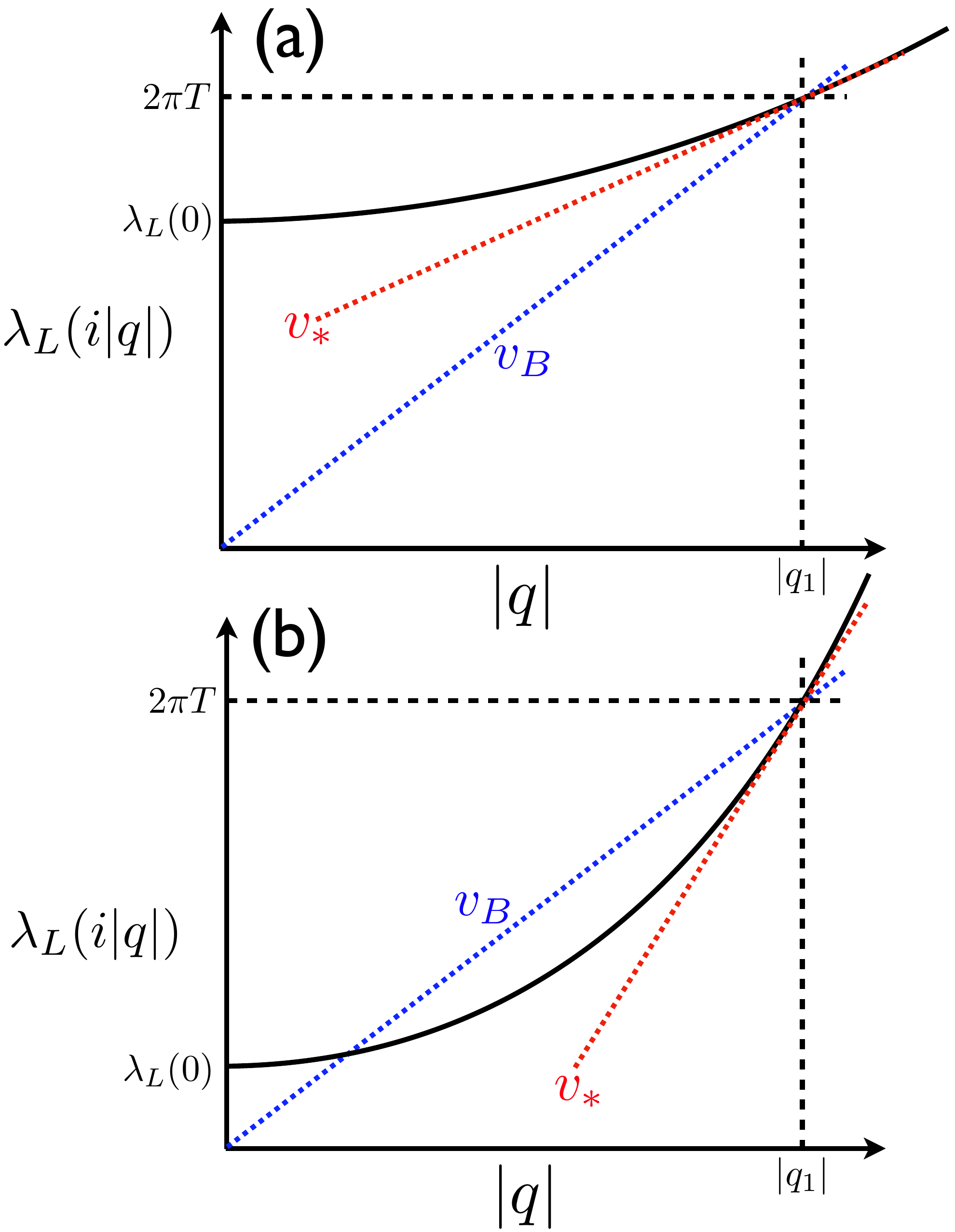}
\caption{
Schematic behavior of Lyapunov exponent $\lambda_L(q)$ on the imaginary $q$-axis at strong and weak interaction respectively (from Ref.~\onlinecite{Gu2}). The pole in the prefactor of the OTOC sits at $q_1$, where $\lambda_L(q_1)=2\pi T$. The butterfly velocity $v_B$ \eqref{OTOC3} is the slope of blue lines. The threshold velocity $v_*$ is the tangent slope (red lines) of $\lambda_L(q)$ at $q_1$. At strong interaction (a) $v_*<v_B$ and at weak interaction (b) $v_*>v_B$.}
  \label{fig:chaos_domains2} 
\end{figure}

At large $t$ and $x$, the above integral can be evaluated using saddle point approximation. For convenience we set $d=1$ but the following discussion easily generalizes to higher dimensions. Demanding the exponent in \eqref{eq:OTOCx} be stationary with respect to $q$, we obtain the saddle point $q_*$ on the imaginary $q$-axis defined by
\begin{equation}\label{eq:q_*}
    \lambda_L'(q_*)=-ix/t,
\end{equation}
and the saddle point yields
\begin{equation}\label{eq:OTOC_*}
    \rOTOC_*(x,t)\sim \frac{1}{N}e^{\lambda_L(q_*)t+iq_*|x|}.
\end{equation}

Recalling the definition of $q_1$ in Eq.~(\ref{OTOC2}), we note that if $|q_1|<|q_*|$, {\it i.e.\/} the pole sits between the saddle point and the real $q$-axis, we will hit the pole when we deform the integration contour, and we must include the pole contribution to the OTOC:
\begin{equation}
    \label{eq:OTOC_1}
    \rOTOC_1(x,t)\sim \frac{1}{N}e^{2\pi T(t-|x|/v_B)}\,,
\end{equation}
where $v_B$ was defined in Eq.~(\ref{OTOC3}).
As shown in Fig.~\ref{fig:chaos_domains2}, $\lambda_L(i|q|)$ is a convex function of $|q|$, and therefore we can rewrite the condition $|q_1|<|q_*|$ as $i\lambda_L'(q_1)<i\lambda_L'(q_*)$, {\it i.e.\/} 
\begin{equation}
    \label{eq:v_*}
    |x|/t>v_*\equiv i\lambda_L'(q_1).
\end{equation}
We refer to $v_*$ as the threshold velocity with the following meaning: If $|x/t|>v_*$, $q_1$ will be hit during the deformation of integration contour, so the pole will contribute to the OTOC and vice versa.


We now proceed to discuss behaviors of the OTOC in different regimes. Let 
\begin{equation}
t_\text{scr}\approx(\ln N)/\lambda_L
\end{equation}
denote the scrambling time after which, the OTOC saturates to $O(1)$. Then the propagation of the OTOC has the following behavior (see Fig.~\ref{fig:chaos_domains1}):

\begin{enumerate}
  \item In the incoherent metal regime $T\gg E_c$ , $\lambda_L(0)$ is close to maximal and we have $v_*<v_B$ (see (a) of Fig.~\ref{fig:chaos_domains2}). 
  
  At short distances $v_B(t-t_\text{scr})<|x|<v_*t$ and $|q_*|<|q_1|$, so the OTOC receives contribution only from the saddle point, and it has a diffusive behavior in region A of Fig.~\ref{fig:chaos_domains1}. We further assume that $q_*$ is small so that a Taylor expansion of $\lambda_L(q_*)$ is valid, and we arrive at a diffusive OTOC:
\begin{equation}\label{eq:OTOC_diffusive}
    \rOTOC(x,t)\sim \frac{1}{N}e^{\lambda_L (0) t-x^2/(4D_* t)},
\end{equation}
  where the diffusion coefficient $D_*$ is defined by
\begin{equation}\label{eq:D_*}
    \lambda_L(q)=\lambda_L(0)-D_*q^2+O(q^4).
\end{equation}
The form of the $\lambda_L(q)$ is similar to the one appears in the linearized reaction-diffusion equation (also known as the Fisher-Kolmogorov-Petrovskii-Piskunov equation, for its relation to OTOC see {\it e.g.\/} Ref.~\onlinecite{aleiner2016microscopic}). However, the comparison is merely formal, the diffusion coefficient $D_*$ here does not necessarily correspond to physical transport, in particular it does not agree with the energy diffusion as will be shown in Fig.~\ref{fig:Dcompare}. In contrast, for the weakly coupled theories\cite{aleiner2016microscopic,Stanford,patelPRX} where a quasi-particle picture still applies, the coefficient $D_*$ could be related to the diffusion pole of the Green's function. For the incoherent metal, we do not know the exact microscopic origin of $D_*$ since it is strongly interacting. We may speculate that the appearance of $D_*$ as well as the slow down of the butterfly effect (comparing to the maximal chaos at long distance discussed below) are attributed to the ``conformal matters'' inherited from the SYK model.

At long distances $\max\{v_*t,v_B(t-t_\text{scr})\}<|x|<v_B t$ and $|q_*|>|q_1|$, so the OTOC contains contribution both from the saddle point and the pole. Since the pole contribution grows with time at maximal chaos rate, it dominates the OTOC. The OTOC shows a wave-front propagation with maximal chaos in region B of Fig.~\ref{fig:chaos_domains1}:
\begin{equation}
    \rOTOC(x,t)\sim \frac{1}{N} e^{2\pi T (t-|x|/v_B)}. \label{eq:OTOCfront}
\end{equation}
    For later comparison to energy transport, it is convenient to introduce a `diffusion coefficient' (as in Eq.~(\ref{OTOC3}))
\begin{equation}
D_{\rm chaos} \equiv v_B^2/(2\pi T)=2\pi T/|q_1|^2.
 \end{equation}

We refer to $D_*$ as short-distance diffusion coefficient and $D_{\rm chaos}$ as long-distance diffusion coefficient.
 
\item 
In the Fermi-liquid regime, $\lambda_L(0)$ is far from maximal and $v_*>v_B$ (see Fig.~\ref{fig:chaos_domains2}(b)). As a consequence, even if the pole contributes to OTOC it is exponentially small relative to the non-growing part, so we always observe a diffusive OTOC as in Eq.\eqref{eq:OTOC_diffusive}. In terms of Fig.~\ref{fig:chaos_domains1}, region A now dominates the chaos and region B completely disappears. 
\end{enumerate}

\section{Scrambling and Butterfly velocity in the $t$-$U$ Model}\label{sec:scrambling_tJ}

\subsection{The $t$-$U$ model}
    In this section, we review the basic properties of the $t$-$U$ model \cite{Song}. In the $t$-$U$ model, there is a SYK-type island on each site of the lattice. Each SYK island consists of $N$ flavors of electrons, and the electrons interact with each other via a four-fermion random interaction $U_{abcd,x}$. Electrons can also hop to adjacent islands with a random amplitude $t^{ab}_{xx'}$.
    The Hamiltonian of the $t$-$U$ model illustrated in Fig.~\ref{fig:SYKlattice} is
\begin{eqnarray}
    H&=&H_\text{hopping}+H_\text{SYK},\label{eq:Hamiltonian}\\
    H_\text{hopping}&=&\frac{1}{\sqrt{zN}}\sum_{<xx'>}\sum_{ab}t^{ab}_{xx'}c_{ax}^\dagger c_{bx'},\label{eq:H_hopping}\\
    H_\text{SYK}&=&\frac{1}{(2N)^{3/2}}\sum_{x}U_{abcd,x}c_{ax}^\dagger c_{bx}^\dagger c_{cx} c_{dx}.
\end{eqnarray}
    Here $c_{ax}$ is the electron annihilation operator, where $x$ labels lattice site, and $a$ labels flavor. $t^{ab}_{xx'}$ and $U_{abcd,x}$ are Gaussian random couplings satisfying $t^{ab}_{xx'}=(t^{ba}_{x'x})^*$, $U_{abcd,x}=-U_{bacd,x}=-U_{abdc,x}=U_{cdab,x}^*$, $\overline{|t^{ab}_{xx'}|^2}=t_0^2$, $\overline{|U_{abcd,x}|^2}=U^2$. $z$ is the coordination number of the lattice. We will work in the limit of $U\gg t_0$ and large $N$.

    The analysis of Ref.~\onlinecite{Song} shows the theory has a coherence energy scale $E_c=t_0^2/U$. Various properties of the system, such as the entropy $S$, the conductivity $\sigma$, and the thermal conductivity $\kappa/T$, are universal functions of $T/E_c$. When $T<E_c$, the system demonstrates a heavy-Fermi liquid like behavior: the entropy is proportional to $T$, with a large slope compared to free fermions. The resistivity $\rho=1/\sigma$ and the inverse thermal conductivity $T/\kappa$ grows quadratically in $T$, suggesting the existence of quasiparticle excitations. When $T>E_c$, the system behaves as an incoherent metal, where the entropy saturates to a constant value predicted by the SYK model, and both $\rho$ and $T/\kappa$ grows linearly with $T$.

\subsection{Green's Function}
    In this section we review Green's functions of the $t$-$U$ model in both imaginary time and real time. The imaginary time Green's function is useful for thermodynamics, and the real time Green's function is useful for transport and scrambling.
\subsubsection{Imaginary Time}
    We start with the imaginary time Green's function. We first perform disorder averaging over $t$ and $U$, and due to the self-averaging property of SYK model, we can do this with a single replica. After that we introduce the Green's function bilinear $G_x(\tau_1,\tau_2)=-({1}/{N})\sum_a c_{ax}(\tau_1)c_{ax}^\dagger(\tau_2)$, and the self-energy $\Sigma(\tau_1,\tau_2)$ as a Lagrange multiplier to impose the definition, and we can obtain the imaginary time action $S_\beta[G,\Sigma]$ as
\begin{equation}\label{eq:action_imaginary}
\begin{split}
    &\frac{S_\beta[G,\Sigma]}{N}=-\sum_x\mathrm{Tr}\ln(\partial_\tau+\Sigma_x)\\&-\sum_x\int\rd^2\tau \Sigma_x(\tau_1,\tau_2)G_x(\tau_2,\tau_1)\\
                            &-\int\rd^2\tau \left[\sum_x\frac{U^2}{4}G_x(\tau_1,\tau_2)^2G_x(\tau_2,\tau_1)^2\right. \\
                            &\left.-\sum_{<xx'>}\frac{t_0^2}{2z}G_x(\tau_1,\tau_2)G_{x'}(\tau_2,\tau_1)\right].
\end{split}
\end{equation}
In the large $N$ limit, we can obtain the equation of motion from saddle point expansion:
\begin{equation}\label{eq:EoM_imaginary}
\begin{split}
   & G^{-1}(i\omega_n)=i\omega_n-\Sigma(i\omega_n),\\
    &\Sigma(\tau)=t_0^2G(\tau)-U^2G(\tau)^2G(-\tau).
\end{split}
\end{equation}
    The above equations are solved numerically by combining iteration and fast Fourier transform (FFT). See Appendix~\ref{sec:numerics} for more details.
\subsubsection{Real Time}
    Next, we turn to the computation of real time Green's function. We use the Keldysh formalism (see Appendix~\ref{sec:keldysh_supp}) to compute the retarded and the advanced Green's functions. In the Keldysh formalism, the time contour for path integral is doubled to a forward branch $s=+$ and a backward branch $s=-$. The Keldysh action is $S_K=S[\psi_+]-S[\psi_-]$, where $S$ is the original action and $\psi_+,\psi_-$ are fields supported on the forward and the backward branch respectively. 

    We perform the disorder average over $t,U$ of the Keldysh action, and then introduce the Green's function bilinear $iG_{ss'x}(t,t')=({1}/{N})\sum_a c_{axs}(t)c_{axs'}^\dagger(t')$ and the Lagrange multiplier $\Sigma_{ss'x}(t,t')$ to impose the definition. As a result, we obtain the Keldysh action $S_K$ which reads
\begin{equation}
\begin{aligned}
    &\frac{iS_K[G,\Sigma]}{N}=\sum_x \Tr \ln(i\partial_t\sigma^z_{ss'}-\Sigma_{ss'x})\\
    & \qquad +\sum_{ss'} \sum_x  \int\rd^2 t \Sigma_{s'sx}(t',t)G_{ss'x}(t,t')\\
   & -\sum_{ss'}ss'\int\rd^2 t \bigg[\sum_x \frac{U^2}{4}G_{ss'x}(t,t')^2G_{s'sx}(t',t)^2\\
   &\qquad \qquad +\sum_{<xx'>}\frac{t_0^2}{2z}G_{ss'x}(t,t')G_{s'sx'}(t',t)\bigg].
\end{aligned}\label{eq: Keldysh action tU}
\end{equation}
    Here $s,s'=\pm 1$ denote the two branches, and the Pauli matrix $\sigma^z$ acts on $ss'$ indices. The Green's functions $G_{ss'}$ can be combined into retarded, advanced and Keldysh Green's functions using Keldysh rotation 
\begin{equation}
\begin{aligned}
    G_R&=\frac{1}{2}(-G_{--} + G_{-+} - G_{+-} + G_{++}),\\
    G_K&=\frac{1}{2}(G_{--} + G_{-+} + G_{+-} + G_{++}),\\
    G_A&=\frac{1}{2}(-G_{--} - G_{-+} + G_{+-} + G_{++}).
\end{aligned}
\end{equation}

    The equations of motion can be obtained by varying the Keldysh action $S_K$. However there are multiple solutions corresponding to different temperatures. We fix the temperature by supplementing the fluctuation dissipation relation
\begin{equation}
    G_K(\omega)=(G_R(\omega)-G_A(\omega))\tanh{\frac{\omega}{2T}}.
\end{equation}
    The equations of motion now take the following form
\begin{equation}\label{eq:EoM_Keldysh}
\begin{aligned}
    G_R(\omega)^{-1}=&\omega-t_0^2G_R(\omega)-\Sigma_R(\omega),\\
    G_K(\omega)=& 2i\tanh\frac{\omega}{2T}\Im G_R(\omega),\\
    \Sigma_R(t)=&\frac{1}{2}U^2G_K(-t)G_K(t)G_R(t)\\
    +&\frac{1}{4}U^2G_K(t)^2G_R^*(t)+\frac{1}{4}U^2G_R(t)^2G_R^*(t).
\end{aligned}
\end{equation}
The above equations can also be solved using iteration and FFT (see Appendix~\ref{sec:numerics}).

\subsection{Computation of the OTOC}

We will use the kinetic equation method to numerically obtain the scrambling rate $\lambda_L(q)$ as a function of momentum $q$ and then use the ladder identity to compute diffusion coefficients $D_*,D_{\rm chaos}$ which were defined in Section~\ref{sec:scrambling}.
    
We first derive equations for the following retarded OTOCs
\begin{equation}\label{eq:OTOC}
\begin{aligned}
    &f_1(x;t_1,t_3;t_2,t_4)=\frac{1}{N^2}\theta(t_{24})\theta(t_{13}) \\ &\sum_{ab} \Tr\left( y^2\{c_{ax}(t_1),c_{b0}^\dagger(t_3)\}^\dagger y^2\{c_{ax}(t_2),c_{b0}^\dagger(t_4)\}\right),\\
    &f_2(x;t_1,t_3;t_2,t_4)=\frac{1}{N^2}\theta(t_{24})\theta(t_{13})\\
    & \sum_{ab} \Tr\left( y^2\{c_{ax}(t_1),c_{b0}^\dagger(t_3)\}y^2\{c_{ax}(t_2),c_{b0}^\dagger(t_4)\}\right).
\end{aligned}
\end{equation}
    Here, we have introduced two types of OTOCs $f_1,f_2$ because in the complex SYK model there are two ways to arrange fermionic arrows, as shown in the following diagramatical representation:
\begin{equation}
\begin{aligned}
   f_1(q,\Omega,\Omega',\omega)&= \begin{tikzpicture}[scale=1,baseline={(v0.base)}]
\begin{feynman}[scale=1.5,large,transform shape]
    \vertex (v1);
    \vertex [right=0.7 of v1] (v2);
    \vertex [right=0.6 of v2] (v3);
    \vertex [right=0.7 of v3] (v4);
    \vertex [below=1 of v1] (v5);
    \vertex [right=0.7 of v5] (v6);
    \vertex [right=0.6 of v6] (v7);
    \vertex [right=0.7 of v7] (v8);
    \vertex [below=0.6 of v1] (v0);
    \diagram
    {
        (v1)--[fermion,edge label=\tiny $\Omega+\omega/2$](v2)--(v3)--[fermion, edge label=\tiny $\Omega'+\omega/2$](v4),
        (v8)--[fermion, edge label=\tiny $\Omega'-\omega/2$](v7)--(v6)--[fermion, edge label=\tiny $\Omega-\omega/2$](v5),
    };
    \draw (v2) rectangle (v7);
    \node at ($(v2)!.5!(v7) $) {$f_1$};
\end{feynman}
\end{tikzpicture},
\\
   f_2(q,\Omega,\Omega',\omega)&= \begin{tikzpicture}[scale=1,baseline={(v0.base)}]
\begin{feynman}[scale=1.5,large,transform shape]
    \vertex (v1);
    \vertex [right=0.7 of v1] (v2);
    \vertex [right=0.6 of v2] (v3);
    \vertex [right=0.7 of v3] (v4);
    \vertex [below=1 of v1] (v5);
    \vertex [right=0.7 of v5] (v6);
    \vertex [right=0.6 of v6] (v7);
    \vertex [right=0.7 of v7] (v8);
    \vertex [below=0.6 of v1] (v0);
    \diagram
    {
        (v1)--[anti fermion,edge label=\tiny $\Omega-\omega/2$](v2)--(v3)--[fermion, edge label=\tiny $\Omega'+\omega/2$](v4),
        (v8)--[fermion, edge label=\tiny $\Omega'-\omega/2$](v7)--(v6)--[anti fermion, edge label=\tiny $\Omega+\omega/2$](v5),
    };
    \draw (v2) rectangle (v7);
    \node at ($(v2)!.5!(v7) $) {$f_2$};
\end{feynman}
\end{tikzpicture}.
\end{aligned}
\label{eq:f1f2graph}
\end{equation}
    The retarded OTOCs $f_{1}$ and $f_2$ have the same Lyapunov exponent as the OTOC defined in \eqref{eq:OTOC_nonretarded} but have simpler diagrammatic expansions. 
    
	The Fourier transform $f_1(q,\Omega,\Omega',\omega)$ is defined similarly as in Eq.\eqref{eq:f_1_omega}, and the Fourier transform of $f_2$ is defined as
	
\begin{equation}
\begin{split}
    f_2(q,\Omega,\Omega',\omega)=&\int\rd^d x \rd^3 t f_2(x;t_1,t_3;t_2;t_4)\\
\times &e^{-i q\cdot x+i\Omega t_{43}+i\Omega't_{21}+i\omega t},
\end{split}
\end{equation}	
where the signs of $t_{43}$ and $t_{21}$ are opposite to Eq.\eqref{eq:f_1_omega}.
	
	To get the Lyapunov exponent, we search for poles of $f_1$ and $f_2$ at $\omega=i\lambda_L(q)$. The retarded OTOCs $f_1$, $f_2$ have the same Lyapunov exponent $\lambda_L(q)$ as the non-retarded version \eqref{eq:OTOC_nonretarded}, but it does not have the pole structure as Eq.\eqref{OTOC1}, so it only has diffusive propagation.

\begin{figure}[htb!]
\centering
\begin{tikzpicture}[scale=1.5]
\begin{scope}[thick, every node/.style={sloped,allow upside down}]
  \draw (0,0)--node{\midarrow}(-3,0) arc (90:270:1mm) --node{\midarrow} (0,-0.2) --node{\midarrow} (0, -1.5) --node{\midarrow}(-3,-1.5) arc (90:270:1mm) --node{\midarrow}(0,-1.7)--node{\midarrow}(0,-3);
 \node[above] at (-0.,0) { $-\infty$};
 \node[left] at (-3.1,-0.1) { $1:t$};
 \node[left] at (-3.1,-1.6) { $2:t-i\beta/2$};
 \node[below] at (0,-3) { $-\infty-i\beta$};
\end{scope}
\end{tikzpicture}
\caption{\label{fig:keldysh_contour} The time contour $C$ for calculating OTOC. The contour is drawn such that the real time goes to the left, which is convenient when
acting by operators on left.}
\end{figure}
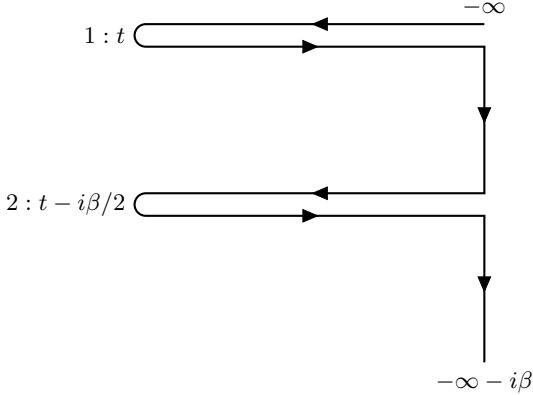

    As OTOCs, $f_1$ and $f_2$ can be conveniently computed using the Keldysh perturbation theory, where the path integral is adapted to include multiple time folds, as is shown in Fig.~\ref{fig:keldysh_contour}. The time contour $C$ now consists of two real folds and two imaginary segments, with $t=0$ identified with $t=-i\beta$.

    It is convenient to relabel fields on each real fold using $r$-$a$ variables
\begin{equation}
\begin{aligned}
    \psi_{ri}&= \displaystyle \frac{\psi_{+i}+\psi_{-i}}{\sqrt{2}},\\
    \psi_{ai}&=\displaystyle \frac{\psi_{+i}-\psi_{-i}}{\sqrt{2}},
\end{aligned}
\end{equation}
where $i=1,2$ labels the two real folds as in Fig.~\ref{fig:keldysh_contour}, and $+/-$ labels the future/past directing segment. More details of Keldysh perturbation theory is included in Appendix \ref{sec:keldysh_supp} and \ref{sec:frules}.

    The OTOCs \eqref{eq:OTOC} can now be written as a path integral over the contour $C$, for example, 
\begin{equation}\label{eq:f1_perturbtive}
\begin{split}
    f_1(x,t_1,t_3&;t_2,t_4)=\int \mathcal{D}c~e^{i\int_C\rd t\rd^d x \mathcal{L}(t,x)} \\ 
    \times& c_{r2,x}^\dagger(t_1)c_{a2,0}(0,t_3)c_{r1,x}(t_2)c_{a1,0}^\dagger(t_4)
\end{split}
\end{equation}and there is a similar expression for $f_2$.

 We then expand Eq.\eqref{eq:f1_perturbtive} in powers of interaction vertices, and sum over all the ladder diagrams to obtain the Bethe-Salpeter equation for $f_1$ and $f_2$ at leading $1/N$ order.

    To begin with, it is convenient to sum over the hopping vertices Eq.~\eqref{eq:H_hopping}, because they contain all the momentum dependence. This step is justified by the fact that the hopping vertex is a momentum-dependent number $t_0^2 \mu(q)$ (see \eqref{eq:L=GRGA}) in the functional space and therefore it commutes with other operators. Diagrammatically, the sum $L(q,\Omega,\omega)$ is
\begin{equation}\label{eq:L_diagram}
\begin{tikzpicture}[scale=1,baseline={(v0.base)}]
\begin{feynman}[scale=1,large,transform shape]
    \vertex (v1);
    \vertex [right=0.7 of v1] (v2);
    \vertex [right=0.6 of v2] (v3);
    \vertex [right=0.7 of v3] (v4);
    \vertex [below=1 of v1] (v5);
    \vertex [right=0.7 of v5] (v6);
    \vertex [right=0.6 of v6] (v7);
    \vertex [right=0.7 of v7] (v8);
    \vertex [below=0.6 of v1] (v0);
    \diagram
    {
        (v1)--[fermion,edge label=\scriptsize $\Omega+\omega/2$](v2)--(v3)--[fermion, edge label=\scriptsize $\Omega+\omega/2$](v4),
        (v8)--[fermion, edge label=\scriptsize $\Omega-\omega/2$](v7)--(v6)--[fermion, edge label=\scriptsize $\Omega-\omega/2$](v5),
    };
    \draw (v2) rectangle (v7);
    \node at ($(v2)!.5!(v7) $) {$L$};
\end{feynman}
\end{tikzpicture}
=
\begin{tikzpicture}[scale=1,baseline={(v0.base)}]
\begin{feynman}[scale=1,large,transform shape]
    \vertex (v1);
    \vertex [right=0.7 of v1] (v2);
    \vertex [right=0.6 of v2] (v3);
    \vertex [right=0.7 of v3] (v4);
    \vertex [below=1 of v1] (v5);
    \vertex [right=0.7 of v5] (v6);
    \vertex [right=0.6 of v6] (v7);
    \vertex [right=0.7 of v7] (v8);
    \vertex [below=0.6 of v1] (v0);
    \vertex [right=0.3 of v2] (vtm);
    \vertex [below= 1 of vtm] (vbm);
    \diagram
    {
        (v1)--[fermion](v2)--(v3)--[fermion](v4),
        (v8)--[fermion](v7)--(v6)--[fermion](v5),
    };
\end{feynman}
\end{tikzpicture}
+
\begin{tikzpicture}[scale=1,baseline={(v0.base)}]
\begin{feynman}[scale=1,large,transform shape]
    \vertex (v1);
    \vertex [right=0.7 of v1] (v2);
    \vertex [right=0.6 of v2] (v3);
    \vertex [right=0.7 of v3] (v4);
    \vertex [below=1 of v1] (v5);
    \vertex [right=0.7 of v5] (v6);
    \vertex [right=0.6 of v6] (v7);
    \vertex [right=0.7 of v7] (v8);
    \vertex [below=0.6 of v1] (v0);
    \vertex [right=0.3 of v2] (vtm);
    \vertex [below= 1 of vtm] (vbm);
    \diagram
    {
        (v1)--[fermion](v2)--(v3)--[fermion](v4),
        (v8)--[fermion](v7)--(v6)--[fermion](v5),
        (vtm)--[scalar](vbm),
    };
\end{feynman}
\end{tikzpicture}
+\cdots
\end{equation}

In terms of propagators and couplings, this is

\begin{align}\label{eq:L=GRGA}
    L(q,\Omega,\omega)&=G_R(\Omega+\omega/2)G_A(\Omega-\omega/2)\nonumber\\
                    &+\mu(q)t_0^2(G_R(\Omega+\omega/2)G_A(\Omega-\omega/2))^2+\cdots \nonumber\\
                    &=\frac{G_R(\Omega+\omega/2)G_A(\Omega-\omega/2)}{1-\mu(q)t_0^2G_R(\Omega+\omega/2)G_A(\Omega-\omega/2)},
\end{align}
where $\mu(q)=\frac{1}{z}\sum_{a}e^{i q\cdot a}$ is a summation over lattice neighbors. At long wavelength, $\mu(q)=1-\alpha q^2+\cdots$, and in practice we take $\mu(q)=\cos(q)$, {\it i.e.} taking the system to be an 1D chain with unit lattice spacing.

    We move on to include other diagrams. The Bethe-Salpeter equation can be written as
\begin{align}\label{eq:bethe_salpeter_diagram}
\begin{tikzpicture}[scale=1,baseline={(v0.base)}]
\begin{feynman}[scale=1,large,transform shape]
    \vertex (v1);
    \vertex [right=0.7 of v1] (v2);
    \vertex [right=0.6 of v2] (v3);
    \vertex [right=0.7 of v3] (v4);
    \vertex [below=1 of v1] (v5);
    \vertex [right=0.7 of v5] (v6);
    \vertex [right=0.6 of v6] (v7);
    \vertex [right=0.7 of v7] (v8);
    \vertex [below=0.6 of v1] (v0);
    \vertex [right=0.3 of v2] (vtm);
    \vertex [below= 1 of vtm] (vbm);
    \diagram
    {
        (v1)--[fermion](v2)--(v3)--[fermion](v4),
        (v8)--[fermion](v7)--(v6)--[fermion](v5),
    };
    \draw (v2) rectangle (v7);
    \node at ($(v2)!.5!(v7) $) {$f_1$};
\end{feynman}
\end{tikzpicture}
&=\begin{tikzpicture}[scale=1,baseline={(v0.base)}]
\begin{feynman}[scale=1,large,transform shape]
    \vertex (v1);
    \vertex [right=0.7 of v1] (v2);
    \vertex [right=0.6 of v2] (v3);
    \vertex [right=0.7 of v3] (v4);
    \vertex [below=1 of v1] (v5);
    \vertex [right=0.7 of v5] (v6);
    \vertex [right=0.6 of v6] (v7);
    \vertex [right=0.7 of v7] (v8);
    \vertex [below=0.6 of v1] (v0);
    \vertex [right=0.3 of v2] (vtm);
    \vertex [below= 1 of vtm] (vbm);
    \diagram
    {
        (v1)--[fermion](v2)--(v3)--[fermion](v4),
        (v8)--[fermion](v7)--(v6)--[fermion](v5),
    };
    \draw (v2) rectangle (v7);
    \node at ($(v2)!.5!(v7) $) {$L$};
\end{feynman}
\end{tikzpicture}\nonumber
\\
&+\begin{tikzpicture}[scale=1,baseline={(v0.base)}]
\begin{feynman}[scale=1,large,transform shape]
    \vertex (v1);
    \vertex [right=0.7 of v1] (v2);
    \vertex [right=0.6 of v2] (v3);
    \vertex [right=0.7 of v3] (v4);
    \vertex [below=1 of v1] (v5);
    \vertex [right=0.7 of v5] (v6);
    \vertex [right=0.6 of v6] (v7);
    \vertex [right=0.7 of v7] (v8);
    \vertex [below=0.6 of v1] (v0);
    \vertex [right=0.3 of v2] (vtm);
    \vertex [below= 1 of vtm] (vbm);
    \diagram
    {
        (v1)--[fermion](v2)--(v3)--(v4),
        (v8)--(v7)--(v6)--[fermion](v5),
        (vtm)--[fermion, quarter left](vbm)--[fermion, quarter left](vtm),
    };
    \draw (v3) rectangle (v8);
    \node at ($(v3)!.5!(v8) $) {$f_1$};
    \draw (v1) rectangle (v6);
    \node at ($(v1)!.5!(v6) $) {$L$};
\end{feynman}
\end{tikzpicture}
+\begin{tikzpicture}[scale=1,baseline={(v0.base)}]
\begin{feynman}[scale=1,large,transform shape]
    \vertex (v1);
    \vertex [right=0.7 of v1] (v2);
    \vertex [right=0.6 of v2] (v3);
    \vertex [right=0.7 of v3] (v4);
    \vertex [below=1 of v1] (v5);
    \vertex [right=0.7 of v5] (v6);
    \vertex [right=0.6 of v6] (v7);
    \vertex [right=0.7 of v7] (v8);
    \vertex [below=0.6 of v1] (v0);
    \vertex [right=0.3 of v2] (vtm);
    \vertex [below= 1 of vtm] (vbm);
    \diagram
    {
        (v1)--[fermion](v2)--(v3)--(v4),
        (v8)--(v7)--(v6)--[fermion](v5),
        (vtm)--[fermion, quarter left](vbm),
        (vtm)--[fermion, quarter right](vbm),
    };
    \draw (v3) rectangle (v8);
    \node at ($(v3)!.5!(v8) $) {$f_2$};
    \draw (v1) rectangle (v6);
    \node at ($(v1)!.5!(v6) $) {$L$};
\end{feynman}
\end{tikzpicture}, \nonumber\\
\begin{tikzpicture}[scale=1,baseline={(v0.base)}]
\begin{feynman}[scale=1,large,transform shape]
    \vertex (v1);
    \vertex [right=0.7 of v1] (v2);
    \vertex [right=0.6 of v2] (v3);
    \vertex [right=0.7 of v3] (v4);
    \vertex [below=1 of v1] (v5);
    \vertex [right=0.7 of v5] (v6);
    \vertex [right=0.6 of v6] (v7);
    \vertex [right=0.7 of v7] (v8);
    \vertex [below=0.6 of v1] (v0);
    \vertex [right=0.3 of v2] (vtm);
    \vertex [below= 1 of vtm] (vbm);
    \diagram
    {
        (v1)--[anti fermion](v2)--(v3)--[fermion](v4),
        (v8)--[fermion](v7)--(v6)--[anti fermion](v5),
    };
    \draw (v2) rectangle (v7);
    \node at ($(v2)!.5!(v7) $) {$f_2$};
\end{feynman}
\end{tikzpicture}
&=
\begin{tikzpicture}[scale=1,baseline={(v0.base)}]
\begin{feynman}[scale=1,large,transform shape]
    \vertex (v1);
    \vertex [right=0.7 of v1] (v2);
    \vertex [right=0.6 of v2] (v3);
    \vertex [right=0.7 of v3] (v4);
    \vertex [below=1 of v1] (v5);
    \vertex [right=0.7 of v5] (v6);
    \vertex [right=0.6 of v6] (v7);
    \vertex [right=0.7 of v7] (v8);
    \vertex [below=0.6 of v1] (v0);
    \vertex [right=0.3 of v2] (vtm);
    \vertex [below= 1 of vtm] (vbm);
    \diagram
    {
        (v1)--[anti fermion](v2)--(v3)--(v4),
        (v8)--(v7)--(v6)--[anti fermion](v5),
        (vtm)--[fermion, quarter left](vbm)--[fermion, quarter left](vtm),
    };
    \draw (v3) rectangle (v8);
    \node at ($(v3)!.5!(v8) $) {$f_2$};
    \draw (v1) rectangle (v6);
    \node at ($(v1)!.5!(v6) $) {$\tilde{L}$};
\end{feynman}
\end{tikzpicture}
+\begin{tikzpicture}[scale=1,baseline={(v0.base)}]
\begin{feynman}[scale=1,large,transform shape]
    \vertex (v1);
    \vertex [right=0.7 of v1] (v2);
    \vertex [right=0.6 of v2] (v3);
    \vertex [right=0.7 of v3] (v4);
    \vertex [below=1 of v1] (v5);
    \vertex [right=0.7 of v5] (v6);
    \vertex [right=0.6 of v6] (v7);
    \vertex [right=0.7 of v7] (v8);
    \vertex [below=0.6 of v1] (v0);
    \vertex [right=0.3 of v2] (vtm);
    \vertex [below= 1 of vtm] (vbm);
    \diagram
    {
        (v1)--[anti fermion](v2)--(v3)--(v4),
        (v8)--(v7)--(v6)--[anti fermion](v5),
        (vtm)--[anti fermion, quarter left](vbm),
        (vtm)--[anti fermion, quarter right](vbm),
    };
    \draw (v3) rectangle (v8);
    \node at ($(v3)!.5!(v8) $) {$f_1$};
    \draw (v1) rectangle (v6);
    \node at ($(v1)!.5!(v6) $) {$\tilde{L}$};
\end{feynman}
\end{tikzpicture},
\end{align}
where each right/left propagator is a retarded/advanced propagator, each vertical propagator is a Wightman propagator $G_W$ (see Appendix.~\ref{sec:wightman}), and $\tilde{L}$ is obtained from $L$ by reversing all arrows.

    The Bethe-Salpeter equations above can be simplified in the following manners:
\begin{enumerate}
  \item The two legs on the right of $f_1,f_2$ are not relevant, so we can suppress the $\Omega'$ dependence.
 \item At half filling, $G_R(t)$ and $G_A(t)$ are pure imaginary, and it follows that $L=\tilde{L}$.
    \item At half filling, the electron Wightman propagator $G_W(t)$ is an even function, and it follows that all the electron rung-diagrams agree up to symmetry factor.
\end{enumerate}

    With the above simplifications, the Bethe-Salpeter equation can be written as (we have suppressed the $\Omega'$ argument, and numerical factors are explained in Appendix \ref{sec:frules})

\begin{equation}\label{eq:f1f2_BS}
\begin{split}
    &f_1(q,\Omega,\omega)=L(q,\Omega,\omega)\times\\
    &\left[1+U^2\int\frac{\rd \tilde{\Omega}}{2\pi}K(\Omega-\tilde{\Omega})(f_1(q,\tilde{\Omega},\omega)+\frac{1}{2}f_2(q,-\tilde{\Omega},\omega))\right],\\
   & f_2(q,-\Omega,\omega)=L(q,\Omega,\omega)\times\\
    &\left[U^2\int\frac{\rd \tilde{\Omega}}{2\pi}K(\Omega-\tilde{\Omega})(f_2(q,-\tilde{\Omega},\omega)+\frac{1}{2}f_1(q,\tilde{\Omega},\omega))\right],
\end{split}
\end{equation}

    where
\begin{equation}\label{eq:rung_kernel}
    K(\Omega)=\int\frac{\rd \mu}{2\pi}G_W(\mu)G_W(\mu+\Omega).
\end{equation}
    Note here the argument in $f_2$ has a minus sign to match the convention of Fourier transform. The $\frac{1}{2}$ factor is due to combiatorics. 

    The Bethe-Salpeter equations \eqref{eq:f1f2_BS} is a coupled equation for $f_1,f_2$. To obtain the Lyapunov exponent, we add the two equations together and we obtain an equation for the sum $F=f_1+f_2$:
    
    
\begin{equation}\label{eq:f=Kf}
\begin{split}
    F(q,\Omega,\omega)&=L(q,\Omega,\omega)\\&+
    \frac{3J^2}{2}L(q,\Omega,\omega)\int\frac{\rd \tilde{\Omega}}{2\pi}K(\Omega-\tilde{\Omega})F(q,\tilde{\Omega},\omega).
\end{split}
\end{equation}
As a sanity check, we can take the $t_0\to 0$ limit, and we will recover the Bethe-Salpeter equation for the original SYK model\cite{Maldacena}. If we repeat the exercise for the difference $f_1-f_2$, we found that it has no exponential growth in $t_0\to 0$ limit, and we conclude that the sum $F$ is the correct place to look for Lyapunov exponent.

    Following the approaches of Ref.~\onlinecite{Patel2016}, we will numerically extract $\lambda_L(q)$ from Eq.\eqref{eq:f=Kf}. The equation can be written as a matrix equation $F=L+(LK)_{\omega,q}F$ and hence $F=(1-(LK)_{\omega,q})^{-1}L$. As discussed earlier, the exponential growth in time translates to a pole at $\omega=i\lambda_L(q)$, and this implies that the matrix $M_{\omega,q}=1-(LK)_{\omega,q}$ is singular at $\omega=i\lambda_L(q)$. Our algorithm sweeps $\omega$ on the imaginary axis and searches for the point where the smallest eigenvalue of $M_{\omega,q}$ vanishes. The details of numerical implementation are in Appendix \ref{sec:numerics}.

    Before moving to the results, we will make a few comments on the relation between the retarded OTOCs $f_{1,2}$ and the regular OTOC defined in \eqref{eq:OTOC_nonretarded}.
    \begin{itemize}
        \item The retarded OTOCs contain linear combinations of regular OTOCs with the same growing exponent. Therefore, to obtain $\lambda_L(q)$ for each momentum $q$, one can choose to work on any type of OTOCs. As we have mentioned, we decide to work on retarded OTOCs because they have simpler diagrams. More explicitly, there is no interaction vertex on the imaginary time circle for the retarded OTOCs due to the cancellation among the terms contained in $f_1$ and $f_2$. 
        
        \item 
        We are also interested in the spatial propagation of the scrambling which relies on the pole structure of OTOC in the momentum space. For this purpose, we need to determine the prefactors that are beyond the kinetic equation method and depend on the types of OTOCs we choose. We will apply the ladder identity to achieve the prefactor for the regular OTOC. The decision is explained as follows. Physically, the regular OTOC \eqref{eq:OTOC_nonretarded}  contains all the contributions to the scrambling. In contrast, the retarded OTOCs may exclude some degrees of freedom, {\it e.g.} in the single site SYK model, the contributions from the reparametrization modes ({\it i.e.} Schwarzian modes) are excluded and the retarded OTOC only contains the ``stringy'' modes as discussed in Ref.~\onlinecite{Gu2} section 5. 
         
         \item As a consequence of the ladder identity\cite{Gu2}, the regular OTOC \eqref{eq:OTOC_nonretarded} has a pole $[\cos (\lambda_L(q)/(4T))]^{-1}$ in the prefactor. This pole leads to a sharp butterfly wavefront discussed in Section~\ref{sec:chaos_propagation}; however the retarded OTOCs do not have a similar pole due to an additional factor $\cos (\lambda_L(q)\beta/4)$ in the numerator. This may be explained by the following simple observation. Let us try to expand {\it e.g}. $f_1$ in \eqref{eq:OTOC} and collect the OTOCs in the expression: there are two terms with different regularization comparing to the definition in \eqref{eq:OTOC_nonretarded}. More exactly, they differ by an imaginary time evolution
         $\pm \beta/4$, and in total give a $e^{i \lambda_L \beta/4 }+e^{-i \lambda_L \beta/4}= 2 \cos (\lambda_L\beta/4)$ factor. This factor exactly cancels the same factor in the denominator coming from the regular OTOC.  
    \end{itemize}
    
    In the end, let us also comment on the finite $N$ effects in OTOCs. In this paper, we will only work on the leading order in $1/N$, namely within the validity that ${\rm OTOC} \sim {e^{\lambda_L t}}/{N}$. Physically, this corresponds to the early time regime (i.e. long before the scrambling time/saturation, ${e^{\lambda_L t}}/{N}\ll 1$) where the important physics is the initial growth of scrambling and the propagation of chaos wavefront. In this regime, the scrambling is simple and can be determined by linear equations we derived. On the contrary, the physics in late time  generally requires the knowledge of non-linearity, namely the higher order in ${e^{\lambda_L t}}/{N}$ effects. In other words, the scrambling time is the time scale when we have to worry about the finite $N$ effect. 
    In terms of diagrams, that means 
    we need to include non-melonic diagrams and diagrams with multiple ladders. The detailed discussion of the non-linear effects in SYK-like models is an interesting future direction.

\subsection{Numerical Result for Scrambling, Comparison to Energy Transport}

\begin{figure}[h]
  \centering
  \includegraphics[width=0.95 \columnwidth]{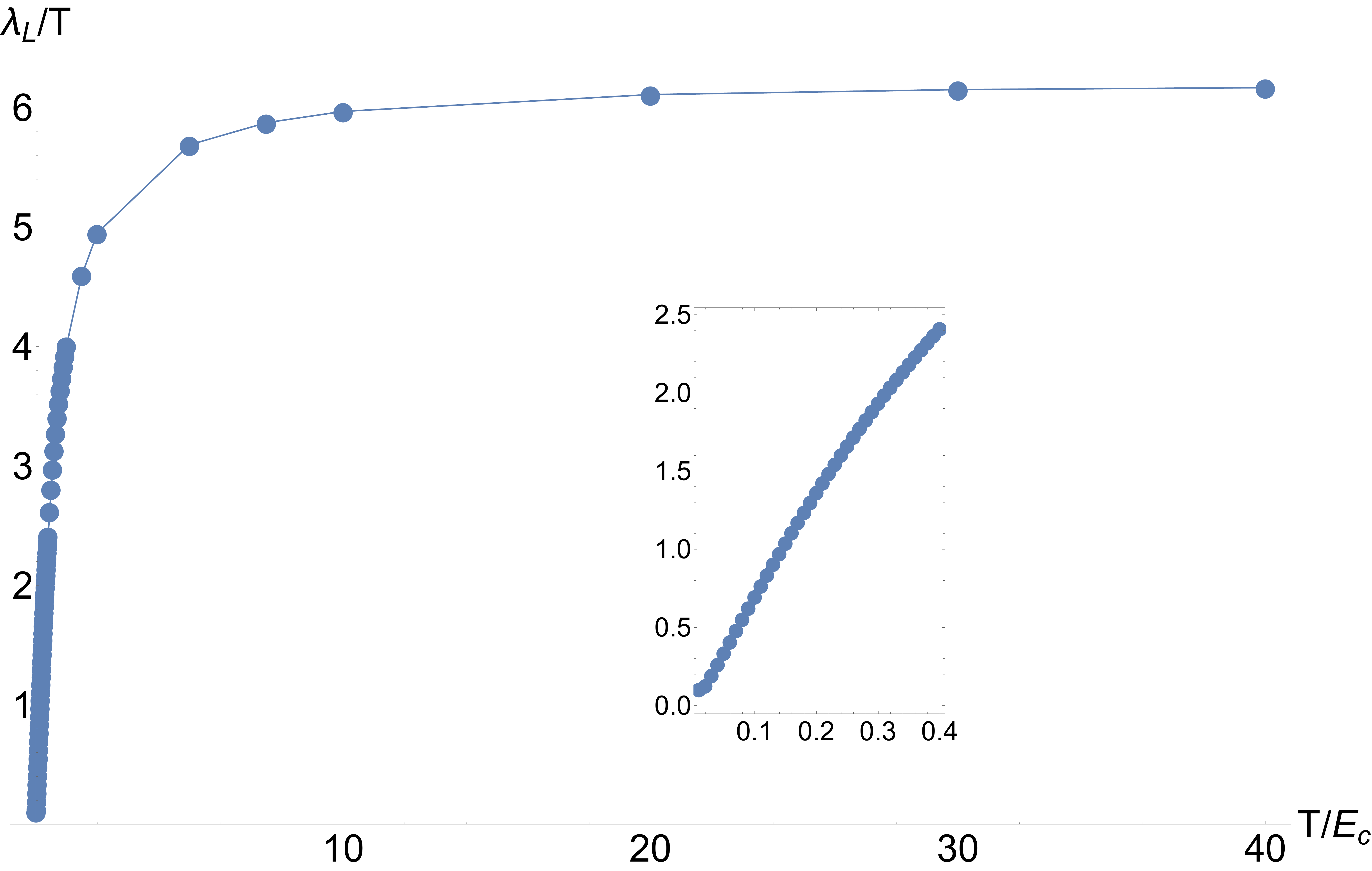}
  \caption{The zero momentum scrambling rate $\lambda_L(q=0)$ plotted versus temperature $T/E_c$ ($U=200t_0$). The inset zooms into the low temperature region. The scrambling rate $\lambda_L$ grows as $T^2/E_c$ at low temperatures, and saturates to a $T$-linear curve  at high temperatures.\label{fig:scrambling0}}
\end{figure}

\begin{figure}
\centering
  \includegraphics[width=0.95\columnwidth]{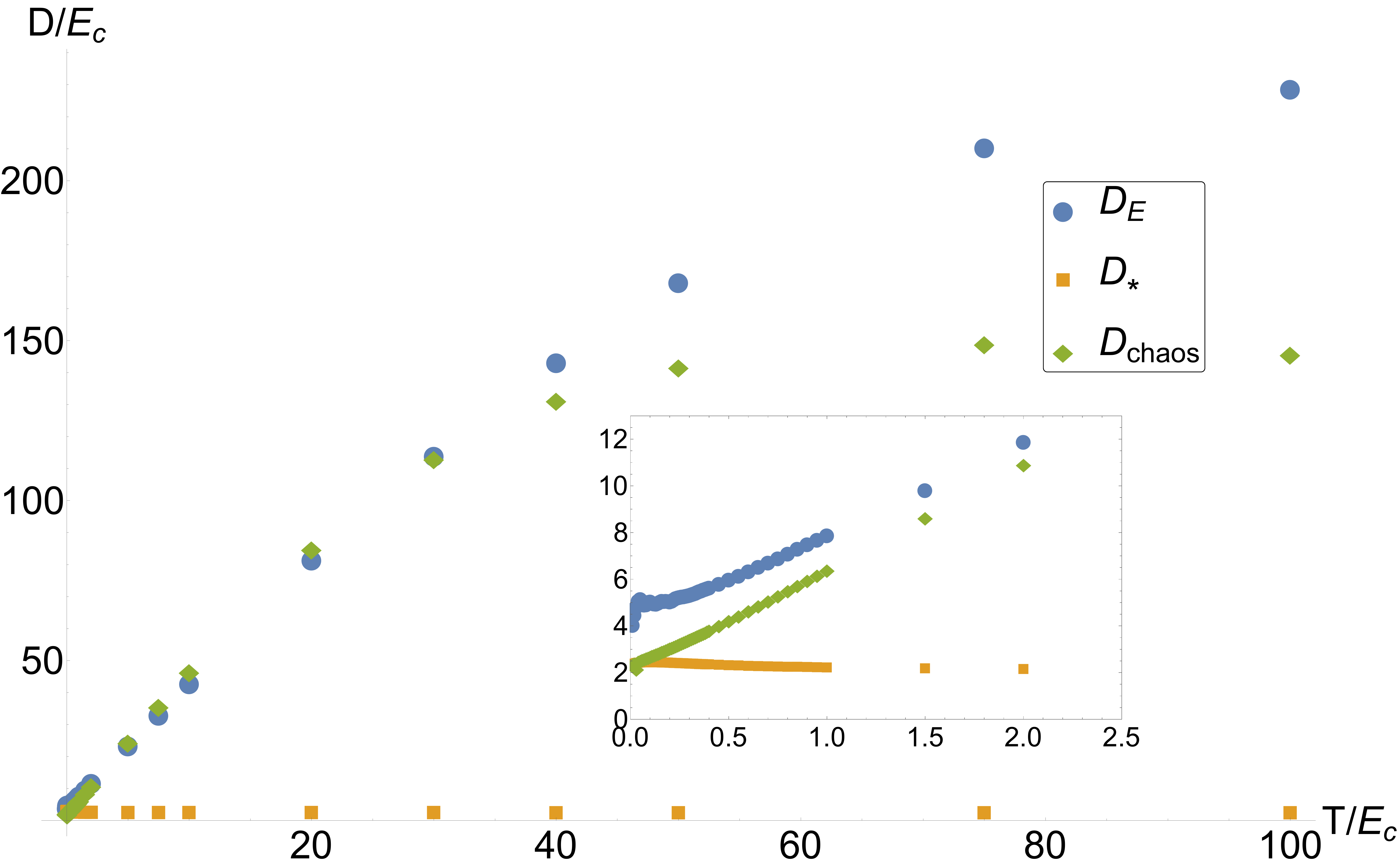}
  \caption{The scrambling diffusion coefficients $D_*,D_{\rm chaos}$ and the energy diffusion coefficient $D_E$ plotted against temperature $T/E_c$ ($U=200t_0$). The inset shows the low temperature region. $D_E$ and $D_{\rm chaos}$ are roughly equal while $D_*$ differs significantly. \label{fig:Dcompare}}
\end{figure}

\begin{figure}
\centering
  \includegraphics[width=0.95\columnwidth]{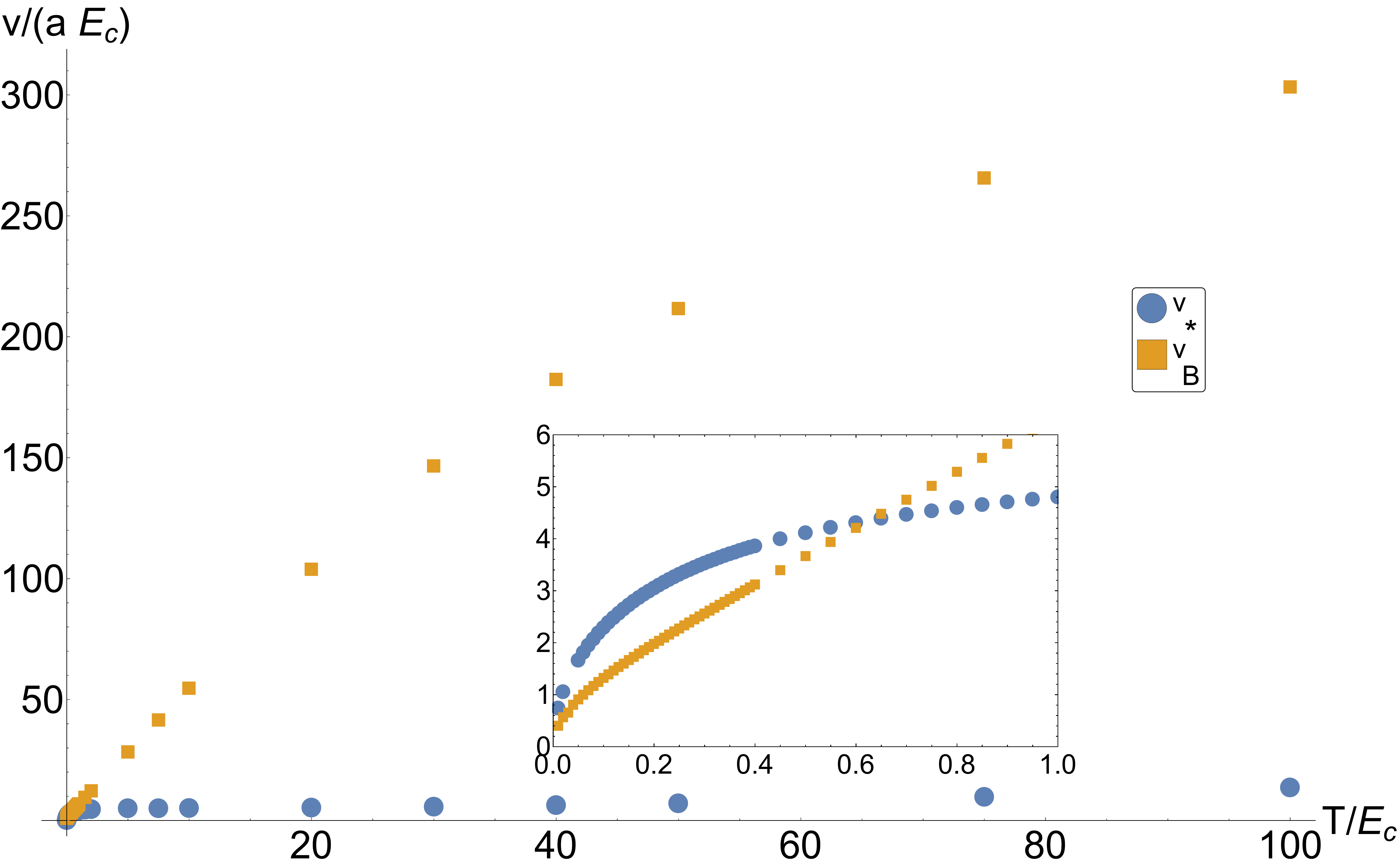}
  \caption{The two characteristic velocities $v_*$ and $v_B$ plotted against temperature, in the unit of $a E_c$ where $a=1$ is lattice spacing. $U/t_0=200$. At low temperatures (see inset) $v_*>v_B$ and at high temperatures $v_*<v_B$. \label{fig:vplot_g=0}}
\end{figure}

    In this section, we present the numerical results for $\lambda_L\equiv\lambda_L(q=0),D_*$ and $D_{\rm chaos}$, and compare $D_*,D_{\rm chaos}$ to energy transport. The scrambling rate is plotted in Fig.~\ref{fig:scrambling0} and the scrambling diffusion coefficients are plotted in Fig.~\ref{fig:Dcompare}. In Fig.~\ref{fig:vplot_g=0}, we compare the two characteristic velocites $v_*$ and $v_B$.

    We first comment on the $U/t_0$ dependence of our result. We found that although the qualitative features are the same, the overall magnitude of $\lambda_L$ does depend on $U/t_0$, especially at high temperatures. The reason might be the conformal symmetry breaking by temperature as in the original SYK model. Anyway, we will focus on the results of largest $U$ value in practice which is $U/t_0=200$.

    The numerical results show the following features:
\begin{enumerate}

\item    At low temperatures, the scrambling rate $\lambda_L$ grows quadratically as $T^2/E_c$, which matches with expectations from Fermi liquid theory. It is reported in Ref.~\onlinecite{Aurelio2018} that in a Majorana version of the $t$-$U$ model, $\lambda_L$ vanishes identically below some critical temperature, but our results do not support this.

\item    At high temperatures, the scrambling rate $\lambda_L$ approaches to a linear growth. Our numerical result yields $\lambda_L\approx6.2T$, close to the chaos bound. 

\item    The short-distance scrambling diffusion coefficient $D_*$ is roughly the order of $E_c$, with some weak $T$ dependence.

\item The long-distance scrambling diffusion coefficient $D_{\rm chaos}$ increases with temperature, and surpasses $D_*$ at temperature $T/E_c\sim O(1)$.

\item At low temperatures $v_*>v_B$ and at high temperatures $v_*<v_B$. The intersection is at at $T/E_c\sim O(1)$. This agrees with the qualitative features discussed in Sec.~\ref{sec:chaos_propagation}.

\end{enumerate}

\begin{figure}
    \centering
    \includegraphics[width=0.95\columnwidth]{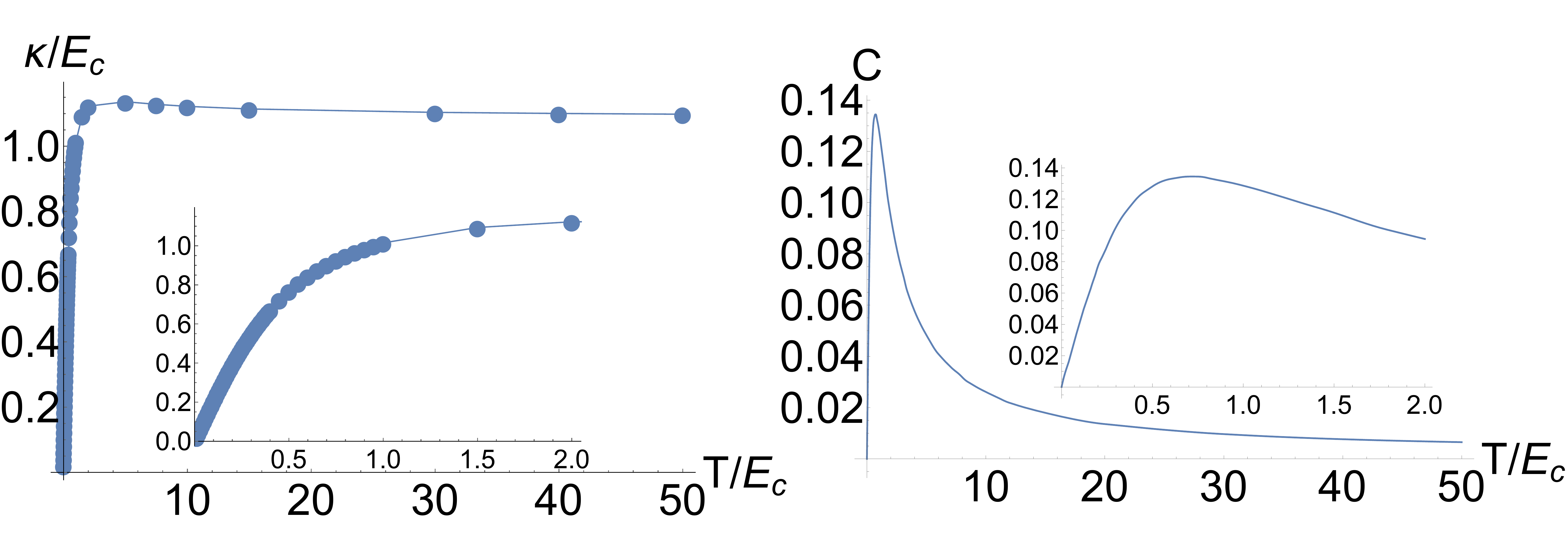}
    \caption{The thermal conductivity $\kappa$ (left) and the heat capacity $C$ (right) plotted against temperature $T/E_c$. The insets show the low temperature region. The thermal conductivity is computed at $U=200t_0$. The heat capacity is computed by combining data of various $U/t_0$. \label{fig:kappa_C}}
\end{figure}

    Next, we compare the scrambling diffusion coefficients $D_*,D_{\rm chaos}$ to the energy diffusion coefficient $D_E\equiv\kappa/C$, where $\kappa$ is the thermal conductivity and $C$ is the heat capacity. $\kappa$ and $C$ are plotted in Fig.~\ref{fig:kappa_C}. Both $\kappa$ and $C$ have been computed in Ref.~\onlinecite{Song}, and will be reproduced later in this paper. We found that at low temperatures $T\ll E_c$, $D_*\approx D_{\rm chaos}\approx D_E$. However, at elevated temperatures, the long-distance scrambling diffusion coefficient $D_{\rm chaos}$ closely follows the energy diffusion coefficient $D_E$, while the short-distance scrambling diffusion coefficient $D_*$ has a totally different behavior. This suggests that $D_{\rm chaos}$ may arise from the degrees of freedom that are also responsible for energy transport. Both $D_E$ and $D_{\rm chaos}$ grow linearly at low temperatures and saturate at high temperatures. The linear growth parts have approximately the same slope, but $D_{\rm chaos}$ saturates at lower temperature than $D_E$. The difference in saturation might be due to lattice details, because at high temperature $q_1$ is comparable to inverse lattice spacing. This finding agrees with the conjectured equivalence between energy transport and scrambling propagation \cite{Blake16,Blake16a,Gu,Patel2016,Blake2017,Gu2}.

    We also comment on the relation to charge diffusion coefficient $D_C=\sigma/K$ where $\sigma$ is conductivity and $K$ is charge compressibility. According to Ref.~\onlinecite{Song}, $\sigma$ is an order-one number at low temperatures and $\sigma\sim E_c/T$ at high temperatures; $K$ is of the order $1/U$ at all temperatures. This implies that $D_C\sim U$ at low temperatures and $D_C\sim t_0^2/T$ at high temperatures. As a result, $D_C$ has totally different behavior from $D_E,D_\text{chaos},D_*$. 

    To summarize, we have confirmed the following features of the $t$-$U$ model as summarized in Fig.~\ref{fig:crossovers}: First, both the scrambling rate $\lambda_L$ and the chaos propagation show a crossover behavior from Fermi liquid to SYK maximal chaos, consistent with the qualitative picture in Sec.~\ref{sec:scrambling}. Second, the long distance scrambling diffusion coefficient $D_{\rm chaos}$ approximately equals the energy diffusion coefficient $D_E$ in a wide range of temperature.

\section{Introducing Phonons}
\label{sec:phonons}

Phonons play an important role in understanding properties of strange metals. In this section, we propose a modification to the $t$-$U$ model to include effects of phonons, following Werman {\it et al.} \cite{Werman,Werman2}. For simplicity, we use the Einstein model of phonons, {\it i.e. } dispersionless phonon. To explore physics above the Mott-Ioffe-Regel (MIR) limit, we send the Debye frequency $\omega_0$ to zero. As the dispersionless phonons are not propagating, we can model them by harmonic oscillators residing on each site. To reflect the fact that in cuprates there are a large number of phonon bands, we add $N(N+1)/2$ types of phonons and let them couple to electrons through the Yukawa coupling $X_{abx}c_{ax}^\dagger c_{bx}$. Note that our phonon field is complex and there are $N^2$ real degrees of freedom.

    The Hamiltonian 
\begin{equation}
    H=H_\text{hopping}+H_\text{SYK}+H_\text{ph}+H_\text{e-ph}\label{eq:Hamiltonian_ph}
\end{equation}
    consists of four terms:
    the random hopping term
    \begin{equation}
    H_\text{hopping}=\frac{1}{\sqrt{zN}}\sum_{<xx'>}\sum_{ab}t^{ab}_{xx'}c_{ax}^\dagger c_{bx'},\label{eq:H_hopping_ph}
    \end{equation}
    the onsite SYK-interaction term
    \begin{equation}
    H_\text{SYK}=\frac{1}{(2N)^{3/2}}\sum_{x}U_{abcd,x}c_{ax}^\dagger c_{bx}^\dagger c_{cx} c_{dx},
    \end{equation}
    the phonon Hamiltonian
    \begin{equation}
    H_\text{ph}=\frac{M}{2}\sum_{x} \sum_{ab}\left(|\partial_t X_{abx}|^2+\omega_0^2|X_{abx}|^2\right),
    \end{equation}
    and the electron-phonon coupling term
    \begin{equation}
    H_\text{e-ph}=-\frac{\alpha}{\sqrt{N}}\sum_{x}\sum_{ab} X_{abx}c_{ax}^\dagger c_{bx}.
    \end{equation}
Here $X_{abx}$ is phonon field satisfying $X_{abx}=X_{ba x}^\dagger$. Other quantities have the same meaning as the $t$-$U$ model. 

In this model the dimensionless phonon coupling is defined as $g={\alpha^2}/({M \omega_0^2 t_0})$. We are interested in the physics when the temperature is much higher than Debye frequency, so we will consider the limit $\omega_0\rightarrow 0$, but keep $g$ fixed.

Also, the system has been tuned to half filling by setting chemical potential $\mu$ to zero. Readers may be concerned that the electron-phonon interaction can shift $\mu$. However, our model has the property that the electron-phonon interaction conserves the flavor indexed by $a$,  and it follows that the tadpole diagrams in the self-energy are $1/N$ suppressed. Consequently, at leading $1/N$ order the chemical potential shift is zero.

If $U=0$, the system is analytically soluble (see Appendix \ref{sec:U=0}) and it reduces to the electron-phonon system described in Ref.~\onlinecite{Werman}: for $T \ll t_0/g$ the electron-phonon scattering is weak and the electronic quasiparticles are well-defined; for $T \gg t_0/g$, the phonons act like static impurities of density proportional to $T$ (from the phonon Bose factor), and this leads to a linear-in-$T$ resistivity.
If $g=0$, the system is described  by a heavy Fermi liquid to SYK crossover \cite{Song}, which happens at temperature $T\sim E_c= t_0^2/U$. When both $U\neq0$ and $g\neq0$, as we will see below, the competition between electron-phonon and electron-electron interactions is set by the ratio $gt_0/U$, as illustrated in Fig.~\ref{fig:crossovers}. If $gt_0/U \gg 1$ the system will first enter the electron-phonon chaos regime as we raise the temperature, and vice versa.

\subsection{Keldysh Action and Equations of Motion}

    In this section we discuss the Keldysh formalism for the above Hamiltonian, and explain how to solve for the Green's function. It turns out that despite the addition of a large number of phonons, the problem is still as tractable as the $t$-$U$ model we started with.
    
 We perform disorder average over $t,U$ of the Keldysh action, and integrate out the quadratic phonon field $X$. Next we introduce the flavor averaged onsite Green's function $iG_{ss'x}(t,t')=\frac{1}{N}\sum_a c_{axs}(t)c_{axs'}^\dagger(t')$ and the Lagrange multiplier $\Sigma_{ss'x}(t,t')$ to impose the constraint. We then obtain the action 
 \begin{equation}
 S_K=S_\text{ph}+S_K^{tU}+S_\text{e-ph},
 \label{eq:SKphonon}
  \end{equation}
  where 
  \begin{equation}
      iS_\text{ph}=-\frac{N^2}{2}\sum_x \Tr \ln{i(\partial_t^2+\omega_0^2)}
  \end{equation}
  is the free phonon contribution, and $S_K^{tU}$ is the Kelydsh action \eqref{eq: Keldysh action tU} for $t$-$U$ model and the last term $S_\text{e-ph}$ is the electron-phonon interaction:
\begin{equation}
\begin{aligned}
    \frac{iS_\text{e-ph}}{N}&= -\frac{i\alpha^2}{2}\sum_{x, ss'}ss'\\
    &\int\rd^2 t G_{ss'x}(t,t')G_{s'sx}(t',t) D_{s's}(t',t).
\end{aligned}
\end{equation}
    Here $D_{ss'}(t,t')=[(-\partial_t^2-\omega_0^2)^{-1}]_{ss'}(t,t')$ is the free-phonon propagator. They can also be written in terms of the R,A,K components using Keldysh rotation.
In thermal equilibrium, the phonon Green's functions above are given by
\begin{eqnarray}
    D_{R/A}(\omega)&=&\frac{1}{M}\frac{1}{(\omega\pm i0)^2-\omega_0^2},\\
    D_K(\omega)&=&\coth(\frac{\omega}{2T})(D_R(\omega)-D_A(\omega)),
\end{eqnarray}
    where the second equation is the fluctuation-dissipation theorem.

    In the limit $T\gg\omega_0$, the $D_K$ component is dominant, we can approximate
\begin{equation}
    D_{ss'}(t)=\frac{1}{2}D_K(t)=\frac{-iT}{ M\omega_0^2}.
\end{equation}
    Therefore the action $S_{\text{e-ph}}$ reduces to
\begin{equation}\label{eq:Seph}
    \frac{iS_{\text{e-ph}}}{N}= -\frac{g t_0 T}{2}\sum_{x,ss'}ss'\int\rd^2 t G_{ss'x}(t,t')G_{s'sx}(t',t),
\end{equation}
which has the same form as on-site random hopping term. 

    Variation of the above action yields the saddle point equations
\begin{equation}\label{eq:EoM_Keldysh_ph}
\begin{aligned}
    G_R(\omega)^{-1}=&\omega-(t_0^2+g t_0 T)G_R(\omega)-\Sigma_R(\omega),\\
    G_K(\omega)=& 2i\tanh\frac{\omega}{2T}\Im G_R(\omega),\\
    \Sigma_R(t)=&\frac{1}{2}U^2G_K(-t)G_K(t)G_R(t)\\
    +&\frac{1}{4}U^2G_K(t)^2G_R^*(t)+\frac{1}{4}U^2G_R(t)^2G_R^*(t)\,.
\end{aligned}
\end{equation}

    We see that the equation of motion \eqref{eq:EoM_Keldysh_ph} have the same structure as the $t$-$U$ model counterpart, except that the electron hopping term $t_0^2$ is enhanced by phonons to $t_0^2+g t_0 T$. Physically, the similarity is due to the observation that in the limit $\omega_0\ll T$, the electron-phonons term plays a similar role as a random on-site $t$ term in the $t$-$U$ model.
    
    The above equations can be written into a dimensionless form in Appendix \ref{sec:dimensionless}, and the two important dimensionless parameters are $gt_0/U$ and $T/E_c$.

\section{Transport in the Phonon Model}

\subsection{Deriving Transport Coefficients}
    To leading order, the transport coefficients can be derived by expanding the action \eqref{eq:SKphonon} around the saddle point solution with $U(1)$ phase fluctuations $\phi$ and time-reparameterization fluctuations $\epsilon$ to quadratic order, and the transport coefficients can be directly read out as the coefficients of $\phi^2$ and $\epsilon^2$~ \cite{Song}. However, for the purpose of diagrammatic expansion, we will use Kubo formulas.

    Although the Hamiltonian Eq.\eqref{eq:Hamiltonian} breaks spatial translation symmetry, it respects time translation and $U(1)$ rotation. Consequently we can still use Noether procedure to extract the current operator from the hopping term Eq.\eqref{eq:H_hopping}, but the current operator now has explicit spatial dependence.

    If we perform an infinitesimal inhomogeneous symmetry transformation, for example, a $U(1)$ rotation $c_x\to c_x e^{i\varepsilon_x}$, the action will change by
\begin{equation}
    \delta S=\sum_{<xx'>}j_{xx'}(\varepsilon_x-\varepsilon_{x'}),
\end{equation}
    and we obtain the current operator $j_{xx'}$ on the link $xx'$.

    Following that procedure, we get the charge current $j_C$ and the heat current $j_E$:
\begin{eqnarray}
    j_{Cxx'}&=&i\sum_{ab}t_{xx'}^{ab}c_{ax}^\dagger c_{bx'}-t_{x'x}^{ba}c_{bx'}^\dagger c_{ax},\\
    j_{Exx'}&=&\sum_{ab}t_{xx'}^{ab}c_{ax}^\dagger \partial_t c_{bx'}+t_{x'x}^{ba}\partial_t c_{bx'}^\dagger c_{ax}.\label{eq:j_E}
\end{eqnarray}
Note that because the phonon modes are purely local, there is no direct contribution above from the phonon degrees of freedom.
    The on-site current-current polarization function in imaginary time is as usual defined as
\begin{equation}
    \Pi^{jj}_{xx'}(\tau)=-\langle j_{xx'}(\tau) j_{xx'}(0)\rangle,
\end{equation}where the average is over states and disorder.
The conductivities per flavor are given by Kubo formulas
\begin{eqnarray}
    N\sigma=\lim_{\omega\to 0}\frac{\Im\Pi_{xx'}^{j_Cj_C}(i\omega\to\omega+i\delta)}{-\omega},\\
    N\kappa=\lim_{\omega\to 0}\frac{\Im\Pi_{xx'}^{j_Ej_E}(i\omega\to\omega+i\delta)}{-\omega T}.
\end{eqnarray}

    For reference, we list the leading order formulas for DC electrical conductivity, DC thermal conductivity, and optical conductivity respectively.
\begin{eqnarray}
   && \sigma_\text{DC}= \frac{2t_0^2}{\pi}\int\rd \omega (\Im G_R(\omega))^2 \beta n_F(\omega)n_F(-\omega),\label{eq:sigma_dc}\\
&&    \frac{\kappa_0}{T}= \frac{2t_0^2}{\pi}\int\rd \omega (\beta\omega\Im G_R(\omega))^2 \beta n_F(\omega)n_F(-\omega),\label{eq:kappa}\\
  &&  \Re\sigma_\text{opt}(\nu)=\frac{2t_0^2}{\pi}\int\rd \omega \Im G_R(\omega)\Im G_R(\omega+\nu)\nonumber \\ && \hspace{100pt} \times\frac{n_F(\omega)-n_F(\omega+\nu)}{\nu}.
\end{eqnarray}
\subsection{DC Resistivity}

\begin{figure}[htb]
\centering
\includegraphics[width=0.95\columnwidth]{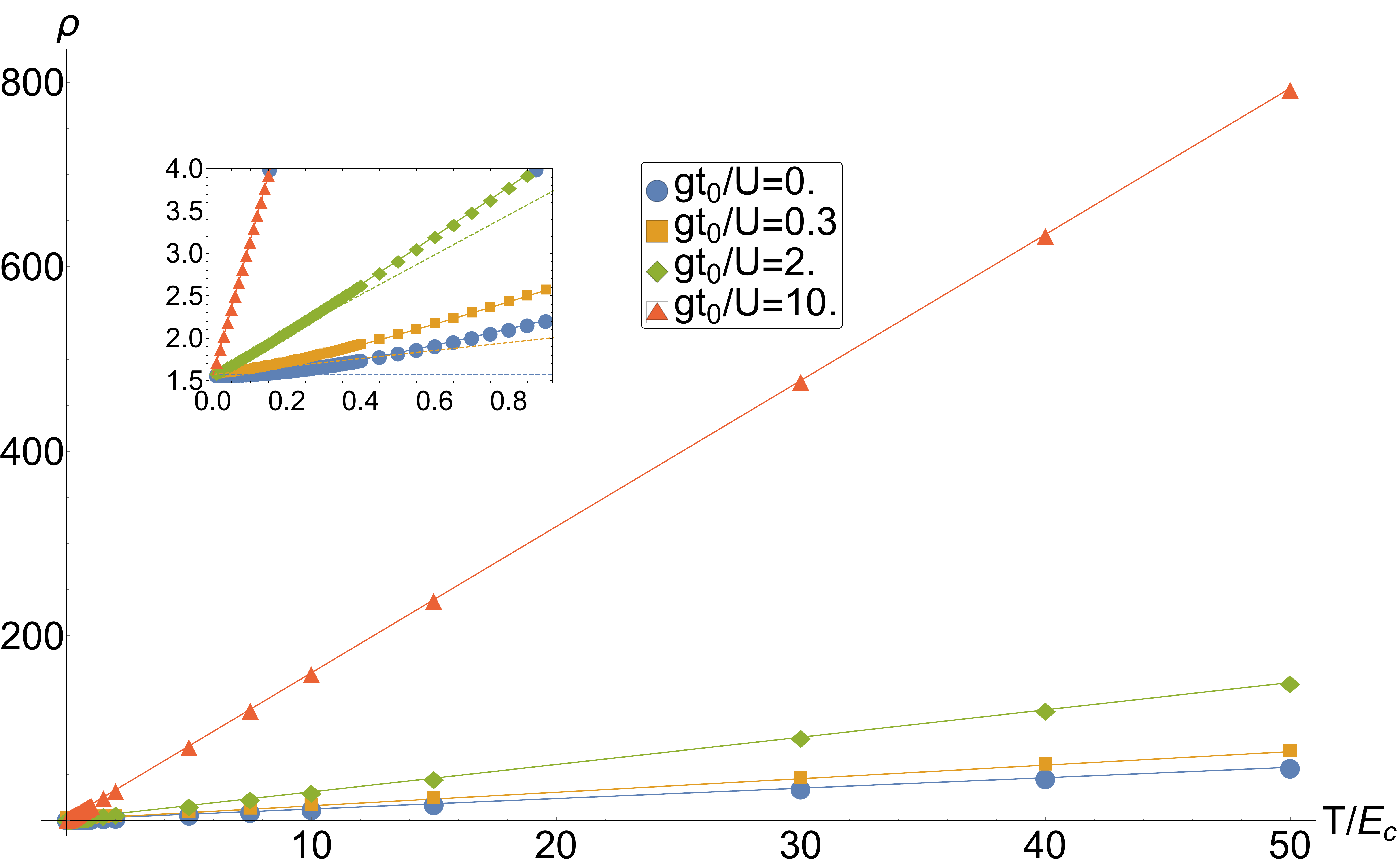}
\caption{\label{fig:rhoplot1} The resistivity $\rho=1/\sigma_\text{DC}$ plotted as a function of dimensionless temperature $T/E_c$, for different value of $g$ at $U/t_0=200$. The solid lines are guides to eyes. The dashed lines show $U=0$ values for comparison. The inset zooms into the low temperature region. }
\end{figure}

\begin{figure}[htb]
\centering
\includegraphics[width=0.95\columnwidth]{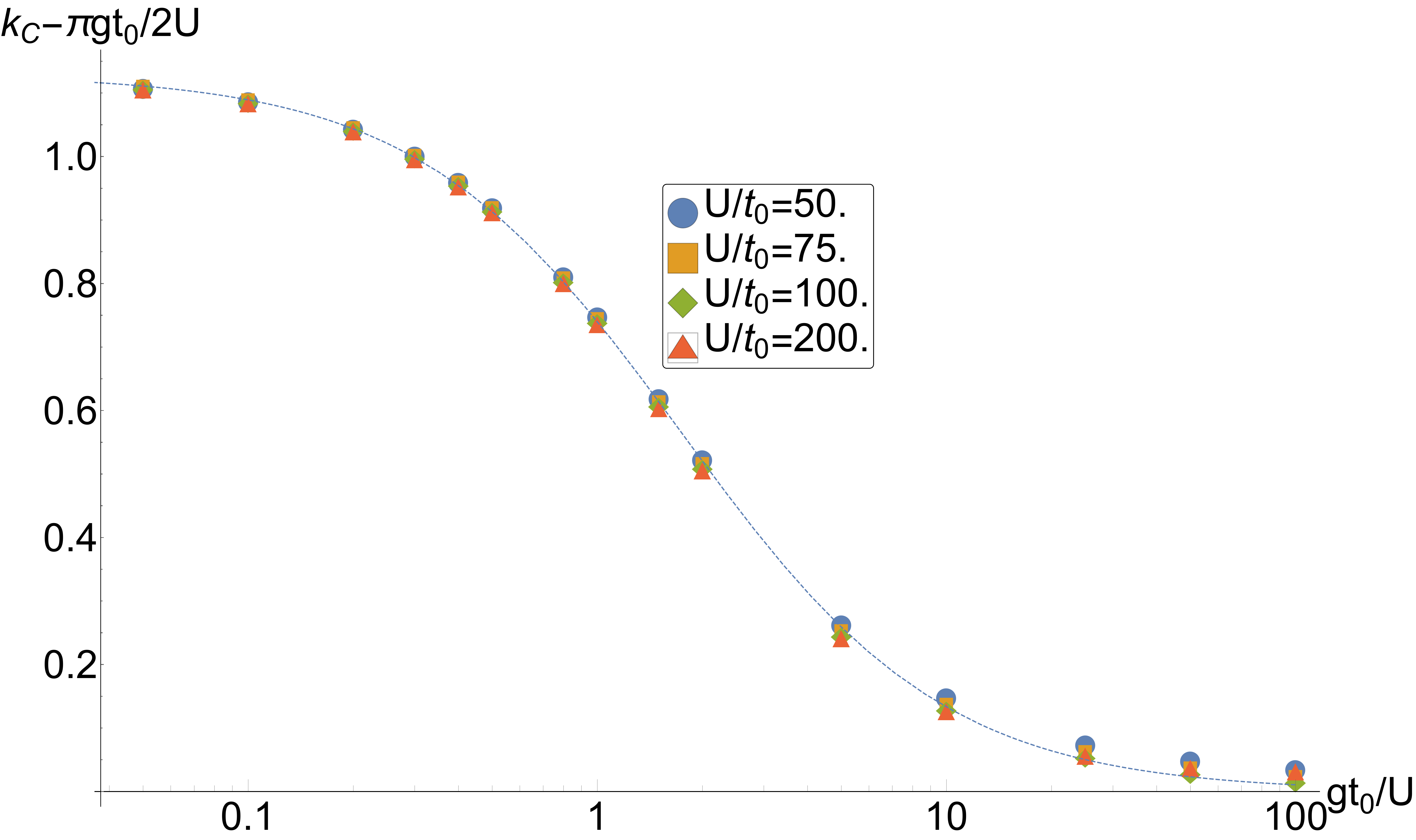}
\caption{\label{fig:rhoplot2} The slope $k_C$ plotted against $gt_0/U$ for different values of $U$. The curves of different $U$ collapses into a single curve, confirming the scaling property. The dashed line is a fit discussed in the main text.}
\end{figure}

    We calculate the DC resistivity using the saddle point Green's function and Eq.~\eqref{eq:sigma_dc}.
    The resistivity results are shown in Fig.~\ref{fig:rhoplot1} and Fig.~\ref{fig:rhoplot2}.

    The resistivity is a dimensionless quantity, so on dimensional grounds, it should be a function of $gt_0/U$ and $T/E_c$ only. As a sanity check, we calculated $\rho$ at fixed $gt_0/U$, $T/E_c$, while varying $U/t_0$, and we found that the results are independent of $U/t_0$ to good precision.

    For $g=0$, the resistivity increases quadratically at low temperature, and becomes linear after the coherence scale $E_c=t_0^2/U$. This reproduces the result in Ref.~\onlinecite{Song}.

     In the regime $T\ll E_c$, the resistivity curve can be approximated by the $U=0$ result. The approximation works better for larger $g$.

     Now shift attention to the high temperature regime $T>E_c$, we see from Fig.~\ref{fig:rhoplot1} that the resistivity is linear in $T$ in this regime. We denote the slope of the curve as $k_C$, that is
     $$
       \lim_{T\rightarrow\infty} \frac{\rd\rho}{\rd (T/E_c)}\rightarrow k_C,
     $$
     and plot $k_C$ in Fig.~\ref{fig:rhoplot2}. When $gt_0/U\gg1$, we get the $U=0$ result $k_C=\frac{\pi}{2}\frac{gt_0}{U}$. When $gt_0/U=0$, we get $k_C=1.129$, which agrees with the pure SYK result $k_C=2/\sqrt{\pi}=1.128$ in Ref.~\onlinecite{Song}.
     We notice that as $g$ increases, $k_C$ gets closer to the $U=0$ value. The effect of SYK interaction is suppressed by phonon.

     Interestingly, we find that $k_C$ can be fitted pretty well with the function $k_C=f(gt_0/U;a,b,c)+\frac{\pi}{2}\frac{gt_0}{U}$, where
     \begin{equation}\label{eq:fit}
        f(x;a,b,c)=\frac{c}{1+b x^a},
     \end{equation}
     with $a=1.15,b=0.526,c=1.130$.

\subsection{Optical Conductivity}
\begin{figure}[htb]
\centering
\includegraphics[width=0.95\columnwidth]{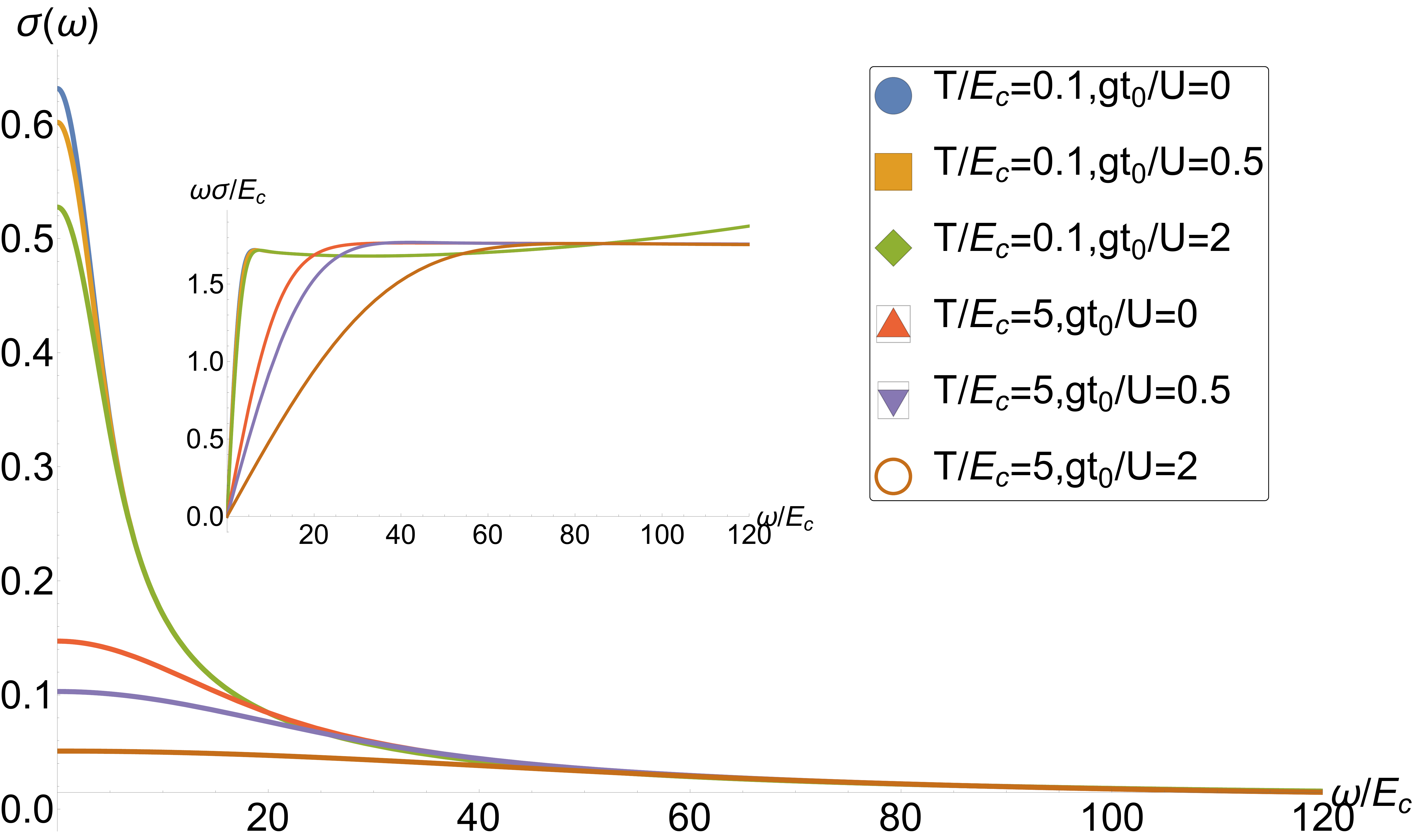}
\caption{\label{fig:optical_conductivity} The real part of the optical conductivity at $U=200t_0$. The large frequency part approximately follows a $1/\omega$ trend.}
\end{figure}

    The real part of the optical conductivity is shown in Fig.~\ref{fig:optical_conductivity}.

    By dimensional analysis, the optical conductivity $\sigma(\omega)$ should be a function of $\omega/E_c$, $gt_0/U$ and $T/E_c$, and our result confirms that.

    The optical conductivity has a peak at $\omega=0$, but instead of Lorentzian decay (a character of the Drude peak), has $1/\omega$ decay at large $\omega$, which reflects the $1/\sqrt{U\omega}$ SYK spectral weight at high frequencies. Nonzero temperature $T$ and phonon coupling $g$ can cause the peak to get broadened and lowered, but the curve eventually follows the $1/\omega$ behavior. We see that the low frequency behavior of $\sigma(\omega)$ is dominated by phonon, but the high frequency behavior still follows from SYK.

\begin{figure}[htb]
\centering
\includegraphics[width=0.95\columnwidth]{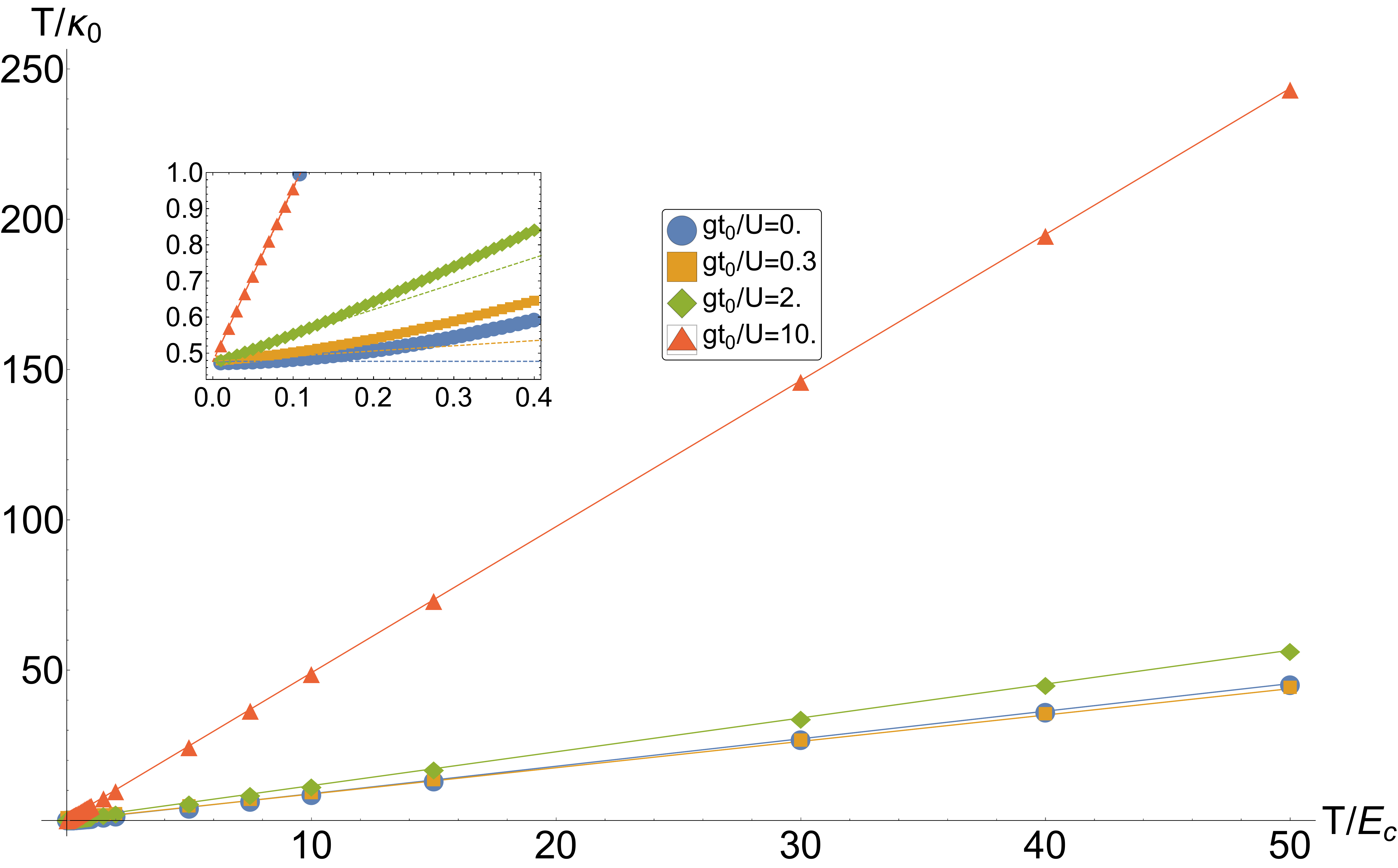}
\caption{\label{fig:kappaplot1} The inverse thermal conductivity (electron part) $T/\kappa_0$ plotted as a function of dimensionless temperature $T/E_c$, for different value of $g$ at $U/t_0=200$. The solid lines are guides to eyes. The dashed lines show $U=0$ values for comparison. The inset zooms into the low temperature region. }
\end{figure}

\begin{figure}[htb]
\centering
\includegraphics[width=0.95\columnwidth]{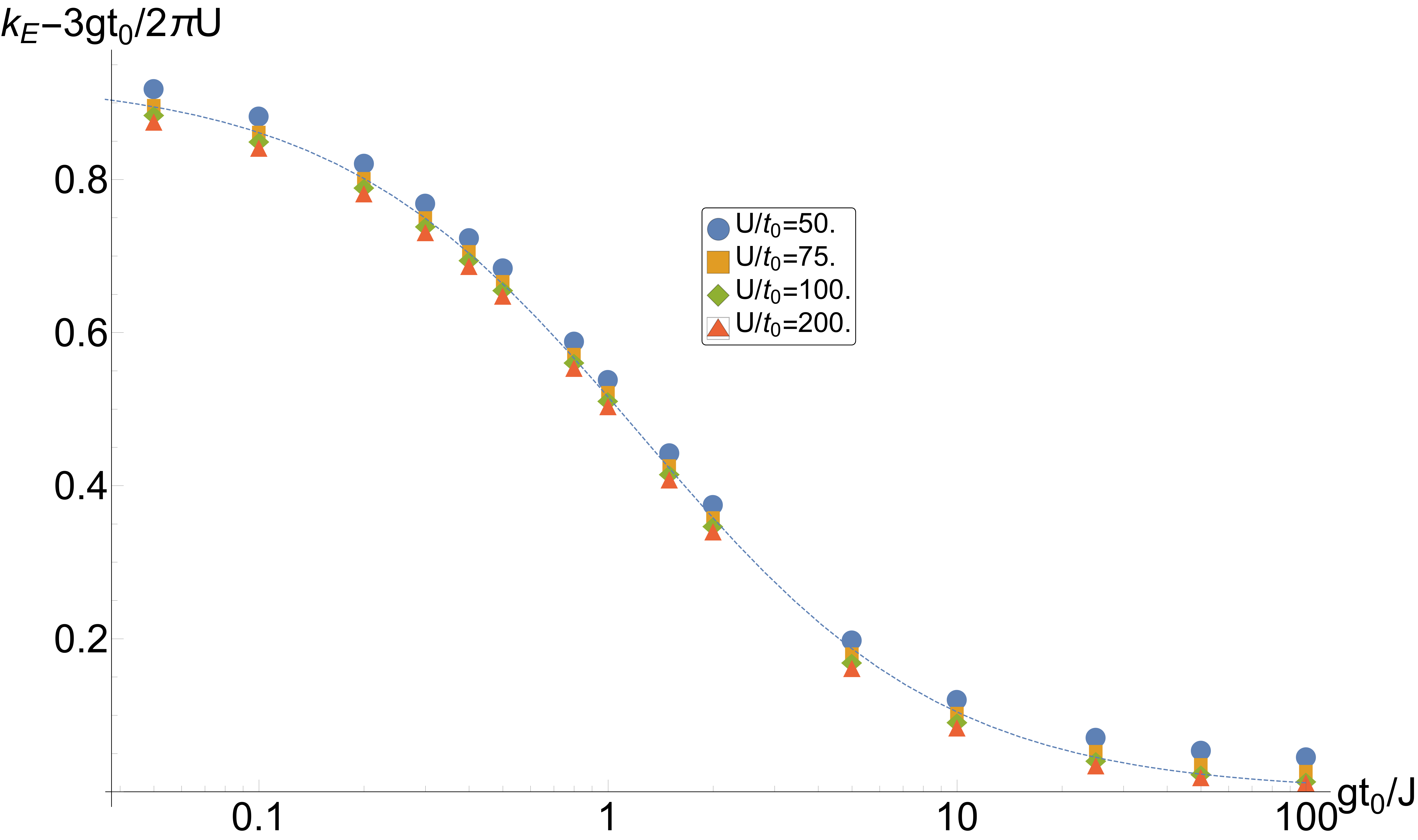}
\caption{\label{fig:kappaplot2} The slope $k_E$ plotted against $gt_0/U$ for different values of $U$. The curves of different $U$ collapses into a single curve, confirming the scaling property. The dashed line is a fit discussed in the main text.}
\end{figure}

\subsection{Resistivity Saturation}
\label{sec:saturation}
    In Ref.~\onlinecite{Werman2}, it was found that if we put phonons on bonds between lattice sites, and let them couple to electrons at the two ends of the bonds, there will be resistivity saturation effects. This can be easily seen in our model.

We replace the electron-phonon coupling by
\begin{equation}
    H_\text{e-ph}=-\frac{\alpha}{\sqrt{zN}}\sum_{<xx'>}\sum_{ab} X_{ab}^{xx'}c_{ax}^\dagger c_{bx'},
\end{equation}
    which is normalized to reproduce the saddle point equations in the previous section.

    The only difference now is that this term couples electrons at different sites, so it should enter the conductivity calculation. One can see that this is a simple replacement of the pre-factor $t_0^2\to t_0^2+gt_0T$. The conductivity formula is now
\begin{equation}
    \sigma= \frac{2(t_0^2+gt_0 T)}{\pi}\int\rd \omega (\Im G_R(\omega))^2 \beta n_F(\omega)n_F(-\omega),
\end{equation}
    which implies that
\begin{equation}
    \rho_\text{bond}=\frac{\rho_\text{site}}{1+\frac{g T}{t_0}}.
\end{equation}
    Since $\rho_\text{site}$ is linear in $T$ at high temperature, $\rho_\text{bond}$ saturates. The value of saturation is
\begin{equation}
    \rho_\infty=\frac{k_C}{gt_0/U}.
\end{equation}

    Hence, if $gt_0/U\gg1$, $\rho_\infty=\pi/2$. However, for $gt_0/U\ll 1$, $\rho_\infty=k_{C}U/gt_0\gg 1$. Both SYK interaction and electron-phonon interaction give rise to linear-in-$T$ electron scattering rate, but they differ on whether the MIR limit is exceeded. For electron-phonon interaction, the resistivity saturates to an $O(1)$ value which is the MIR limit in natural units, but for SYK interaction, the resistivity can reach values much larger than the MIR limit.

\subsection{Thermal Conductivity and Phonon Drag Effect}
\label{sec:phonondrag}
      
    Naively, the thermal conductivity can be calculated using the leading order result \eqref{eq:kappa} which only includes effects of electrons. The results are shown in Fig.~\ref{fig:kappaplot1} and Fig.~\ref{fig:kappaplot2} (here $\kappa_0$ means $\kappa_{\rm electron}$ in Fig.~\ref{fig:crossovers}). 
    
   The results of electron thermal conductivity are similar to the DC resistivity. We can as well define the slope
   $$
       \lim_{T\rightarrow\infty} \frac{\rd(T/\kappa_0)}{\rd (T/E_c)}\rightarrow k_E.
   $$
    When $U=0$, $k_E=3gt_0/(2\pi U)$. When $gt_0/U=0$, we get $k_E=0.92$, agreeing with $k_E=16/\pi^{5/2}=0.915$ of Ref.~\onlinecite{Song}.

    We also found that $k_E$ can be fitted by $k_E=f(gt_0/U;a,b,c)+3gt_0/(2\pi U)$, where $f$ is defined by Eq.\eqref{eq:fit} and $a=0.99,b=0.81,c=0.932$.
    
    However, the thermal conductivity that we have calculated does not include the contribution of phonons. Electrons are strongly incoherent due to electron-phonon interactions and the SYK interaction. However, because there are many more phonons, $O(N^2)$, than electrons, $O(N)$, the interaction effects on phonons are diluted and phonons are still well defined quasiparticles with a long lifetime of order $O(N)$. If we excite an electron in the system, it quickly decays and transfers its energy to phonons. Because phonons are long-lived quasiparticles, we expect them to have a significant contribution to transport. This phenomenon is called ``phonon drag", and has been studied in systems without SYK interaction \cite{Werman}. Results there show that phonon drag is important in energy transport but not in charge transport. In the rest of this section, we will work out the phonon drag correction (denoted as $\kappa_1$ here and $\kappa_{\rm phonon}$ in Fig.~\ref{fig:crossovers}) to DC thermal conductivity.

We recall that there is no intrinsic phonon conductivity, of $O(N^2)$, in our model because the phonons are purely local. Then phonons contribute only via the drag effects noted above.
\subsubsection{Phonon Self Energy}

    As a preparation, we need the phonon self energy to obtain the phonon life-time. Because there are more phonons than electrons, the phonon self energy is of order $O(1/N)$, so is the phonon decay rate.

    The dressed phonon propagator is
\begin{equation}
    D(\omega)=\frac{1}{M}\frac{1}{\omega^2-\omega_0^2-\Sigma_\text{ph}(\omega)}.
\end{equation}

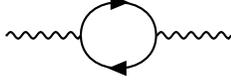
\begin{figure}[htb!]
\centering
\begin{tikzpicture}
\begin{feynman}
\diagram[large, horizontal=v1 to v2] {
v1--[photon] v2--[fermion, half left, looseness=1.5] v3--[photon] v4,
v3--[fermion, half left, looseness=1.5]v2
};
\end{feynman}
\end{tikzpicture}
\caption{\label{fig:phonon_self_energy} The diagram for phonon self energy. }
\end{figure}
    The leading term in $1/N$ is given in Fig.~\ref{fig:phonon_self_energy}. The expression for $\Sigma_\text{ph}$ is
\begin{equation}
     \Sigma_\text{ph}(i\omega)=\frac{\alpha^2}{NM}T\sum_{i\nu}G(i\omega+i\nu)G(i\nu).
\end{equation}
    Perform Matsubara summation, analytically continue to real frequency $i\omega\to \omega+i\delta$, and take the imaginary part, we get
\begin{equation}
\begin{split}
    \Im\Sigma_\text{ph}(\omega)&=-\frac{2\alpha^2}{MN}\int\frac{\rd{\nu}}{2\pi}\Im G_R(\nu)\Im G_R(\omega+\nu)\\
    &\times\left[n_F(\nu)-n_F(\omega+\nu)\right]\\
    &=-\frac{\alpha^2}{2MN}\frac{\omega\sigma(\omega)}{t_0^2}\\
    &=-\frac{1}{2N}g\omega_0^2\frac{\omega\sigma(\omega)}{t_0}\equiv-2\omega\Gamma(\omega),
\end{split}
\end{equation}
    where we have used the definitions of the phonon coupling $g$ and the optical conductivity $\sigma(\omega)$, and in the last line we defined the phonon decay rate $\Gamma$.

    Here we see that the $\Im\Sigma_\text{ph}$ is small, implying that phonons are long-lived (lifetime$\sim O(N)$) quasiparticles.

\subsubsection{Correction to Thermal Conductivity}

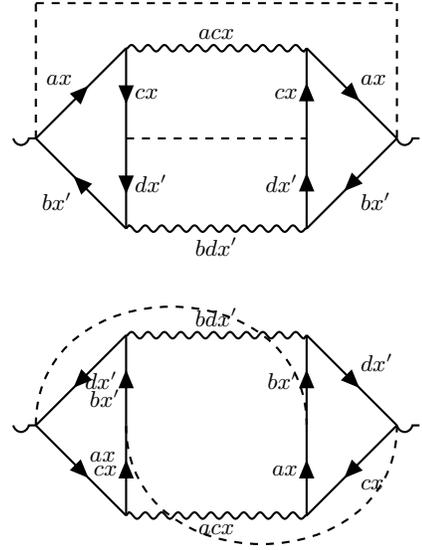
\begin{figure}[htb!]
\centering
\begin{tikzpicture}[scale=1]
\begin{feynman}[scale=0.6,large,transform shape]

\vertex (v1);
\vertex [right=2cm of v1] (v7);
\vertex [above=2cm of v7] (v2);
\vertex [right=4cm of v2] (v3);
\vertex [right=4cm of v7] (v8);
\vertex [right=2cm of v8] (v4);
\vertex [below=2cm of v7] (v6);
\vertex [right=4cm of v6] (v5);
\vertex [above=3cm of v1] (a);
\vertex [above=3cm of v4] (b);
\vertex [left=0.5cm of v1] (i);
\vertex [right=0.5cm of v4] (o);
\diagram[scale=1.6] {
(v1)--[fermion, edge label=$ax$](v2)--[photon, edge label=$acx$](v3)--[fermion, edge label=$ax$](v4)--[fermion, edge label=$bx'$] (v5)--[photon, edge label=$bdx'$](v6)--[fermion,edge label=$bx'$](v1),
(v2)--[fermion,edge label=$cx$](v7)--[fermion, edge label=$dx'$](v6),
(v5)--[fermion,edge label=$dx'$](v8)--[fermion,edge label=$cx$](v3),
(v7)--[scalar](v8),
(v1)--[scalar](a)--[scalar](b)--[scalar](v4),
(i)--[gluon](v1),
(v4)--[gluon](o),
};
\end{feynman}
\end{tikzpicture}
\begin{tikzpicture}
\begin{feynman}[scale=0.6,large,transform shape]

\vertex (v1);
\vertex [right=2cm of v1] (v7);
\vertex [above=2cm of v7] (v2);
\vertex [right=4cm of v2] (v3);
\vertex [right=4cm of v7] (v8);
\vertex [right=2cm of v8] (v4);
\vertex [below=2cm of v7] (v6);
\vertex [right=4cm of v6] (v5);
\vertex [above=3cm of v1] (a);
\vertex [above=3cm of v4] (b);
\vertex [left=0.5cm of v1] (i);
\vertex [right=0.5cm of v4] (o);
\diagram[scale=1.6] {
(v2)--[photon, edge label=$bdx'$](v3)--[fermion, edge label= $dx'$](v4)--[fermion, edge label=$cx$] (v5)--[photon, edge label=$acx$](v6),
(v1)--[fermion,edge label=$ax$](v6)--[fermion,edge label=$cx$](v7)--[fermion, edge label=$dx'$](v2)--[fermion,edge label=$bx'$](v1),
(v5)--[fermion,edge label=$ax$](v8)--[fermion,edge label=$bx'$](v3),
(v1)--[scalar, half left, looseness=1.5](v8),
(v7)--[scalar, half right, looseness=1.5](v4),
(i)--[gluon](v1),
(v4)--[gluon](o),
};
\end{feynman}
\end{tikzpicture}
\caption{\label{fig:phonon_drag} The diagrams for phonon drag. Solid arrowed lines denote fermions. Wavy lines denote phonons. Dashed lines denote contractions of $t^{ab}_{xx'}$. }
\end{figure}

    In this section we calculate the phonon drag contribution to thermal conductivity, whose diagrams are given by Fig.~\ref{fig:phonon_drag}. Naively, these diagrams are sub-leading in $1/N$, but because the phonons are long-lived with a decay rate $\Gamma\sim O(1/N)$, the product of two phonon propagators has an $O(N)$ enhancement:
\begin{equation}\label{eq:DRDA}
    D_R(\varepsilon)D_A(\varepsilon+\omega)\approx\frac{1}{M^2\omega_0}\frac{\pi}{2\Gamma(\varepsilon)+i\omega}\delta(\varepsilon^2-\omega_0^2).
\end{equation}
    Eq.\eqref{eq:DRDA} says that $D_RD_A$ is strongly on-shell, and has a magnitude of $O(N)$ at the DC limit $\omega=0$. Taking this into account, the two diagrams in Fig.~\ref{fig:phonon_drag} are of the same order as the leading order diagrams.

    The current-current correlation function is
\begin{equation}\label{eq:Pi_E}
\begin{split}
    \Pi(i\omega)=-\frac{2\alpha^4t_0^4}{N}T\sum_{\Omega} &V(i\Omega+i\omega,i\Omega)V(i\Omega,i\Omega+i\omega)\\
    &D(i\Omega+i\omega)D(i\Omega)\,.
\end{split}
\end{equation}
    Here, the vertex function $V$ is defined as
\begin{equation}\label{eq:V}
\begin{split}
    V(i\omega_1,i\omega_2)=&T\sum_{\mu}  \big[G(i\mu+i\omega_1)G(i\mu+i\omega_2)(i\mu+i\omega_1) \\
 -&G(i\mu-i\omega_1)G(i\mu-i\omega_2)(i\mu-i\omega_1)\big] G(i\mu)^2 \,.
\end{split}
\end{equation}

    In Eq.\eqref{eq:Pi_E} and \eqref{eq:V}, the prefactors are obtained as the following:
\begin{enumerate}
  \item Each internal phonon vertex is associated with a factor $(-\alpha)$. Each internal hopping vertex is associated with a factor $(-t_0)$.
  \item The time derivative in the current operator Eq.\eqref{eq:j_E} yields an extra $i$ due to Wick rotation $t\to -i\tau$, which leads to an overall minus sign in Eq.\eqref{eq:Pi_E}.
  \item The two diagrams in Fig.~\ref{fig:phonon_drag} should differ by a minus sign. The first diagram comes from contraction $\langle\partial_\tau c^\dagger\partial_\tau c\rangle+h.c.$, and the second diagram comes from $\langle\partial_\tau c\partial_\tau c\rangle+h.c.$. The minus sign is due to opposite frequency sign conventions for $c$ and $c^\dagger$.
  \item $\langle j_Ej_E\rangle$ contains four terms. By simple inspection we see that each term in $\langle j_Ej_E\rangle$ leads to two diagrams, so in total there should be eight diagrams, but the $VV$ product in Eq.\eqref{eq:Pi_E} only contains four diagrams. In fact, the missing four terms are those obtained by exchanging $\omega_1,\omega_2$ in Eq.\eqref{eq:V}. The exchange doesn't affect the value of Eq.\eqref{eq:V} in the DC limit, so we simply account for it by a factor of 2.
\end{enumerate}

    We now start evaluating Eq.\eqref{eq:Pi_E} by performing the Matsubara summation over $\Omega$, and we get
\begin{equation}
\begin{split}
    \Pi(\omega)=&\frac{2i\alpha^4t_0^4}{N} \int \frac{\rd\Omega}{2\pi}[n_B(\Omega)-n_B(\Omega-\omega)]\\
                &\times V(\Omega+i\delta,\Omega-\omega-i\delta)V(\Omega-\omega-i\delta,\Omega+i\delta)\\
                &\times D_R(\Omega)D_A(\Omega-\omega).
\end{split}
\end{equation}

    Next, we use Eq.\eqref{eq:DRDA}, apply the Kubo formula, expand to lowest order in $\omega_0$ and rewrite various constants to obtain:
\begin{equation}
\begin{split}
    \kappa_1=&\sum_{\Omega=\pm \omega_0} \frac{g t_0^7}{\omega_0^2\sigma}|V(\Omega+i\delta,\Omega-i\delta)|^2\\
          =&\frac{2g t_0^7}{\omega_0^2\sigma}|V(\omega_0+i\delta,\omega_0-i\delta)|^2,
\end{split}
\end{equation}
    where $\sigma$ is the DC conductivity.

    The final step is to evaluate the vertex function $|V(\omega_0+i\delta,\omega_0-i\delta)|$ to linear order in $\omega_0$.

    When performing Matsubara summation in $V$, the resulting integrand has three branch cuts in the complex $\mu$-plane, the contribution we seek is to integrate along the upper half of the highest cut, and the lower half of the lowest cut, which amounts to making all propagators retarded or advanced:

\begin{equation}\label{eq:kappa1}
\begin{split}
    V(\omega_0+i\delta,&\omega_0-i\delta)=\int\frac{\rd\varepsilon}{2\pi i} n_F(\varepsilon)G_R(\varepsilon)^2\\
    &\times \left[G_R(\varepsilon-\omega_0)^2-G_R(\varepsilon+\omega_0)^2\right]+c.c. \\
    =& -\omega_0\int\frac{\rd\varepsilon}{2\pi}
    \Im [G_R(\varepsilon)^4] \\
    &\times \left[\tanh\left(\frac{\beta\varepsilon}{2}\right)-2\beta n_F(\varepsilon)n_F(-\varepsilon)\varepsilon\right],
\end{split}
\end{equation}
where in the second line, we expanded in $\omega_0$, integrated by parts and symmetrized the integral because $\Im [G_R^4]$ is odd.

    In the above integral, the $\tanh(\beta\epsilon/2)$ term is dominant over the $\beta n_F(\varepsilon)n_F(-\varepsilon)\varepsilon$ term, because the first has support over the whole spectrum $[-\Lambda,\Lambda],~\Lambda\sim\min(U,\sqrt{t_0^2+g T t_0})$, while the second only has support over $[-T,T]$, so it is suppressed by higher powers of $T/t_0$.

    The contributions to $V$ from other contours in the complex-$\mu$ plane, in fact, can all be arranged to have a factor $n_F(\varepsilon+\omega_0)-n_F(\omega_0)$, and are suppressed by the same reason as above.

    In summary, the phonon-drag contribution to the thermal conductivity is
\begin{equation}
    \kappa_1=\frac{2gt_0^7}{\sigma}\left[\int\frac{\rd\varepsilon}{2\pi}\tanh(\frac{\beta\varepsilon}{2})\Im [G_R(\varepsilon)^4]\right]^2.
\end{equation}

    As a sanity check, we inspect the $U=0,T\ll t_0$ limit. In this limit,
\begin{eqnarray}
    &\sigma\sim\frac{t_0^2}{t_0^2+g T t_0},\\
    &\left(\int\frac{\rd\varepsilon}{2\pi}\tanh(\frac{\beta\varepsilon}{2})\Im [G_R(\varepsilon)^4]\right)^2\sim\frac{1}{(t_0^2+gTt_0)^3},
\end{eqnarray}so
\begin{equation}\label{eq:kappa1_ex}
    \kappa_1\sim\frac{gt_0^5}{(t_0^2+gTt_0)^2}.
\end{equation}
    Further assuming $g T\gg t_0$, we get $\kappa_1\sim\frac{t_0}{g}(\frac{t_0}{T})^2$, which agrees with Ref.~\onlinecite{Werman}.
\subsubsection{Results}

\begin{figure}[htb]
\centering
\includegraphics[width=0.95\columnwidth]{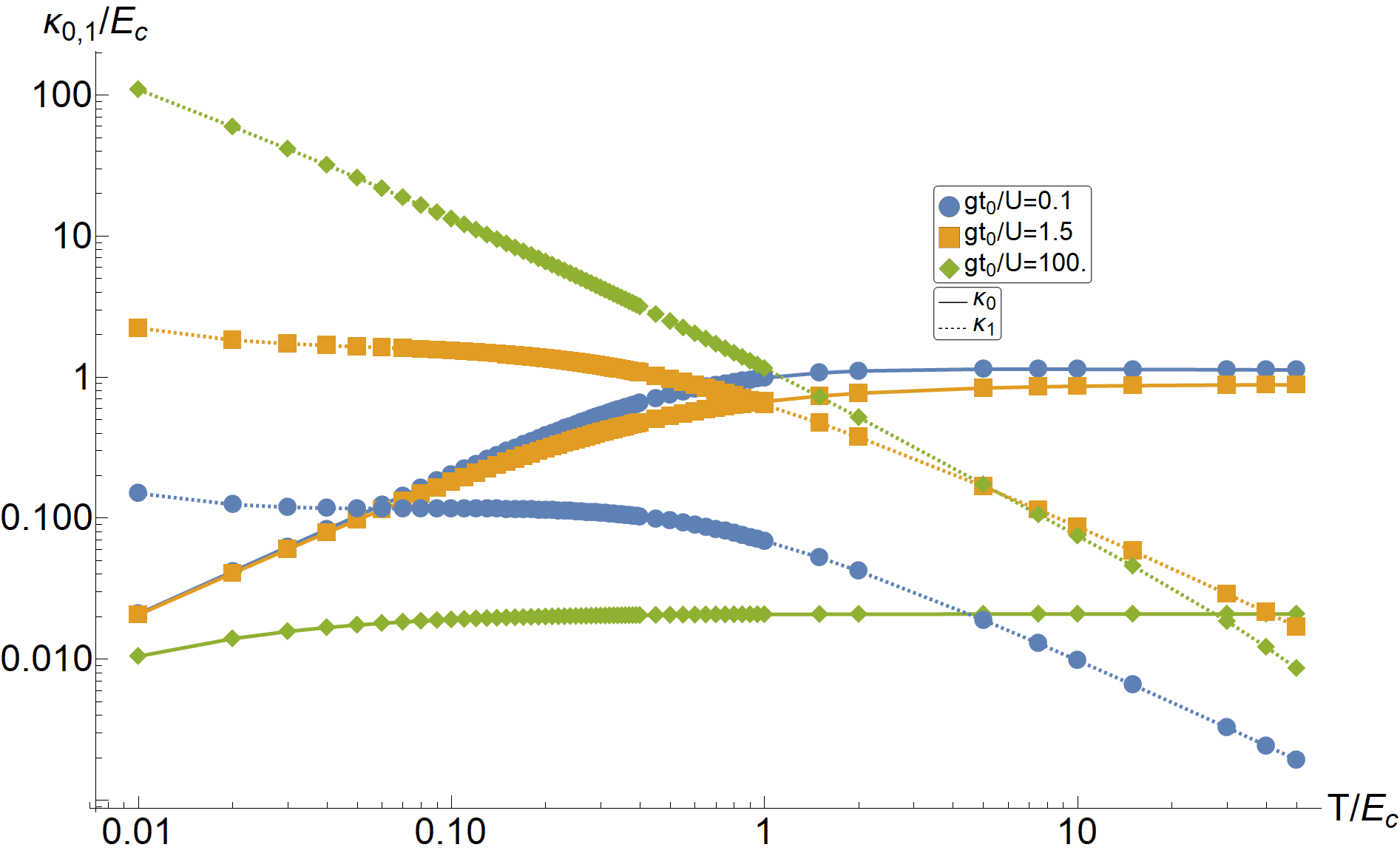}
\caption{\label{fig:kappaplot4} A comparison between the electron thermal conductivity $\kappa_0$ and the phonon-drag thermal conductivity $\kappa_1$. At low temperatures $\kappa_1$ dominates, while at high temperature $\kappa_0$ dominates. $U/t_0=200$.}
\end{figure}

    Now we present the numerical results for our model. In Fig.~\ref{fig:kappaplot4}, we compare the phonon-drag thermal conductivity $\kappa_1$ to the leading order value $\kappa_0$ at several $g$'s. We see that at low temperatures $\kappa_1\gg\kappa_0$, while at higher temperatures $\kappa_0\gg\kappa_1$.

\begin{figure}[htb]
\centering
\includegraphics[width=0.95\columnwidth]{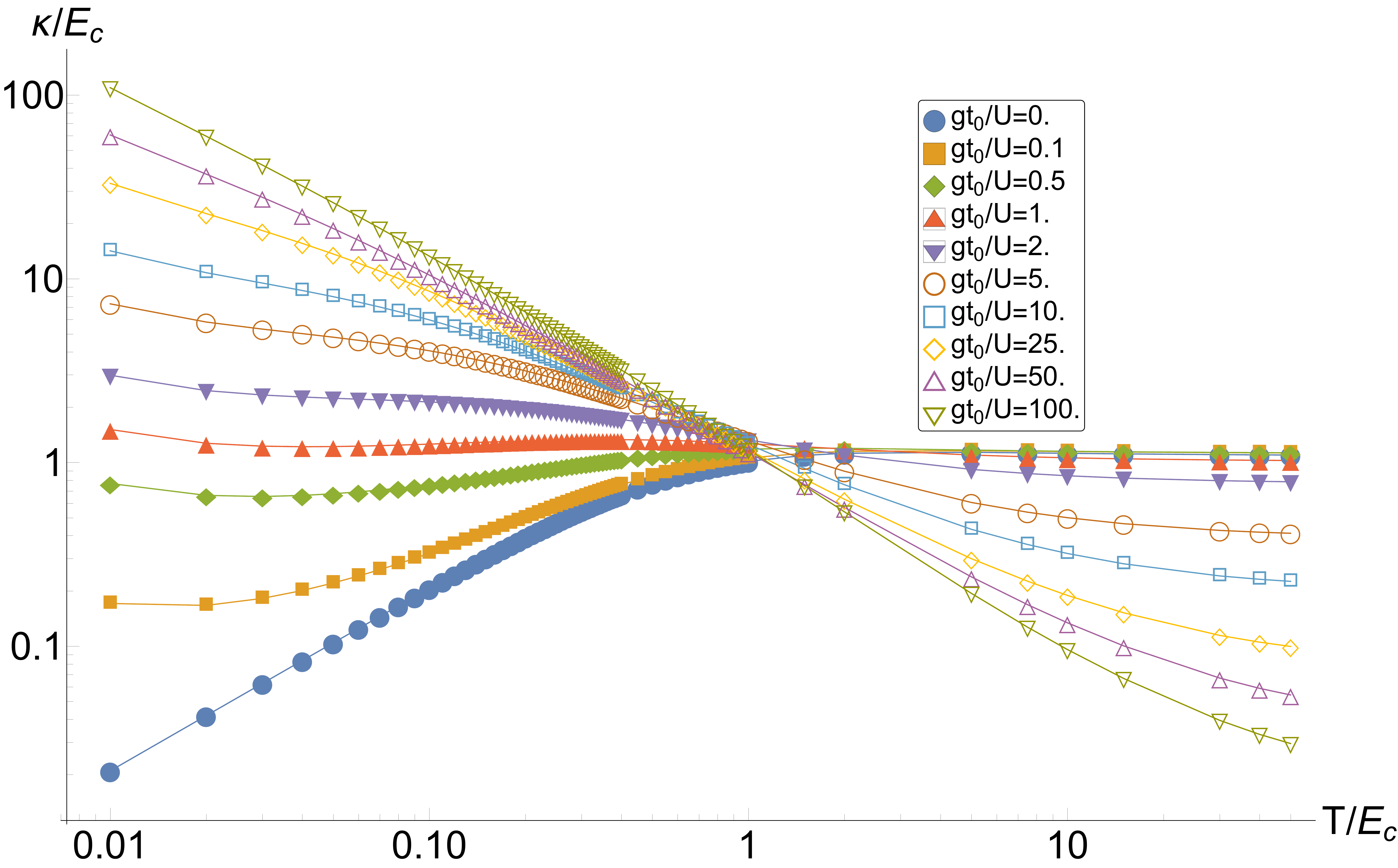}
\caption{\label{fig:kappaplot3} The total thermal conductivity $\kappa=\kappa_0+\kappa_1$ plotted against temperature at different $g$'s in log-log scale. $U=200t_0$.}
\end{figure}

    In Fig.~\ref{fig:kappaplot3}, we plot the total thermal conductivity $\kappa=\kappa_0+\kappa_1$. At low temperature, there is a hierarchy that $\kappa$ is positively correlated to $g$, further inspections show that $\kappa$ is roughly linear in $g$ for large $gt_0/U$.

    At very low temperatures $(gt_0/U)(T/E_c)\ll1$, this linear-in-$g$ behavior might be partially understood using Eq.\eqref{eq:kappa1_ex},which says $\kappa_1\sim g t_0$. However, it is peculiar that this hierarchy can persist to higher temperature ranges where $(gt_0/U)(T/E_c)>1$.


    As temperature rises, the phonon drag contributions die off, and the hierarchy inverts at around $T/E_c\sim 1$. The high temperature behavior is mostly dominated by the electron contribution.

\begin{figure}[htb]
\centering
\includegraphics[width=0.95\columnwidth]{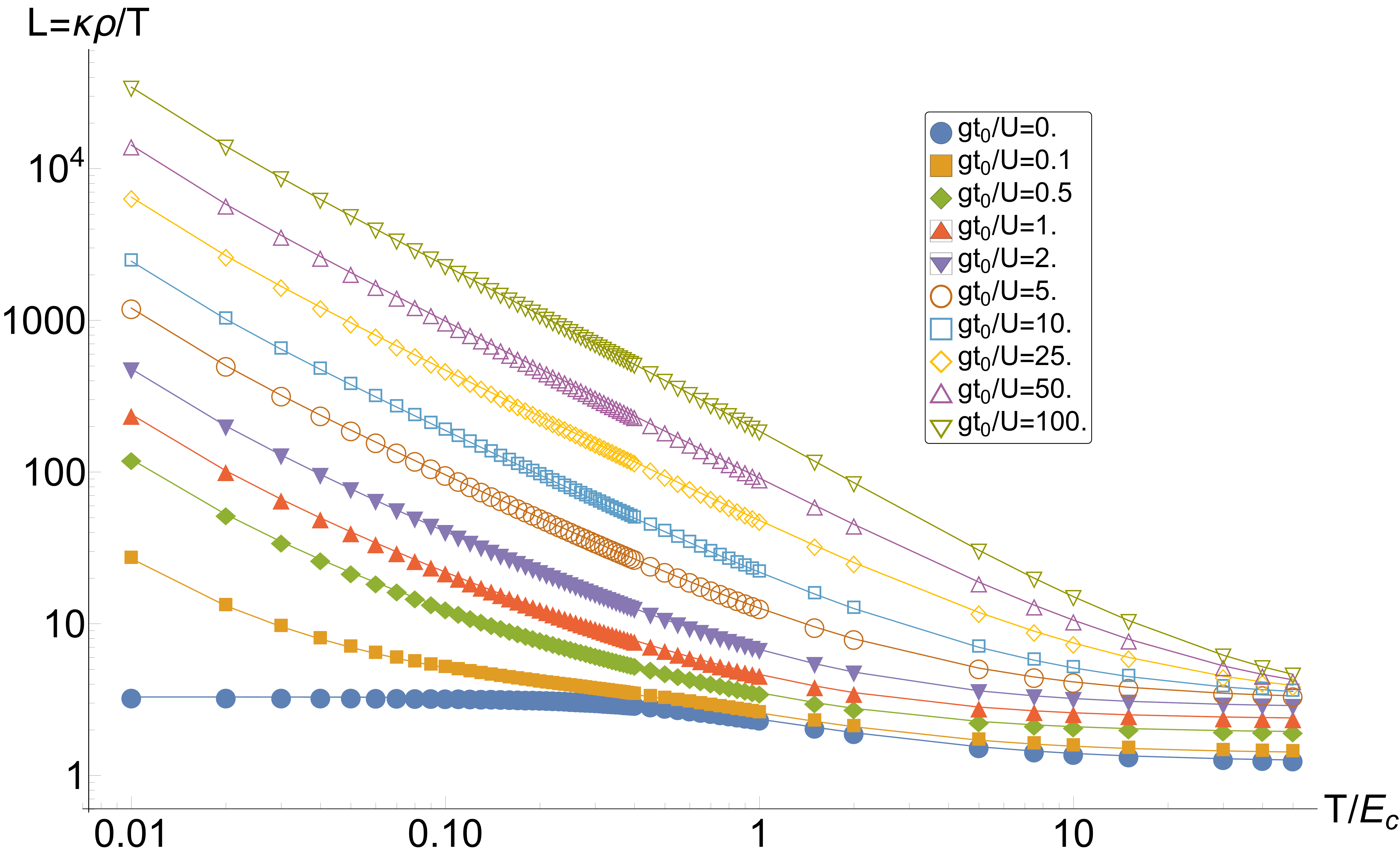}
\caption{\label{fig:Lplot} The Lorentz ratio $L=\kappa\rho/T$, plotted versus temperature for different $g$ at $U=200t_0$ in log-log scale. As a reference, the $g=0$ curve has $L=\pi^2/3$ at low temperatures and $L=\pi^2/8$ at high temperatures.}
\end{figure}

    In Fig.~\ref{fig:Lplot}, we inspect the violation of Wiedermann-Franz law by plotting the Lorentz ratio $L=\kappa\rho/T$. For $g=0$, we get a crossover from $L=\pi^2/3$ at low temperatures to $L=\pi^2/8$ at high temperatures, which agrees with Ref.~\onlinecite{Song}. For non-zero $g$, we see a huge enhancement of the Lorentz ratio and the Wiedermann-Franz law is strongly violated.

\section{Thermodynamics of the Phonon Model}

    In this section, we compute the entropy and the heat capacity using the imaginary time formalism. At zeroth order, the entropy is dominated by $O(N^2)$ species of nearly free phonons, but on top of it there is the $O(N)$ piece contributed by electrons. We will calculate this electron contribution.
    The grand potential (action) of the system consists of two parts. The first part is the potential of $N(N+1)/2$ species of free phonons. The second part is the potential of electrons, which is given by

\begin{equation}\label{eq:grandpotential}
\begin{split}
   & \frac{\mathcal{G}}{TNN_\text{site}}=-\mathrm{Tr}\ln(\partial_\tau+\Sigma)\\
                            &+\int\rd^2\tau\bigg[-\frac{U^2}{4}G(\tau_1,\tau_2)^2G(\tau_2,\tau_1)^2\\
                            &-\Sigma(\tau_1,\tau_2)G(\tau_2,\tau_1)+\frac{t_0^2+gt_0T}{2}G(\tau_1,\tau_2)G(\tau_2,\tau_1)\bigg],
\end{split}
\end{equation}
where $N_\text{site}$ is the number of lattice sites.
     Variation of Eq.\eqref{eq:grandpotential} yields the saddle point equation of motion:
\begin{equation}
    \begin{aligned}
    G^{-1}(i\omega_n)=&i\omega_n-\Sigma(i\omega_n),\\
    \Sigma(\tau)=&(t_0^2+gt_0T)G(\tau)-U^2G(\tau)^2G(-\tau).    
    \end{aligned}
\end{equation}
    The above equation can be solved numerically by iteration with fast Fourier transform, see Appendix \ref{sec:numerics}.

    In principle, we can then plug the saddle point solution back to Eq.\eqref{eq:grandpotential} and compute the grand potential. Note that we are working at zero chemical potential, so the grand potential coincides with free energy $\mathcal{F}$, and we might compute the entropy using $S=-\partial\mathcal{F}/\partial T=-\partial\mathcal{G}/\partial T$. However, if we want to compute the heat capacity $C=T\partial S/\partial T$, we would have to do a second numerical differentiation, which has large error. The solution is to derive an analytic expression for the entropy, so that we can directly evaluate the entropy and differentiate only once to get the heat capacity.

    The grand potential $\Omega$ is a function of $(T,U,t_0,g)$, and has the dimension of [energy]. It follows from dimensional analysis that
\begin{equation}
    (U\frac{\partial}{\partial U}+t_0\frac{\partial}{\partial t_0}+T\frac{\partial}{\partial T}|_\text{exp.}+T\frac{\partial}{\partial T}|_\text{imp.})\mathcal{G}=\mathcal{G}.
\end{equation}

    Here we have separated the $T$ derivative into the explicit part and the implicit part. The explicit part only differentiates the temperature dependent coupling $gt_0T$, and the other differentiations are called implicit. To satisfy the third law of thermodynamics, the entropy should be defined as
\begin{equation}
    S\equiv-\frac{\partial}{\partial T}|_\text{imp.}\mathcal{G}=\frac{1}{T}(1-U\frac{\partial}{\partial U}-t_0\frac{\partial}{\partial t_0}-T\frac{\partial}{\partial T}|_\text{exp.})\mathcal{G}.
\end{equation}
    We evaluate the above derivative at the saddle point solution, the only contribution is through the explicit dependence of $\mathcal{G}$ on $(U^2,t_0^2,gt_0T)$, and we obtain
\begin{equation}\label{eq:entropy}
\begin{split}
    S=&\ln 2-\sum_n\ln\frac{G(i\omega_n)}{G_0(i\omega_n)}+\frac{5}{4}\sum_n G(i\omega_n)\Sigma(i\omega_n)\\
+&\frac{t_0^2+gt_0T}{4}\sum_nG(i\omega_n)^2,
\end{split}
\end{equation}where we have used the saddle point solution, and regularized the sum using free fermion propagator $G_0(i\omega_n)=1/i\omega_n$.

\begin{figure}
\centering
  \includegraphics[width=0.95\columnwidth]{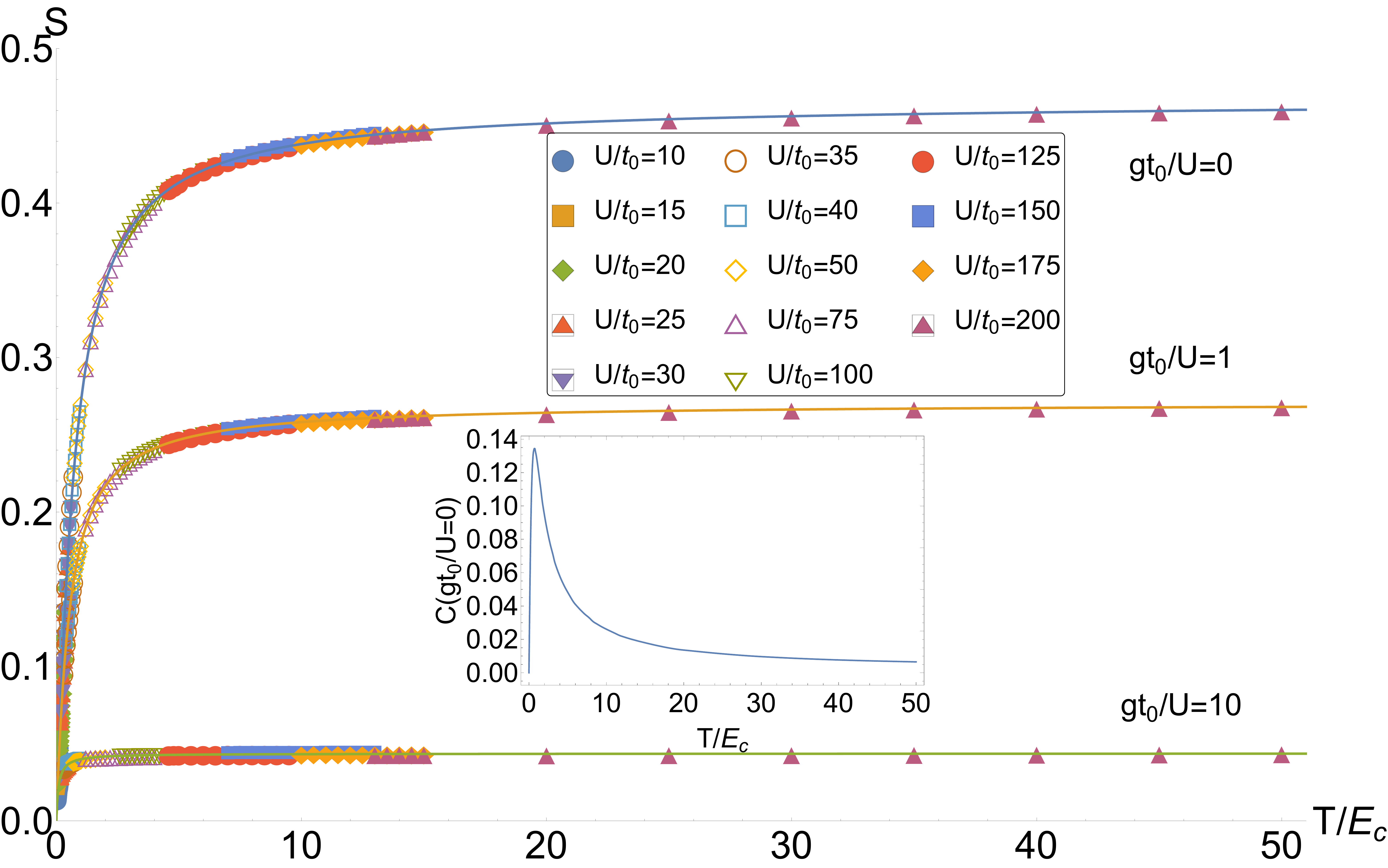}
  \caption{The entropy $S$ plotted against temperature $T$ for various $g$. The inset is the heat capacity at $g=0$. The behavior of the entropy follows a universal function $\mathcal{S}\left(\frac{T/E_c}{1+(gt_0/U)(T/E_c)}\right)$ represented by curves. \label{fig:entropy}}
\end{figure}

   The numerical result for entropy and heat capacity is shown in Fig.~\ref{fig:entropy}. We found that the entropy follows a universal function $S=\mathcal{S}\left(\frac{T/E_c}{1+(gt_0/U)(T/E_c)}\right)$. The universal function $\mathcal{S}$ is the same as the one in Ref.~\onlinecite{Song}, which grows linearly at low temperatures and saturates at high temperatures. With the phonons introduced, the entropy is reduced.
   
   A possible reason for entropy reduction could be the following. In the original SYK model, the non-zero entropy comes from the exponentially many low lying states of electrons. However, in our model, phonons are well-defined quasiparticles so in the low energy sector there are polynomially many phonon states. Generically, coupling between electrons and phonons makes the low lying states sparser and thus reduces the entropy.

\section{Scrambling in the Phonon Model}
\label{sec:phononscrambling}

    In this section, we discuss the scrambling properties of the phonon model.
    
\subsection{Electron Scrambling}

\begin{figure}[htb!]
\centering
\begin{tikzpicture}[scale=1,baseline={(v0.base)}]
\begin{feynman}[scale=1,large,transform shape]
    \vertex (v1);
    \vertex [right=0.7 of v1] (v2);
    \vertex [right=0.6 of v2] (v3);
    \vertex [right=0.7 of v3] (v4);
    \vertex [below=1 of v1] (v5);
    \vertex [right=0.7 of v5] (v6);
    \vertex [right=0.6 of v6] (v7);
    \vertex [right=0.7 of v7] (v8);
    \vertex [below=0.6 of v1] (v0);
    \vertex [right=0.3 of v2] (vtm);
    \vertex [below= 1 of vtm] (vbm);
    \diagram
    {
        (v1)--[fermion](v2)--(v3)--(v4),
        (v8)--(v7)--(v6)--[fermion](v5),
        (vtm)--[photon](vbm),
    };
    \draw (v3) rectangle (v8);
    \node at ($(v3)!.5!(v8) $) {$f_1$};
    \draw (v1) rectangle (v6);
    \node at ($(v1)!.5!(v6) $) {$L$};
\end{feynman}
\end{tikzpicture} 
\caption{A new diagram for computation of OTOC. \label{fig:phonon_vert}}
\end{figure}
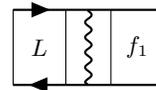

    First, we consider scrambling of electrons which can be computed from the electron retarded OTOC \eqref{eq:OTOC}. Much of the discussion of the $t$-$U$ model carries over to the phonon model. The only addition is to include a diagram with a vertical phonon line, such as Fig.~\ref{fig:phonon_vert}. In the zero Debye frequency limit, this new diagram is simply multiplying a constant $gt_0T$.
    As a result, the fast growing part of the OTOC satisfies the following integral equation:

\begin{equation}\label{eq:f=Kf_ph}
\begin{split}
    &F(q,\Omega,\omega)=L(q,\Omega,\omega)\Bigg[1+ g t_0 T F(q,\Omega,\omega)\\
   & 
    +\frac{3J^2}{2}\int\frac{\rd \tilde{\Omega}}{2\pi}K(\Omega-\tilde{\Omega})F(q,\tilde{\Omega},\omega)\Bigg].
\end{split}
\end{equation}

\begin{figure}
  \centering
  \includegraphics[width=0.95\columnwidth]{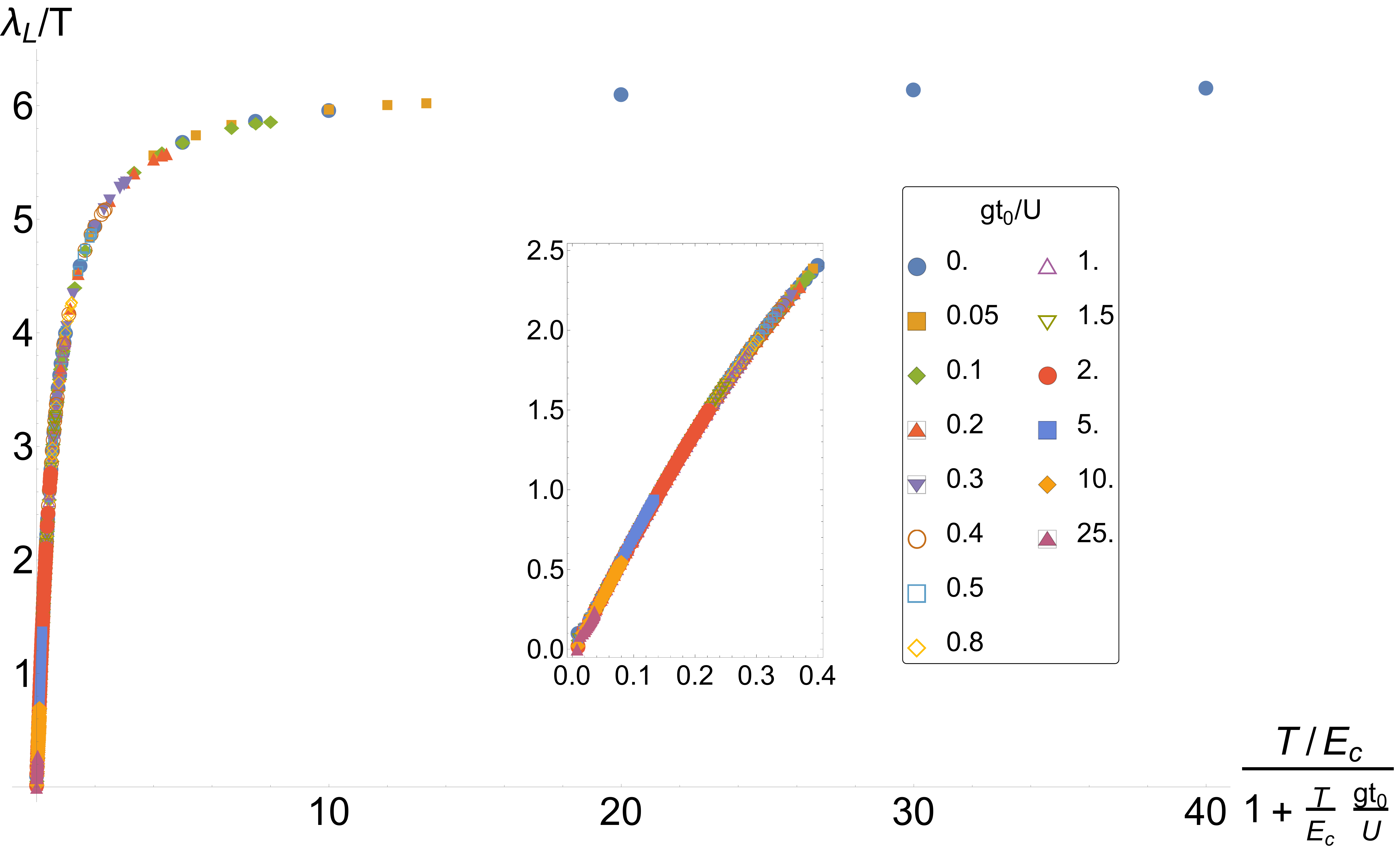}
  \caption{The scrambling rate $\lambda_L$ plotted versus rescaled temperature for different $g$ at $U=200t_0$. The inset zooms into the low temperature sector. After rescaling, data points of different $g$ collapse onto a universal curve.\label{fig:scrambling1}}
\end{figure}

\begin{figure}
    \centering
    \includegraphics[width=0.95\columnwidth]{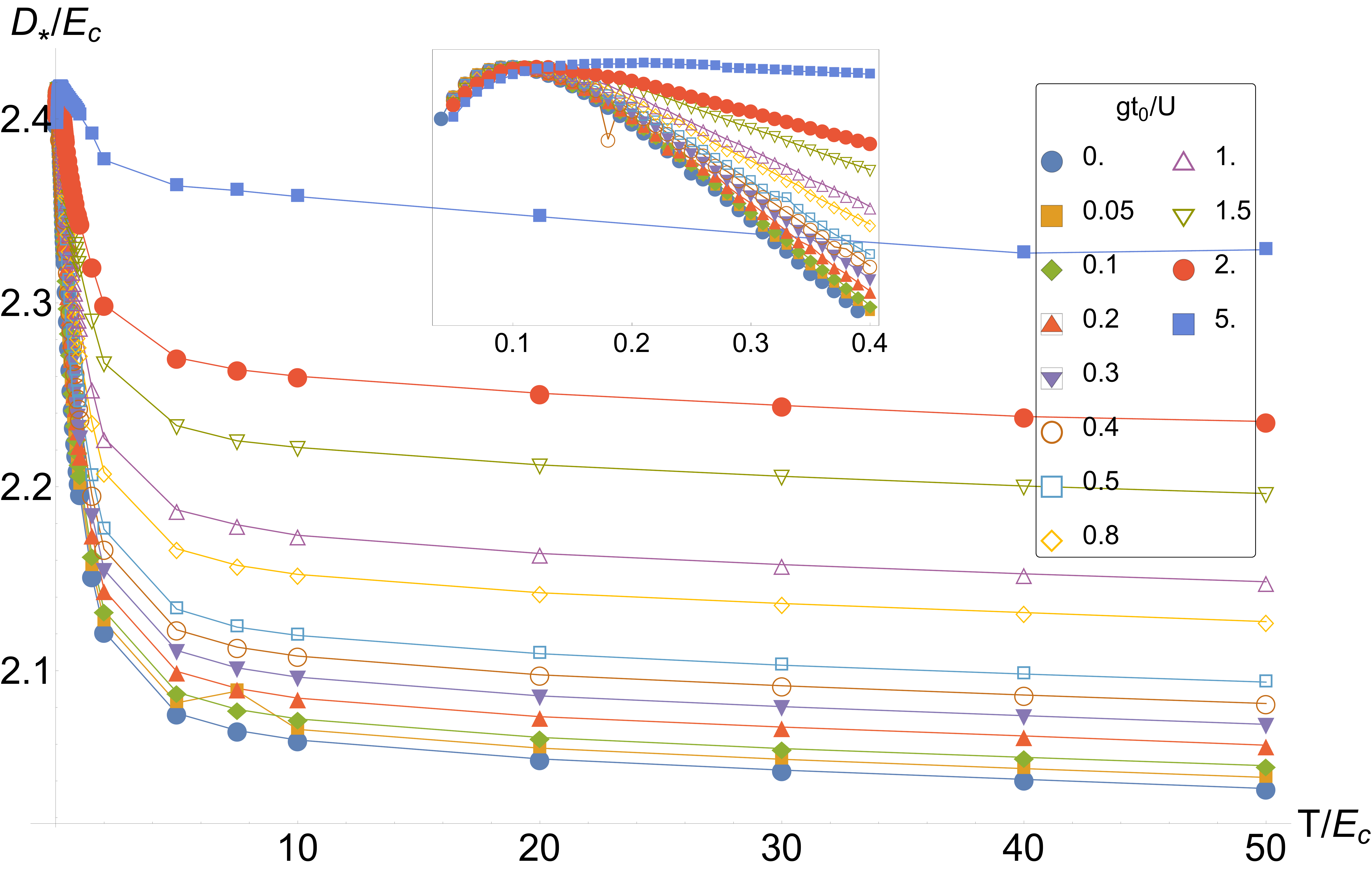}
    \caption{The short-distance scrambling diffusion coefficient $D_*$ plotted versus temperature for different $g$ at $U=200t_0$. The inset zooms into the low temperature sector.\label{fig:scrambling2}}
\end{figure}

\begin{figure}
    \centering
    \includegraphics[width=0.95\columnwidth]{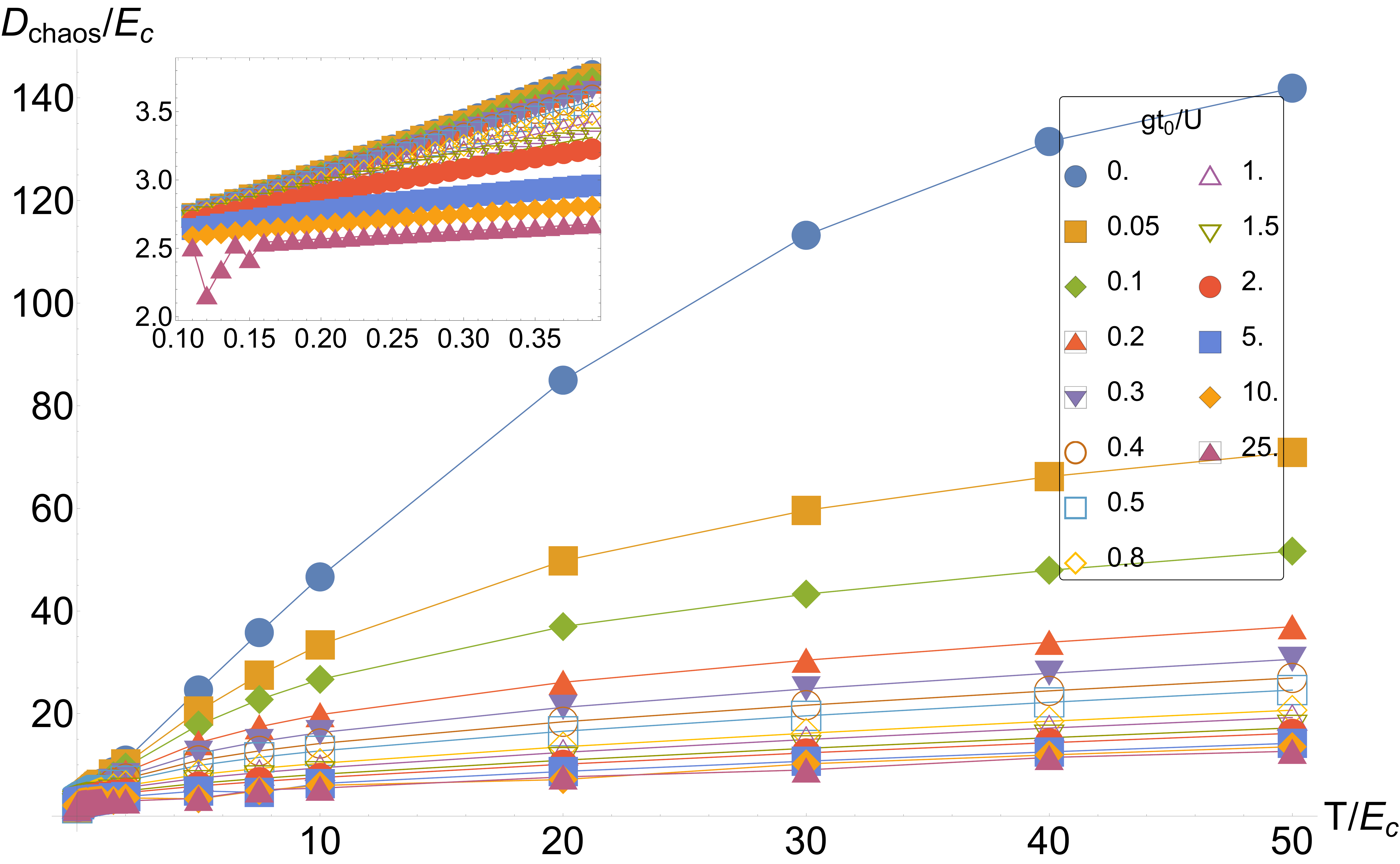}
    \caption{The long-distance scrambling diffusion coefficient $D_{\rm chaos}$ plotted versus temperature for different $g$ at $U=200t_0$. The inset zooms into the low temperature sector.\label{fig:scrambling3}}
\end{figure}

\begin{figure}
\centering
  \includegraphics[width=0.95\columnwidth]{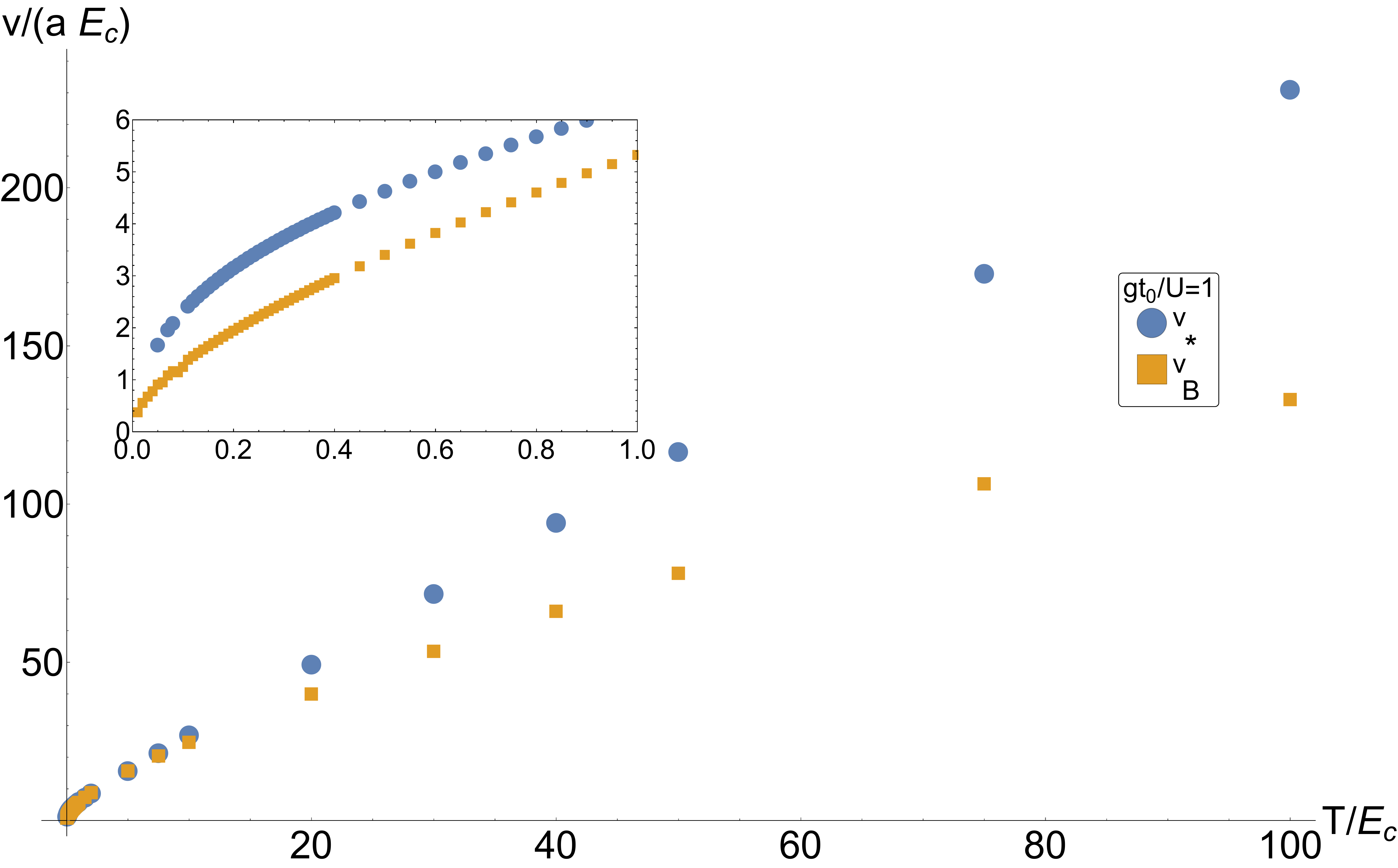}
  \caption{The two characteristic velocities $v_*$ and $v_B$ plotted against temperature, in the unit of $a E_c$ where $a=1$ is lattice spacing. Here $gt_0/U=1$ and $U/t_0=200$. At all temperatures $v_*>v_B$. \label{fig:vplot_g=1}}
\end{figure}

    Following previous procedures, we can compute the scrambling rate $\lambda_L$, short distance diffusion coefficient $D_*$ and long distance diffusion coefficient $D_{\rm chaos}$. The results are shown in Fig.~\ref{fig:scrambling1},\ref{fig:scrambling2},\ref{fig:scrambling3},\ref{fig:vplot_g=1}.
    
    For the scrambling rate $\lambda_L$ (see Fig.~\ref{fig:scrambling1}), we found that $\lambda_L/T$ is a universal function of the combination $(T/E_c)/(1+(T/E_c)(gt_0/U))$; so the value of $\lambda_L$ can be deduced from Fig.~\ref{fig:scrambling0}. This result can be understood from two aspects. First, in the equation of motion \eqref{eq:EoM_Keldysh_ph}, $g$ only appears in the combination $t_0^2+g t_0 T$. Second, $\lambda_L$ is obtained by solving the Bethe-Salpeter equation at zero momentum, where it can be shown that $g$ appears as $t_0^2+g t_0 T$ by re-expanding $L(q,\Omega)$. $\lambda_L$ decreases with increasing $g$ because phonons are still well-defined quasiparticles and thus coupling to phonons slows down scrambling.
    
    As for the scrambling diffusion coefficients (see Fig.~\ref{fig:scrambling2} for $D_*$ and Fig.~\ref{fig:scrambling3} for $D_{\rm chaos}$), we did not find a universal function as the case of $\lambda_L$. This is because the phonon coupling term differs from electron hopping term at non-zero momenta. 
    
    It is interesting to see that the two diffusion coefficients respond to phonons in opposite ways. Phonons help with the short distance diffusion of scrambling but suppress scrambling propagation at long distances.
    
    Because coupling to phonon reduces $\lambda_L$, we expect that $v_*$ is larger compared to the $t$-$U$ model, and hence wave-front OTOC propagation (region B in Fig.~\ref{fig:chaos_domains1}) is diminished at small $g$ and completely suppressed at large $g$. In Fig.~\ref{fig:vplot_g=1}, we show a comparison between $v_*$ and $v_B$ at $gt_0/U=1$ as an example. We see that at this value of $g$, $v_*>v_B$ for all temperatures and thus the OTOC always propagates in a diffusive manner.
     
    A sensible comparison of chaos diffusion constants to energy diffusion constant is not available, because the quasi-free phonons dominate the heat capacity and $D_E\sim O(1/N)$. 
    
\subsection{Phonon Scrambling}

    In this section we consider the scrambling of phonon themselves. Naively one we would expect that phonon scrambles at the same rate as electrons, but we will show that this fast growing term in the phonon OTOC is actually small in $1/N$ power counting. As a result, the phonon only scrambles after electrons are scrambled.

    The phonon retarded OTOC is defined as

\begin{equation}
\begin{split}
   & h(x;t_1,t_3;t_2,t_4)=\frac{\theta(t_{24})\theta(t_{13})}{N^4}\sum_{abcd}\\
   & \Tr \left(y^2[X_{abx}(t_1),X_{cd0}^\dagger(t_3)]^\dagger y^2[X_{abx}(t_2),X_{cd0}^\dagger(t_4)]\right),
\end{split}
\end{equation}
 with Fourier transform
\begin{equation}
\begin{split}
    h(q,\Omega,\Omega',\omega)=&\int\frac{\rd^d q \rd \Omega \rd \omega}{(2\pi)^{d+2}}h(x;t_1,t_3;t_2;t_4)\\
\times&e^{-i q\cdot x+i\Omega t_{31}+i\Omega't_{24}+i\omega(t_{21}+t_{43})/2}.
\end{split}
\end{equation}

    The phonon OTOC can be expressed in terms of the electron OTOC in the following way
\begin{equation}
\begin{aligned}
\begin{tikzpicture}[scale=1,baseline={(v0.base)}]
\begin{feynman}[scale=1,large,transform shape]
    \vertex (v1);
    \vertex [right=0.7 of v1] (v2);
    \vertex [right=0.6 of v2] (v3);
    \vertex [right=0.7 of v3] (v4);
    \vertex [below=1 of v1] (v5);
    \vertex [right=0.7 of v5] (v6);
    \vertex [right=0.6 of v6] (v7);
    \vertex [right=0.7 of v7] (v8);
    \vertex [below=0.6 of v1] (v0);
    \vertex [right=0.3 of v2] (vtm);
    \vertex [below= 1 of vtm] (vbm);
    \diagram
    {
        (v1)--[photon](v2)--(v3)--[photon](v4),
        (v8)--[photon](v7)--(v6)--[photon](v5),
    };
    \draw (v2) rectangle (v7);
    \node at ($(v2)!.5!(v7) $) {$h$};
\end{feynman}
\end{tikzpicture}
&=
\begin{tikzpicture}[scale=1,baseline={(v0.base)}]
\begin{feynman}[scale=1,large,transform shape]
    \vertex (v1);
    \vertex [right=0.7 of v1] (v2);
    \vertex [right=0.6 of v2] (v3);
    \vertex [right=0.7 of v3] (v4);
    \vertex [below=1 of v1] (v5);
    \vertex [right=0.7 of v5] (v6);
    \vertex [right=0.6 of v6] (v7);
    \vertex [right=0.7 of v7] (v8);
    \vertex [below=0.6 of v1] (v0);
    \vertex [right=0.3 of v2] (vtm);
    \vertex [below= 1 of vtm] (vbm);
    \diagram
    {
        (v1)--[photon](v2)--[photon](v3)--[photon](v4),
        (v8)--[photon](v7)--[photon](v6)--[photon](v5),
    };
\end{feynman}
\end{tikzpicture} \\
&+
\begin{tikzpicture}[scale=1,baseline={(v0.base)}]
\begin{feynman}[scale=1,large,transform shape]
    \vertex (v1);
    \vertex [right=0.7 of v1] (v2);
    \vertex [right=0.6 of v2] (v3);
    \vertex [right=0.7 of v3] (v4);
    \vertex [below=1 of v1] (v5);
    \vertex [right=0.7 of v5] (v6);
    \vertex [right=0.6 of v6] (v7);
    \vertex [right=0.7 of v7] (v8);
    \vertex [below=0.6 of v1] (v0);
    \vertex [right=0.3 of v2] (vtm);
    \vertex [below= 1 of vtm] (vbm);
    \vertex [right=0.4 of v1] (u1);
    \vertex [below=1 of u1] (u2);
    \vertex [left=0.4 of v4] (u3);
    \vertex [below=1 of u3] (u4);
    \diagram
    {
        (v1)--[photon](u1)--(v2)--(v3)--(u3)--[photon](v4),
        (v8)--[photon](u4)--(v7)--(v6)--(u2)--[photon](v5),
        (u1)--[anti fermion](u2),
        (u4)--[anti fermion](u3),

    };
    \draw (v2) rectangle (v7);
    \node at ($(v2)!.5!(v7) $) {$f_1$};
\end{feynman}
\end{tikzpicture}
+\begin{tikzpicture}[scale=1,baseline={(v0.base)}]
\begin{feynman}[scale=1,large,transform shape]
    \vertex (v1);
    \vertex [right=0.7 of v1] (v2);
    \vertex [right=0.6 of v2] (v3);
    \vertex [right=0.7 of v3] (v4);
    \vertex [below=1 of v1] (v5);
    \vertex [right=0.7 of v5] (v6);
    \vertex [right=0.6 of v6] (v7);
    \vertex [right=0.7 of v7] (v8);
    \vertex [below=0.6 of v1] (v0);
    \vertex [right=0.3 of v2] (vtm);
    \vertex [below= 1 of vtm] (vbm);
    \vertex [right=0.4 of v1] (u1);
    \vertex [below=1 of u1] (u2);
    \vertex [left=0.4 of v4] (u3);
    \vertex [below=1 of u3] (u4);
    \diagram
    {
        (v1)--[photon](u1)--(v2)--(v3)--(u3)--[photon](v4),
        (v8)--[photon](u4)--(v7)--(v6)--(u2)--[photon](v5),
        (u1)--[fermion](u2),
        (u3)--[fermion](u4),

    };
    \draw (v2) rectangle (v7);
    \node at ($(v2)!.5!(v7) $) {$f_2$};
\end{feynman}
\end{tikzpicture}
\end{aligned} \,
,
\end{equation}
    so the divergent part of $h$ is
\begin{equation}\label{eq:h_growth}
\begin{split}
    h(\Omega,\Omega',\omega)\sim&\frac{\alpha^4}{N^2}\int \frac{\rd\Omega_1\rd\Omega_2 }{(2\pi)^2} D_R(\Omega+\omega/2)D_A(\Omega-\omega/2)\\
\times&G_W(\Omega_1-\Omega)F(\Omega_1,\Omega_2,\omega)\\
    \times& G_R(\Omega_2+\omega/2)G_A(\Omega_2-\omega/2)G_W(\Omega_2-\Omega')\\
\times&D_R(\Omega'+\omega/2)D_A(\Omega'-\omega/2),
\end{split}
\end{equation}where $F=f_1+f_2$. The equation above implies that the poles of $h$ coincides with $F$.

    We have found a pole of $F$ at $\omega=i\lambda_L$, which implies a exponential growth with time $F(t)\sim\frac{1}{N}e^{\lambda_L t}$ and $h(t)\sim\frac{1}{N^3}e^{\lambda_L t}$. We compare this term to the zeroth order result of phonon OTOC $h^{(0)}=\frac{1}{N^2}D_RD_A\sim1/N$, in which the $O(N)$ enhancement comes from Eq.\eqref{eq:DRDA}, and we see that although the $h(t)$ has a fast exponential growth term, it is suppressed by a factor of $1/N^2$ relative to the non-growing part, so it is important only after time $t_*\sim\frac{2\ln(N)}{\lambda_L}$. Consequently, the phonon OTOC does not display exponential growth until $t\sim t_*$. Note that $t_*$ is larger than the electron-scrambling time $t_\text{scr}$, implying that phonons start scrambling after electrons are fully scrambled.

    In an electron-phonon system with only electron-phonon interaction, it is found that both the electrons and the phonons have a scrambling rate of about the phonon decay rate $({1}/{N})({\omega_0^2}/{T})$ \cite{Werman}. It is possible that a similar effect in our model can cause $F$ to develop pole at a small imaginary frequency $\omega=i\lambda',\lambda'\sim O(1/N)$. However, because $t_*\propto\ln(N)$ and $\lambda't_*\propto\ln(N)/N$, this extra pole has no impact on the behavior of $h(t)$ in the large $N$ limit.

\section{Conclusions}
\label{sec:conc}

This work has described the transport and chaos properties of a model electronic system with strong electron-electron and electron-phonon interactions. We did not include phonon-phonon interactions, and the feedback of the electrons on the phonon dynamics was weak: so the phonons act essentially as a heat bath of oscillators at a typical frequency $\omega_0$. All our results here are for $T \gg \omega_0$, as the phonons have little influence at lower $T$.

The electron-electron interactions were described by SYK islands with interactions of strength $U$, and hopping between islands of strength $t_0$ (see Fig.~\ref{fig:SYKlattice}). The electron-phonon interactions were characterized by a dimensionless coupling $g$. The properties of the different regimes of chaos and transport are summarized in Fig.~\ref{fig:crossovers}, which are controlled by the dimensionless ratio $g t_0/U$.

For $g t_0 \ll U$, the phonons have a minor influence, and the electron-electron interaction dominated transport is similar to that described by Song {\it et al.} \cite{Song}.
We computed here the chaos properties across the crossover from the heavy Fermi liquid to the incoherent metal at $T \sim t_0^2/U$. The chaos propagation is controlled by two velocities, $v_\ast$ and $v_B$, as shown in Fig.~\ref{fig:chaos_domains1}. 
\begin{itemize} 
\item
In the heavy Fermi liquid regime, $v_\ast > v_B$, and the chaos propagation is diffusive, as in Eq.~(\ref{eq:OTOC_diffusive}). The Lyapunov rate $\lambda_L \sim T^2/E_c$ is much smaller that the chaos bound of $2 \pi T$. The chaos diffusion is characterized by $D_\ast$, and we found $D_\ast \approx D_E$, the energy diffusion constant.
\item 
In the incoherent metal regime $v_\ast < v_B$, and now there is a wave-front of chaos propagation, as in Eq.~(\ref{eq:OTOCfront}). The Lyapunov rate $\lambda_L$ is close to the chaos bound of $2 \pi T$. The chaos diffusion is characterized by $D_{\rm chaos}$, and we again found that $D_{\rm chaos} \approx D_E$, the energy diffusion constant.
\end{itemize}

With increasing electron-phonon coupling, $g$, the slope of the linear-in-$T$ resistivity in the incoherent metal regime changes. This change is described by the scaling plot in Fig.~\ref{fig:rhoplot2}. The corresponding plot for the electron thermal conductivity is in Fig.~\ref{fig:kappaplot2}. For the Lyapunov rate, $\lambda_L (q=0)$, the entire effect of the electron-phonon coupling is to replace $T/E_c$ by $(T/E_c)/(1 + (T/E_c)(g t_0/U)$ where $E_c = t_0^2/U$; we can then read off the resulting rate from Fig.~\ref{fig:scrambling0}.

For $g t_0 \gg U$, the scattering of electrons off a heat bath of phonons dominates the transport. The density of thermally excited phonons is proportional to $T$, and so this leads to a linear-in-$T$ resitivity. Nevertheless, the properties of this regime are quite different from the incoherent metal described above for small $g t_0/U$. The electron-phonon scattering is essentially elastic, and so does not contribute significantly to chaos. Consequently, we find that that the Lyapunov exponent is much smaller than the chaos bound, and controlled by the weaker electron-electron interactions: $\lambda_L \sim (U/(g t_0)) T$ in the high $T$ regime. We also have $v_\ast > v_B$, and the chaos propagation is diffusive as in Eq.~(\ref{eq:OTOC_diffusive}). The nearly elastic electron-phonon scattering also implies that while the DC conductivity is dominated by electron-phonon scatttering, the optical conductivity is not. As shown in Fig.~\ref{fig:optical_conductivity}, there is a $1/\omega$ behavior in the optical conductivity, which is characteristic of the local incoherent dynamics of the pure SYK model.

The limiting results above for small and large $g t_0/U$ illustrate one of our main results: there is a fundamental distinction between the linear-in-$T$ resistivity between the electron-electron and electron-phonon dominated regimes. Experimentally, this distinction can be detected by comparing the DC and optical conductivities. Strong electron-phonon scattering increases the DC resistivity, but has little effect on the $1/\omega$ optical conductivity; in contrast, the critical electron-electron interactions describe by SYK physics connect the DC and optical responses via $\omega/T$ scaling. If we increase the electron-phonon coupling $g$, the electron scattering rate also increases without an apparent bound, as does the DC resistivity (modulo the saturation effects discussed in Section~\ref{sec:saturation}). But this increased electron scattering rate does not show up in OTOC: the Lyapunov rate is far from maximal at large $g$, with $\lambda_L \sim T (U/(g t_0))$
actually decreasing with increasing $g$. On the other hand, in the SYK physics, the same rate $\sim T$ shows up in the resistivity and the OTOC. Alternatively, the distinction can also be diagnosed via the resistivity saturation effect. If the linear-in-$T$ resistivity is due to phonons, it will saturate to the MIR limit \cite{Werman,Werman2}, as in a generic model there are phonons sitting on both sites and bonds. However, linear-in-$T$ resistivity originated from the SYK interaction can easily surpass the MIR limit. In future work, it would be interesting to treat the electron-phonon and phonon-phonon interactions in a more self-consistent manner; then, we can expect an ``electron-phonon" soup \cite{JC18} in which the strong dependence on the electron-phonon coupling disappears, and both transport and chaos are determined by a common rate $\sim T$.

Finally, it is useful to compare our results with recent studies of operator spreading using random unitary circuits \cite{Nahum2018,Pollman2018,Khemani2018,XuSwingle2018,XuSwingle2018a}, which report a broadening of the chaos wavefront. In our model, the phonons remain essentially free oscillators, and the $N^2$ oscillators on each island can have consequences similar to a random unitary perturbation. And it is notable that we do find a diffusive broadening of the chaos wavefront with $v_\ast > v_B$, as the electron-phonon coupling is increased. Moreover, with strong electron-phonon coupling, the Lyapunov rate is much smaller than the maximal rate, as summarized in Fig.~\ref{fig:crossovers}. However, when the chaos is near maximal, at weak electron-phonon coupling, the sharp chaos wavefront is preserved.

\section*{Acknowledgements}

We thank  Erez Berg, Aharon Kapitulnik, Vedika Khemani, Aavishkar Patel, Xue-Yang Song, Brian Swingle, and Shenglong Xu for useful discussions. This research was supported by the US Department of Energy under Grant No. DE-SC0019030. Y.G. is also supported by
the Gordon and Betty Moore Foundation EPiQS Initiative through Grant (GBMF-4306). This work was performed in part at the Aspen Center for Physics, which is supported by National Science Foundation grant PHY-1607611.

\appendix

\section{\label{sec:numerics}Numerical Implementation}

\subsection{Solving for Green's Function}

    We solve the saddle point equations \eqref{eq:EoM_imaginary},\eqref{eq:EoM_Keldysh} by iteration. We mainly follow the prescription in Ref.~\onlinecite{Song}. We first discuss the real time equation of motion.

     The time domain is $[t=0\dots T_0(1-1/N_t),0]$, where zeros are padded at the tail to ensure fidelity of discretization. We denote the padding ratio by $r$, {\it i.e.} $N=r N_t$. The time interval $T_0=\gamma \beta$ is proportional to inverse temperature $\beta$.

     In frequency domain, we sampled the Green's function at $(2N_w+1)$ points in frequency space: $\Omega_0[-N_w,-N_w+1,...,N_w]$. Here the frequency unit is
     $\Omega_0=\frac{2\pi}{\gamma r\beta}$.
     In practice, we used $N_t=N_w=300000$, $r=4$, $\gamma=40$.

    When calculating $\Sigma_R$, there is a subtlety that $\Sigma_R(t=0)$ cannot be determined directly because it requires high frequency information. Uncertainty in $\Sigma_R(t=0)$ forbids brute force calculation of $\Im\Sigma_R(\omega)$. To circumvent the problem, we instead calculate $\Sigma_K$ first 
    \begin{equation}
    \begin{aligned}
    &\Sigma_K(t)=\frac{1}{2}U^2G_K(t)|G_R(t)|^2+\frac{1}{4}U^2G_K(-t)G_K(t)^2\\
    &\qquad\qquad +\frac{1}{4}U^2G_K(-t)G_R(t)^2 \quad \quad (t>0),\\
    &\Sigma_K(-t)=-\Sigma_K(t)^*\,,
    \end{aligned}
    \end{equation}
    and then obtain $\Im \Sigma_R$ by the FDT 
\begin{equation}
    \Sigma_K(\omega)= 2i\tanh\frac{\omega}{2T}\Im \Sigma_R(\omega)\,.
\end{equation}

     The algorithm goes like the following:
\begin{enumerate}
    \item Initialize $G_R(\omega)$ using the conformal limit value.
    \item Calculate $G_K(\omega)$ using FDT, then Fourier transform to time domain.
    \item Calculate $\Sigma_R(t)$ using EoM, then Fourier transform to frequency domain.
    \item Calculate the new Green's function $\tilde{G}_R$.
    \item Update $G_R$ by $G_R\to (1-\alpha)G_R+\alpha\tilde{G}_R$. In practice $\alpha=0.3$.
    \item Go back to step 2 and repeat until the iteration reaches the convergence criterion  $\max_\omega|G_R(\omega)-\tilde{G}_R(\omega)|<10^{-4}$.
\end{enumerate}

    For the imaginary time EoM, we use the same algorithm as above with the following changes: Instead of arbitrary discretization, we discretized in fermionic Matsubara frequencies. We could directly calculate $\Sigma$ without the subtlety of $\Sigma_R(t=0)$. To compute the entropy, we used $N_w=2^{20}$ and set convergence goal to $10^{-14}$. To facilitate convergence, we adjusted $\alpha$ during the iteration \cite{Maldacena}: We initially put $\alpha=0.5$, and we monitored the error $\max_\omega|G_R(\omega)-\tilde{G}_R(\omega)|$. If the error increases we reduce $\alpha$ by a half.

\subsection{Solving for OTOC}

      
      As discussed in the maintext, we extract $\lambda_L(q)$ by searching for singularities of $M=1-LK$. First, we need to calculate the matrix elements $L(\Omega,\omega)$ and $K(\Omega)$, which includes
\begin{enumerate}
  \item $G_W(\Omega)$: We can construct it from the spectral function we obtained earlier (see Appendix \ref{sec:wightman}). Note that we actually need self-convolution of $G_W(\Omega)$, which is calculated using fast Fourier transform.
  \item $G_R(\Omega+i\lambda/2)=G_A(\Omega-i\lambda/2)^*$: For this, we need to analytically continue it into the complex plane. There are two ways to do this: First, Fourier transform $G_A(t)e^{\lambda t}$, which is quick but not very accurate. The second way is to use the spectral function $G(z)=\int\frac{\rd x}{2\pi}\frac{A(x)}{z-x}$, which is accurate but fairly slow. Fortunately, we found that the error between the two methods for $G_A(z)$ only weakly depends on $\Im z$ and it can be accurately interpolated as a function of $\Re z$ and $\Im z$. Hence, we can use the spectral function method to calibrate the Fourier transform method, and then use the Fourier transform method for calculation.
\end{enumerate}

    After having the matrix elements, discretize $M_{\omega,q}$ in frequency space. The frequency unit $\Omega_0$ is the same as the previous section, and the highest frequency is $L\Omega_0$. We found $L=3000$ sufficient in practice.

    The following algorithm computes the scrambling rate $\lambda_L$ and short-distance scrambling diffusion coefficient $D_*$. For each $\lambda=-i\omega$ and $q$, we compute $H(\lambda,q)$ which is the absolute value of the smallest eigenvalue of $M_{i\lambda,q}$. Then we numerically minimize $H(\lambda,q)$ over $\lambda$ to obtain $\lambda_L$. 
    To extract the chaos diffusion coefficient, we calculate $\lambda_L(q)$ for different values of small $q$, and fit for $D_*$ using Eq.\eqref{eq:D_*}.

    To compute the long-distance scrambling diffusion coefficient $D_{\rm chaos}$, we instead fix $\lambda_L=2\pi T$, and minimizes $H(2\pi T, i|q_1|)$ over $|q_1|$ to obtain $|q_1|$ and $D_{\rm chaos}$. As for $v_*$, we computed it by varying $\lambda_L$ a little bit and then use $v_*=\Delta\lambda_L(q_1)/\Delta q_1$, where $\Delta\lambda_L=0.005\times 2\pi T$.

\subsection{Computing Entropy and Heat Capacity}

    The entropy $S$ is computed using Eq.~\eqref{eq:entropy} and the imaginary time Green's function. We expect there to be non-universal correction at the order of $t_0/U$ and $T/U$, so it is preferred to put $U$ as large as possible. However, we found that Eq.~\eqref{eq:entropy} converges poorly at small $T/U$ ratio. To solve this problem, we observed that both poor convergence and non-universal corrections overestimate $S$, so at a fixed $T/E_c$, we computed $S$ for various $U$ and took the minimum, as shown in Fig.~\ref{fig:entropy}.
    
    To compute the heat capacity $C$, we interpolated $S(T/E_c)$ using third-order spline method, and then performed numerical differentiation.

\section{Keldysh Formalism}\label{sec:keldysh_supp}
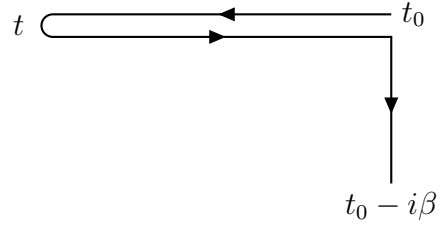
\begin{figure}[h]
\centering
\begin{tikzpicture}[scale=1.5]
\begin{scope}[thick, every node/.style={sloped,allow upside down}]
  \draw (0,0)--node{\midarrow}(-3,0) arc (90:270:1mm) --node{\midarrow} (0,-0.2) --node{\midarrow} (0,-1.5);
 \node at (0.2,0) {\large $t_0$};
 \node at (-3.3,-0.1) {\large $t$};
 \node at (0,-1.7) {\large $t_0-i\beta$};
\end{scope}
\end{tikzpicture}
\caption{\label{fig:keldysh_contour_appx} The time contour $C_\beta$ for Keldysh formalism.}
\end{figure}

    In Keldysh formalism, the time contour of path integral is doubled to have both a forward branch and a backward branch (see Fig.~\ref{fig:keldysh_contour_appx}). The action is thus $S_K=\int_{C_\beta}\rd t\rd^d x\mathcal{L}(\phi,\partial\phi)$, where the field $\phi$ has support over the whole contour $C_\beta$. Since we are only interested in equilibrium physics, we can send the initial time $t_0\to-\infty$ and decouple the imaginary branch. The action can thus be written in terms of fields on the $\pm$ branches as $S_K=S_0[\phi_+]-S_0[\phi_-]$, where $S_0$ is the original action and the minus sign is due to different orientations of time integration.

    We then discuss correlation functions in Keldysh formalism. Correlators computed from the Keldysh path integral are path ordered with respect to $C_\beta$. They are related to usual correlators in the following way:
\begin{equation}
\begin{split}
     \langle\mathcal{P}\phi_+(t_1)\phi_+(t_2)\rangle&=\langle\mathcal{T}\phi(t_1)\phi(t_2)\rangle,\\
     \langle\mathcal{P}\phi_+(t_1)\phi_-(t_2)\rangle&=\langle\phi(t_2)\phi(t_1)\rangle,\\
     \langle\mathcal{P}\phi_-(t_1)\phi_+(t_2)\rangle&=\langle\phi(t_1)\phi(t_2)\rangle,\\ \langle\mathcal{P}\phi_-(t_1)\phi_-(t_2)\rangle&=\langle\tilde{\mathcal{T}}\phi(t_1)\phi(t_2)\rangle,
\end{split}
\end{equation}where $\mathcal{T}$ and $\tilde{\mathcal{T}}$ are time ordering and anti-time ordering respectively.
    Note that the above four correlators are not independent because of the identity
\begin{equation}\label{eq:G_redundancy}
\begin{split}
    &\langle\mathcal{P}\phi_+(t_1)\phi_+(t_2)\rangle+\langle\mathcal{P}\phi_-(t_1)\phi_-(t_2)\rangle\\
   =&\langle\mathcal{P}\phi_+(t_1)\phi_-(t_2)\rangle+\langle\mathcal{P}\phi_-(t_1)\phi_+(t_2)\rangle.
\end{split}
\end{equation}
    Due to the above redundancy in $\pm$ notation, it is more convenient to organize fields in the ``classical-quantum" or ``$r-a$" notation, which is
\begin{equation}
    \phi_{r}=\frac{1}{\sqrt{2}}(\phi_++\phi_-),~\phi_{a}=\frac{1}{\sqrt{2}}(\phi_+-\phi_-).
\end{equation}
    We write correlators with the notation $G_{\alpha_1\dots \alpha_n}=\langle\phi_{\alpha_1}\dots \phi_{\alpha_n}\rangle$, where $\alpha_i=r,a$.
    There are three independent two-point functions, they are
\begin{equation}\label{eq:G_RAK}
\begin{split}
    G_R(t_1,t_2)&=-iG_{ra}(t_1,t_2)=\theta(t_{12})\langle[\phi(t_1),\phi(t_2)]_\mp\rangle,\\
    G_A(t_1,t_2)&=-iG_{ar}(t_1,t_2)=-\theta(t_{21})\langle[\phi(t_1),\phi(t_2)]_\mp\rangle,\\
    G_K(t_1,t_2)&=-iG_{rr}(t_1,t_2)=-i\langle[\phi(t_1),\phi(t_2)]_\pm\rangle,
\end{split}
\end{equation} where the subscripts on the RHS are for boson/fermion respectively. In Eq.~\eqref{eq:G_RAK}, $G_R(G_A)$ is the usual retarded (advanced) Green's function and $G_K$ is called Keldysh Green's function.

    The fourth combination $G_{aa}$ vanishes identically due to Eq.~\eqref{eq:G_redundancy}. This can be generalized to the so-called ``largest time equation"\cite{Defu}, which states that
\begin{equation}
    G_{a\dots}(t_1,\dots)=0,~\hbox{if $t_1$ is the largest time}.
\end{equation}

    In equilibrium, there is a further constraint on $G_R,G_A,G_K$, which is the fluctuation-dissipation theorem:
\begin{equation}\label{eq:FDT}
    G_K(\omega)=(G_R(\omega)-G_A(\omega))\times \left\{
                          \begin{array}{ll}
                            \coth({\omega}/(2T)), & \hbox{boson;} \\
                            \tanh({\omega}/(2T)), & \hbox{fermion,}
                          \end{array}
                        \right.
\end{equation}where $G(\omega)$ denotes the Fourier transform. 

\section{Feynman Rules for OTOC}\label{sec:frules}

    The retarded OTOC is calculated in the time contour Fig.~\ref{fig:keldysh_contour}. The calculation uses generalized Keldysh formalism with two time folds. In this section we try to understand the Feynman rules of Keldysh perturbation theory, especially $i$'s and $\pm$ signs. The rules are the following:
\begin{enumerate}
  \item Each diagram has two horizontal skeletons and the top (bottom) corresponds to fold 1 (fold 2). Bosonic OTOC has an overall minus sign.
  \item Horizontal right(left)-directing propagators are retarded(advanced). Vertical propagators are Wightman propagators.
  \item Each boson retarded, advanced and Keldysh propagator has a factor $i$. Each boson Wightman propagator has a factor of $2$.
  \item Each fermion retarded and Keldysh propagator has a factor $i$, and fermion advanced propagator has a factor $-i$. Each fermion Wightman propagator has a factor of $2$, and if the $c^\dagger$ in the propagator comes from fold 2 (see Fig.~\ref{fig:keldysh_contour}), an extra minus sign is added.
  \item Each interaction vertex has a factor of $i2^{1-n/2}\zeta$, where $n$ is the number of legs (the degree of the vertex). If the vertex involves fermions and sits at fold 2, $\zeta=-1$, otherwise $\zeta=1$.
  \item Each fermion loop has a minus sign.
\end{enumerate}
    The additional factors such as couplings and symmetry factors are the same as the usual perturbation theory.

    Next, we try to show the above rules by examples. We will focus on factors due to Keldysh formalism, and suppress others such as couplings and $N$ dependence. Since there are two time folds, we index the fields with the ``$r-a$'' notation and an extra fold index $i=1,2$, {\it i.e.}$\phi_1(t)=\phi(t)$, $\phi_2(t)=\phi(t-i\beta/2)$.

    We first look at boson OTOC, which is
\begin{equation}
\begin{split}\label{eq:h_appx}
    h(t)&\equiv\theta(t)\langle[\phi(t-i\beta/2),\phi^\dagger(-i\beta/2)]^\dagger[\phi(t),\phi^\dagger(0)]\rangle\\
        &=-\langle\phi_{r2}^\dagger(t)\phi_{a2}(0)\phi_{r1}(t)\phi_{a1}^\dagger(0)\rangle.
\end{split}
\end{equation}In the second line, we obtained the overall minus sign mentioned in rule 1.

    To zeroth order, we can use Wick theorem to evaluate Eq.\eqref{eq:h_appx},
\begin{equation}
  -\langle\phi_{r2}^\dagger(t)\phi_{a2}(0)\rangle\langle\phi_{r1}(t)\phi_{a1}^\dagger(0)\rangle
              =-(iD_A(-t))(iD_R(t)),
\end{equation}which is an example for rule 2 and 3.
    Next we consider what happens when we add interaction vertices. Suppose the original Lagrangian contains a term $$\mathcal{L}\supset\lambda\phi\phi^\dagger,$$ where $\lambda$ is another real scalar field. To get the corresponding term in the Keldysh action, we need to write fields in terms of the $r-a$ variables. In this process, we divide by $\sqrt{2}$ for each field in the vertex, but we also multiply by 2 when we add contributions from the $\pm$ branches, so we have obtain the factor mentioned in rule 5. The final result should be antisymmetric under $\phi_{\pm},\lambda_\pm\to\phi_{\mp},\lambda_{\mp}$, so it should contain odd number of $a$'s. For the example above, the result is
\begin{equation}
    \mathcal{L}_K\supset\frac{1}{\sqrt{2}}\sum_{i=1,2}
\lambda_{ia}\phi_{ir}\phi^\dagger_{ir}+
\lambda_{ir}\phi_{ia}\phi^\dagger_{ir}+
\lambda_{ir}\phi_{ir}\phi^\dagger_{ia}+
\lambda_{ia}\phi_{ia}\phi^\dagger_{ia}.
\end{equation}

    Then we insert one vertex to fold $1,2$ respectively to get the one-rung diagram for $h$: (fields with time argument are external legs)
\begin{equation}\label{eq:h1}
\begin{split}
-\langle\wick[offset=1.5em]{\c1 \phi_{r2}^\dagger(t)\frac{i}{\sqrt{2}}\c2 \lambda_{r2}\c1 \phi_{a2}\c1 \phi_{r2}^\dagger \c1 \phi_{a2}(0)
\c1 \phi_{r1}(t)\frac{i}{\sqrt{2}}\c2 \lambda_{r1}\c1 \phi_{a1}^\dagger\c1 \phi_{r1}\c1 \phi_{a1}^\dagger(0)}\rangle\\
    =-\frac{i^2}{2}(iD_R)(iD_R)(iD_A)(iD_A)(2D_W).
\end{split}
\end{equation} Here, each propagator in fold 1 comes as $\wick{\c\phi_{r1}\c\phi_{a1}^\dagger}=iD_R$, while each propagator in fold 2 comes as $\wick{\c\phi_{r2}^\dagger\c\phi_{a2}}=iD_A$. In Eq.\eqref{eq:h1}, there is a new type of contraction $\langle\lambda_{r2}\lambda_{r1}\rangle$, which evaluates to
\begin{equation}
    \langle\lambda_{r2}(t)\lambda_{r1}(0)\rangle=2\langle\lambda(t-i\beta/2)\lambda(0)\rangle=2D_W(t),
\end{equation}where we have used the definition of Wightman correlator \eqref{eq:GW_def}.

    Now we switch to the fermion case, and the goal is to understand the extra minus signs. The fermion OTOC is
\begin{equation}
\begin{split}
    f(t)=&\theta(t)\langle\{\psi(t-i\beta/2),\psi^\dagger(-i\beta/2)\}^\dagger
    \{\psi(t),\psi^\dagger(0)\}\rangle\\
    =&\langle\psi_{r2}^\dagger(t)\psi_{a2}(0)\psi_{r1}(t)\psi^\dagger_{a1}(0)\rangle.
\end{split}
\end{equation}
    Use $\langle\psi_r(t)\psi_a^\dagger(0)\rangle=iG_R(t)$, $\langle\psi_r^\dagger(t)\psi_a(0)\rangle=-iG_A(-t)$, we get to zeroth order
\begin{equation}
  (iG_R(t))(-iG_A(-t)).
\end{equation} We see that each fermion propagator in fold 2 appears as $\psi_r^\dagger\psi_a$, so they must be exchanged and we get a minus for $G_A$.

    We can repeat the same exercise at one-rung order, the difference from the boson case is the following:
    First, for each vertex in fold 2, it appears as $\psi^\dagger_{r2}\psi_{a2}$, and a typical contraction looks like
$$
    \wick{\c1 \psi^\dagger_{r2}(t)\c2\psi^\dagger_{r2}\c1\psi_{a2}\c2\psi_{a2}(0)},
$$ so we must switch $\psi,\psi^\dagger$ in the vertex, and this yields a minus sign for each fermion vertex in fold 2. Second, there are two ways to give rise to Wightman correlator:
$$
    (1):~\langle\mathcal{P}\psi_{r2}\psi^\dagger_{r1}\rangle,~(2):~\langle\mathcal{P}\psi_{r1}\psi_{r2}^\dagger\rangle.
$$ Here we have explicitly written out path ordering for emphasis. In the second case, the path ordering exchanges the two operators and generates a minus sign.

\section{Wightman Propagators}\label{sec:wightman}

    For a complex field $\mathcal{O}(t)$ (fermion or boson, and we suppress spatial indices), the Wightman propagator is defined as
\begin{equation}\label{eq:GW_def}
    G_W(t)\equiv\langle\mathcal{O}(t-i\beta/2)\mathcal{O^\dagger}(0)\rangle.
\end{equation}
    Here, we choose the two operators to be separated by $i\beta/2$ in imaginary time due to our computation scheme.
    
    In frequency space, $G_W$ can be conveniently expressed in terms of spectral functions as
\begin{equation}
    G_W(\omega)=
    \left\{
      \begin{array}{ll}
       \displaystyle \frac{A(\omega)}{2\cosh(\omega/2T)}, & \hbox{(fermions);} \\
        \displaystyle \frac{B(\omega)}{2\sinh(\omega/2T)}, & \hbox{(bosons).}
      \end{array}
    \right.
\end{equation}

\section{$U=0$ limit}\label{sec:U=0}

    The $U=0$ limit of the phonon model is exactly soluble, the Green's function is
\begin{equation}
G_R(\omega)=\frac{2}{\omega+i\sqrt{4(t_0^2+gTt_0)-(\omega+i0)^2}}.
\end{equation}

    The resistivity at low temperature is
\begin{equation}
    \rho=\frac{\pi}{2}\left(1+\frac{g T}{t_0}+\frac{\pi^2 T^2}{12t_0^2}\right).
\end{equation}

    The inverse of thermal conductivity is
\begin{equation}
    \frac{T}{\kappa_0}=\frac{3}{2\pi}\left(1+\frac{g T}{t_0}+\frac{7\pi^2 T^2}{20t_0^2}\right).
\end{equation}

\section{\label{sec:dimensionless}Dimensionless Form}

    In this section, we demonstrate that the physics of the phonon model is controlled by two dimensionless parameters: the temperature $T/E_c$ ($E_c=t_0^2/U$) and the phonon coupling $gt_0/U$.

    We start with the equation of motion in imaginary time, which reads
\begin{eqnarray}
    G^{-1}(\omega)&=&i \omega-\Sigma(\omega),\\
    \Sigma(\tau)&=&(t_0^2+g t_0 T) G(\tau)+U^2 G(\tau)^2 G(-\tau).
\end{eqnarray}
    Write various quantities in new units,
$$
    \omega=\bar{\omega}E_c,~\tau=\bar{\tau}/E_c,~G(\omega)=\bar{G}(\bar{\omega})/t_0,~\Sigma(\omega)=\bar{\Sigma}(\bar{\omega})t_0,
$$
    and we get
\begin{eqnarray}
    \bar{G}^{-1}(\bar{\omega})&=&i \bar{\omega}\frac{E_c}{t_0}-\bar{\Sigma}(\bar{\omega}),\\
    \bar{\Sigma}(\bar{\tau})&=&(1+\frac{g t_0}{U}\frac{T}{E_c}) \bar{G}(\bar{\tau})+\bar{G}(\bar{\tau})^2 \bar{G}(-\bar{\tau}).
\end{eqnarray}

    As a result, when $E_c \ll t_0$ the relevant dimensionless parameters are $g t_0/U$ and $T/E_c$. From the EoM it may seem that only their product is relevant, but the temperature also has influence through the Matsubara summation, whose step width is $\Delta\bar{\omega}=2\pi T/E_c$.

\bibliographystyle{apsrev4-1_custom}
\bibliography{SYK_phonon}
 
\end{document}